\documentstyle[12pt,psfig]{article}

\begin{document}
\pagenumbering{arabic}
\title{\bf Relativistic transport theory of $N$, $\Delta$ and $N^{*}$(1440)
 interacting through $\sigma$, $\omega$ and $\pi$ mesons \footnote{Supported
by DFG-Graduiertenkolleg Theoretische \& Experimentelle Schwerionenphysik,
GSI, BMBF, DFG, and A.v.Humboldt-Stiftung}} 
\author{Guangjun Mao, L.~Neise, H.~St\"{o}cker, and W.~Greiner  \\
\vspace{-0.5cm}
{\normalsize \it Institut f\"{u}r Theoretische Physik, 
J. W. Goethe-Universit\"{a}t}\\
{\normalsize \it D-60054 Frankfurt am Main, Germany} \\
Zhuxia Li \\
{\normalsize  \it Institute of Atomic Energy, P. O. Box 275(18),
Beijing 102413, P. R. China }}
\date{}
\maketitle
\begin{abstract}
\begin{sloppypar}
A self-consistent relativistic integral-differential equation of the 
Boltzmann-Uehling-Uhlenbeck-type  for
the $N^{*}$(1440) resonance is developed based on an effective Lagrangian
of baryons interacting through mesons. The closed
time-path Green's function technique and semi-classical,
quasi-particle and Born approximations are employed in the derivation. 
The non-equilibrium 
 RBUU-type  equation for the
$N^{*}$(1440) is consistent with that of nucleon's and delta's which we 
derived before. Thus, we obtain a set of coupled equations for the $N$,
$\Delta$ and $N^{*}$(1440) distribution functions. 
 All the $N^{*}$(1440)-relevant in-medium two-body scattering
cross sections within the $N$, $\Delta$ and $N^{*}$(1440) system are derived
from the same effective Lagrangian in addition to the mean field and presented
analytically, which can be directly used in the study of relativistic 
heavy-ion collisions. The theoretical prediction of the free $pp \rightarrow
 pp^{*}(1440)$ cross section is in good agreement with the experimental data.
We calculate the in-medium $N + N \rightarrow 
 N + N^{*}$, $N^{*} + N \rightarrow N + N$ and $N^{*} + N \rightarrow N^{*} + N$
 cross sections in cold nuclear matter  up to twice the nuclear
matter density. The influence of different choices of the $N^{*}N^{*}$ coupling 
strengths, which can not be obtained through fitting certain experimental data,
are discussed.
The results show that the density dependence of predicted
in-medium cross sections are sensitive to the $N^{*}N^{*}$ coupling strengths
used. An evident density dependence will appear when a large scalar 
coupling strength of ${\rm g}_{N^{*}N^{*}}^{\sigma}$ is assumed.

\end{sloppypar}
\bigskip
\noindent {\bf PACS} number(s): 24.10.Cn; 25.70.-z; 21.65.+f
\end{abstract}
\newcounter{cms}
\setlength{\unitlength}{1mm}
\newpage
\begin{center}
{\bf I. INTRODUCTION}
\end{center}
\begin{sloppypar}
One of the central aims of relativistic heavy-ion collisions is to study 
the nuclear equation of state (EOS) under extreme conditions of high
temperature and density \cite{Bau75,Hof76}.
It was recognized twenty years ago that particles emitted in the collisions 
contain important information about the equation of state  
of hot and dense nuclear matter \cite{Cha73,Sto78}.
Since most of the particles such as pion, kaon, dilepton, anti-proton, 
anti-kaon are mainly produced through resonances, the inclusion of resonance
degrees of freedom in transport theories is essential for any realistic
description of relativistic heavy-ion collisions.
Actually, resonances produced in energetic
heavy-ion collisions play as an energy-reservoir in the transport process and
have strong influence on the particle production. The importance of baryon 
resonances on the dynamics of relativistic heavy-ion collisions as well as
its effects on the particle production, especially at subthreshold energies,
have been studied extensively and stressed frequently in the literature
\cite{Hof94}-\cite{BALI94}. 
 Recent theoretical calculations \cite{Ehe93,Hof95} and experimental data 
\cite{Met93} indicated that a gradual transition to {\em resonance matter}
would occur in the collision zone at kinetic energy ranging from SIS 
($\sim$ 1AGeV) up to
AGS ($\sim$ 15AGeV). At an incident energy of 2 GeV/nucleon more than 30\% of 
the nucleons 
are excited to resonance states \cite{Ave94}. 
At intermediate- and high-energy range the most important 
baryonic resonances are $\Delta$(1232), $N^{*}$(1440) and $N^{*}$(1535). 
Theoretical models extended to describe relativistic heavy-ion collisions at 
this energy range should include these resonance degrees of freedom explicitly
and treat them self-consistently. The heart of the problem is to determine 
quantitatively all possible binary collisions relating to  resonances,
such as $N\Delta$, $\Delta\Delta$, $NN^{*}$(1440),
$NN^{*}$(1535) ... collisions.
Unfortunately, very little is known about resonance-relevant in-medium cross
sections in high-density nuclear matter since the experimental determination
of them is inaccessible yet. In most of the transport models they are assumed to
be equal to the free $NN$ scattering cross sections. Since the density changes
drastically in relativistic heavy-ion collisions the medium effects on the
two-body scattering cross sections might be
pronounced. This simple assumption on the resonance-relevant cross sections 
is quite doubtful and should be checked carefully. 

\end{sloppypar}
\begin{sloppypar}
Based on the different 
theoretical models some authors have studied the in-medium cross sections
for $NN \longrightarrow N\Delta$, $N\Delta \longrightarrow NN$ and
$N\Delta \longrightarrow N\Delta$ reactions \cite{Haa87,Ber88,Lee96}.
Different model calculations give rather different results and 
the quantitative estimation of the medium dependence of 
in-medium $\Delta$-relevant cross sections has not been clear yet. 
It is, therefore, very important to include theoretical predicted in-medium
cross sections in the transport model in order to see its effects on the
physical quantities which is now experimental available.
However, these in-medium cross sections, which are not calculated in the       
framework of transport theory,
have the disadvantage that they are 
inconsistent with the other ingredients of the transport model
when applied to the relativistic heavy-ion collision calculations,
and then will cause further uncertainty on the theoretical predications.

\end{sloppypar}
\begin{sloppypar}
On the other hand, the in-medium two-body scattering cross sections can be
studied within the framework of transport theory, i.e.,  
the  Boltzmann-Uehling-Uhlenbeck (BUU) equation. The preliminary version
of BUU-type transport equation was developed in semi-classical and non-relativistic
fashion \cite{Mol84,Kru85,Sto86,BerRep}. It was then extended to the
relativistic form (RBUU) \cite{Elz87,Ko87,Bla88}. The
 BUU/RBUU-type transport equation has been used  extensively by several groups
\cite{Mol84}-\cite{Cas90} to the study of  heavy-ion collisions
and turned out to be very successful.  For increasing kinetic energy,
it is highly desirable to develop a more general version of transport equation
which includes
the $\Delta$ as well as the higher-mass resonances self-consistently, both in
the mean field and in the collision term. With this in mind, several groups
have set out to provide a derivation of such transport equations 
\cite{Wan91,Sch89}.
In Refs. \cite{PRC96,PLB96} we have developed a self-consistent 
RBUU equation for the $\Delta$ distribution 
function within the same framework we have used for the nucleon's \cite{PRC94}-\cite{PRC97}.
In our approach, the $\Delta$ isobars are treated in essentially the same 
way as nucleons. Both mean field and collision term of $\Delta$'s RBUU
equation are derived from the same effective Lagrangian and presented 
analytically. The obtained in-medium cross sections with the $\Delta$
resonance are consistent with the other ingredients of the transport model.
Based on this approach we have studied the medium effects on all the 
$N\Delta$ and $\Delta\Delta$ scattering cross sections and its vice versa 
cross sections within the nucleon+$\Delta$ system. Considerable medium corrections
have been found on the cross sections of certain channels, such as 
$N+N \rightarrow N+\Delta$ and $N+N \rightarrow \Delta + \Delta$ scattering.
 Our results exhibited that in the intermediate- and high-energy
range the in-medium $\Delta$-relevant cross sections deviate substantially
from the Cugnon's parameterization \cite{Cug81} for the free $NN$ scattering cross section 
which is now commonly used in the transport model. It  would be
important to take the in-medium $\Delta$-relevant cross sections into account 
in the study of energetic relativistic heavy-ion collisions rather than replace
it with a free $NN$ scattering cross section. However, up to now, no  
investigation has been made for the in-medium $N^{*}$-relevant cross sections
(the free $N + N \rightarrow N + N^{*}(1440)$ scattering cross section
has been studied by S.~Huber and J.~Aichelin within the one-boson-exchange
model \cite{Hub94}).
While $N^{*}$(1535) is essential for the production of $\eta$-mesons 
\cite{Wol92,Ber94}, $N^{*}$(1440) was reported to  enjoy at least equally 
importance as $\Delta$(1232) for the production of anti-protons
at subthreshold energies \cite{BALI94}.

\end{sloppypar}
\begin{sloppypar}
It is the purpose of this paper to expand our theoretical framework to include
the $N^{*}$(1440) degree of freedom.
We will derive the self-consistent RBUU equation
for the $N^{*}$(1440) distribution function within the framework we have done
for the nucleon's and $\Delta$'s. Special attentions will be paid to the 
$N^{*}$(1440)-relevant in-medium cross sections. Through construction the collision
term of $N^{*}$(1440)'s RBUU equation we will give the analytically expressions
for calculating all the $N^{*}$(1440)-relevant in-medium two-body scattering
cross sections within
the $N$, $\Delta$ and $N^{*}$(1440) system. The presented cross sections are 
consistent with the other ingredients of the transport model and can be used
directly in the study of relativistic heavy-ion collisions. 
The organization of the paper is as follows: in Sect. II we give the model
Lagrangian and derive the RBUU equation for the $N^{*}$(1440) distribution
function by means of the closed time-path Green's function technique. In 
Sect. III we will  construct the collision term and present the analytical
expressions for the in-medium differential cross sections of different channels.
In Sect. IV we introduce the centroid $N^{*}$(1440) mass by taking into 
account the decay width of the $N^{*}$(1440) resonance. It is then used to
calculate the in-medium $N+N \rightarrow N+N^{*}$, $N^{*}+N \rightarrow N+N$
and $N^{*}+N \rightarrow N^{*}+N$ scattering cross sections. We present
numerical results in Sect. V. 
Finally, a summary and outlook are given in Sect. VI.

\end{sloppypar}
\begin{center}
{\bf II. RBUU-TYPE TRANSPORT EQUATION FOR THE $N^{*}$(1440) DISTRIBUTION
 FUNCTION}
\end{center}
\begin{sloppypar}
In this section we will derive the self-consistent RBUU equation for the 
$N^{*}$(1440) distribution function. For consistency with the RBUU equations
of a nucleon and a delta, here we still use the closed time-path Green's
function technique. For a review of the technique we refer to Refs. 
\cite{Dan84,Cho85}.
First of all, let us write down the total effective Lagrangian used in the
model. In the framework of quantum hadrodynamics \cite{Ser86}, the baryon-baryon
interaction is described by the exchange of $\sigma$, $\omega$, $\pi$ and
$\rho$ mesons. While it is well known that the mean field is mainly 
contributed from the $\sigma$ and $\omega$ mesons, some inelastic reaction
channels relating to $\Delta$ production (thus, requiring charge exchange)
can only be described by the exchange of a $\pi$ or $\rho$ meson. However, it has
been found that the pion exchange processes dominate the cross sections of
single-$\Delta$ and double-$\Delta$ production from $NN$ scattering at 
intermediate- and high-energy range which we are interested in, the contribution
of the $\rho$-meson exchange is almost negligible \cite{PRC97,Eng96}. We
think the situation should not be changed substantially while $N^{*}$(1440)
is involved.
 Therefore, in the
following derivation, we take the effective Lagrangian which considers
the $N$, $\Delta$ and $N^{*}(1440)$ system interacting through $\sigma$,
$\omega$ and $\pi$ mesons. The Lagrangian density can be written as

\end{sloppypar}
\begin{equation}
{\cal L}={\cal L}_{\rm F}+{\cal L}_{\rm I}.
\end{equation}
Here ${\cal L}_{F}$ is the Lagrangian density for free nucleon, delta,
$N^{*}$(1440) resonance and meson fields
\begin{eqnarray}
{\cal L}_{\rm F}&=&\bar{\psi}[i\gamma_{\mu}\partial^{\mu}-M_{N}]\psi
+ \bar{\psi}^{*}[i\gamma_{\mu}\partial^{\mu}-M_{N^{*}}]\psi^{*} \nonumber \\
&& +\bar{\psi}_{\Delta \nu}[i\gamma_{\mu}\partial^{\mu}-M_{\Delta}]
\psi^{\nu}_{\Delta} \nonumber + \frac{1}{2}
\partial_{\mu}\sigma\partial^{\mu}\sigma-U(\sigma) \nonumber \\
&& -\frac{1}{4}\omega_{\mu\nu}\omega^{\mu\nu}+U(\omega) 
+ \frac{1}{2}(\partial_{\mu} \mbox{\boldmath $\pi$} \partial^{\mu}
\mbox{\boldmath $\pi$}
-m_{\pi}^{2}\mbox{\boldmath $\pi$}^{2})
    \end{eqnarray}
and U($\sigma$), U($\omega$) are the self-interaction part of the scalar field
\cite{Bog83} and vector field \cite{Bod91,Sug94}
\begin{eqnarray}
  && U(\sigma)=
   \frac{1}{2}m_{\sigma}^{2}\sigma^{2}+\frac{1}{3}b({\rm g}_{NN}^{\sigma}
\sigma)^{3}+\frac{1}{4}c({\rm g}_{NN}^{\sigma}\sigma)^{4}, \\
  && U(\omega)=\frac{1}{2}m_{\omega}^{2}\omega_{\mu}\omega^{\mu}
     (1+\frac{({\rm g}_{NN}^{\omega})^{2}}{2}\frac{\omega_{\mu}\omega^{\mu}}
     {Z^{2}}),
   \end{eqnarray}
respectively.  ${\cal L}_{I}$ is the interaction Lagrangian density
\begin{eqnarray}
 {\cal L}_{I}&=&{\cal L}_{NN}+{\cal L}_{N^{*}N^{*}}
 +{\cal L}_{\Delta \Delta}+{\cal L}_{NN^{*}}+{\cal L}_{\Delta N} 
 +{\cal L}_{\Delta N^{*}} \nonumber \\
     &=&{\rm g}^{\sigma}_{NN}\bar{\psi}(x)\psi(x)\sigma(x)
      - {\rm g}^{\omega}_{NN}\bar{\psi}(x)\gamma
  _{\mu}\psi(x)\omega^{\mu}(x)
   +{\rm g}_{NN}^{\pi}\bar{\psi}(x)\gamma_{\mu}\gamma_{5}\mbox{\boldmath $\tau$}
\cdot \psi(x)\partial^{\mu}\mbox{\boldmath $\pi$}(x) \nonumber \\
     &&+{\rm g}^{\sigma}_{N^{*}N^{*}}\bar{\psi}^{*}(x)\psi^{*}(x)\sigma(x)
      - {\rm g}^{\omega}_{N^{*}N^{*}}\bar{\psi}^{*}(x)\gamma
  _{\mu}\psi^{*}(x)\omega^{\mu}(x)
   +{\rm g}_{N^{*}N^{*}}^{\pi}\bar{\psi}^{*}(x)\gamma_{\mu}\gamma_{5}\mbox{\boldmath $\tau$}
\cdot \psi^{*}(x)\partial^{\mu}\mbox{\boldmath $\pi$}(x) \nonumber \\
 &&+{\rm g}^{\sigma}_{\Delta \Delta}
\bar{\psi}_{\Delta \nu}(x)\psi^{\nu}_{\Delta}(x)\sigma(x)
     -{\rm g}^{\omega}_{\Delta \Delta}\bar{\psi}_{\Delta \nu}(x)\gamma
  _{\mu}\psi^{\nu}_{\Delta}(x)\omega^{\mu}(x)
 +{\rm g}_{\Delta \Delta}^{\pi}\bar{\psi}_{\Delta\nu}(x)\gamma_{\mu}\gamma_{5} {\bf T}
 \cdot \psi^{\nu}_{\Delta}(x) \partial^{\mu}\mbox{\boldmath $\pi$}(x) \nonumber \\
     &&+\lbrack {\rm g}^{\sigma}_{NN^{*}}\bar{\psi}^{*}(x)\psi(x)\sigma(x)
      - {\rm g}^{\omega}_{NN^{*}}\bar{\psi}^{*}(x)\gamma
  _{\mu}\psi(x)\omega^{\mu}(x)
+{\rm g}_{NN^{*}}^{\pi}\bar{\psi}^{*}(x)\gamma_{\mu}\gamma_{5}\mbox{\boldmath $\tau$}
\cdot \psi(x)\partial^{\mu}\mbox{\boldmath $\pi$}(x) \nonumber \\
 &&- {\rm g}_{\Delta N}^{\pi}\bar{\psi}_{\Delta \mu}(x)
 \partial^{\mu}\mbox{\boldmath $\pi$}(x) \cdot {\bf S}^{+}\psi(x)
 - {\rm g}_{\Delta N^{*}}^{\pi}\bar{\psi}_{\Delta \mu}(x)
 \partial^{\mu}\mbox{\boldmath $\pi$}(x) \cdot {\bf S}^{+}\psi^{*}(x) +H.c. \rbrack
 \nonumber \\
 &=& {\rm g}_{NN}^{A}\bar{\psi}(x)\Gamma_{A}^{N} \psi(x) \Phi_{A}(x)
 +{\rm g}_{N^{*}N^{*}}^{A}\bar{\psi}^{*}(x)\Gamma_{A}^{N^{*}} \psi^{*}(x) \Phi_{A}(x)
 + {\rm g}_{\Delta \Delta}^{A} \bar{\psi}_{\Delta \nu}(x)\Gamma_{A}^{\Delta}\psi
 ^{\nu}_{\Delta}(x) \Phi_{A}(x) \nonumber \\
 && +\lbrack {\rm g}_{NN^{*}}^{A}\bar{\psi}^{*}(x)\Gamma_{A}^{N^{*}} \psi(x) \Phi_{A}(x)
 - {\rm g}_{\Delta N}^{\pi}\bar{\psi}_{\Delta \mu}(x)
 \partial^{\mu}\mbox{\boldmath $\pi$}(x) \cdot {\bf S}^{+}\psi(x) \nonumber \\
  && - {\rm g}_{\Delta N^{*}}^{\pi}\bar{\psi}_{\Delta \mu}(x)
 \partial^{\mu}\mbox{\boldmath $\pi$}(x) \cdot {\bf S}^{+}\psi^{*}(x) 
  + h.\, c. \rbrack
   \end{eqnarray}
where $\psi$, $\psi^{*}$ are the Dirac spinors of the nucleon and $N^{*}$(1440)
, and $\psi_{\Delta \mu}$ is the Rarita-Schwinger spinor of the
$\Delta$-baryon. $\mbox{\boldmath $\tau$}$ is the isospin operator of the nucleon and $N^{*}$(1440),
${\bf T}$ is the isospin operator of the $\Delta$, and ${\bf S}^{+}$ is the
isospin transition operator between the isospin 1/2 and 3/2 fields. 
${\rm g}_{NN}^{\pi}=f_{\pi}/m_{\pi}$,
 ${\rm g}_{\Delta N}^{\pi}=f^{*}/m_{\pi}$;
 $\Gamma_{A}^{N}=\Gamma_{A}^{N^{*}}=
 \gamma_{A}\tau_{A}$, $\Gamma_{A}^{\Delta}=\gamma_{A}T_{A}$,
  A=$\sigma$, $\omega$, $\pi$,
 the symbols and notation are defined in Table I.
  \begin{center}
  \fbox{Table I}
  \end{center}

  \begin{sloppypar}
In the language of the closed time-path Green's function technique the 
$N^{*}$(1440) Green's function in the interaction picture                 
 can be defined in the same way as for nucleon's by
 \begin{equation}
iG_{N^{*}}(1,2)=<T[exp(-i \not \!\!\int\,dx H_{I}(x))\psi^{*}(1)\bar{\psi}^{*}(2)]> .
 \end{equation}
Here T is the time ordering operator defined on a time contour. 
The corresponding Dyson equation for $iG_{N^{*}}(1,2)$ can be written as
   \begin{equation}
 iG_{N^{*}}(1,2)=iG_{N^{*}}^{0}(1,2)+\not\!\! \int\,dx_{3} \not\!\! \int\,dx_{4}G_{N^{*}}^{0}(1,4)\Sigma_{N^{*}}(4,3)iG_{N^{*}}(3,2).
     \end{equation}
Here $G_{N^{*}}^{0}(1,2)$ is
the zeroth-order Green's function of $N^{*}$(1440), which 
is similar to that of the nucleon's
zeroth-order Green's function \cite{PRC96,PRC94}.
$G^{0}_{N^{*}}(1,2)$ can be written as
\begin{eqnarray}
&& iG^{0}_{N^{*}}(1,2)=i\int\frac{d^{4}k}{(2\pi)^{4}}G^{0}_{N^{*}}(x,k)e^{-ik(x_{1}-x_{2})}, \\
&& G^{0 \mp \mp}_{N^{*}}(x,k)=(\not\!k+M_{N^{*}}) \left[ \frac{\pm 1}{k^{2}-M_{N^{*}}^{2}\pm i
\epsilon}
 + \frac{\pi i}{E(k)}\delta(k_{0}-E(k))f_{N^{*}}(x,k) \right] , \\
&& G^{0+-}_{N^{*}}(x,k)=-\frac{\pi i}{E(k)}\delta(k_{0}-E(k))[1-f_{N^{*}}(x,k)](\not\! k+M_{N^{*}}),
 \\
 && G^{0-+}_{N^{*}}(x,k)=\frac{\pi i}{E(k)}\delta(k_{0}-E(k))f_{N^{*}}(x,k)(\not\! k+M_{N^{*}}).
   \end{eqnarray}
$f_{N^{*}}(x,k)$ is the distribution function of $N^{*}$(1440).
As in most/all presently used RBUU-type transport models, also here we do not 
take into account
the temperature degree of freedom. Furthermore,
in our theoretical framework the negative-energy states are neglected.
$\Sigma_{N^{*}}(4,3)$  is the $N^{*}$ self-energy. The lowest-order 
self-energies contributing to the collision term come from the Born diagrams.
Through considering the $N^{*}$ self-energy up to the Born approximation and 
adopting the semi-classical approximation 
 (in which the Green's functions and self-energies are assumed to be peaked
 around relative coordinate and smoothly changing with center-of-mass
 coordinate) 
 and quasi-particle approximations (in which we dress the mass and momentum
 in the zeroth-order Green's functions appearing in the self-energy terms
 with the effective mass and 
 momentum) the self-consistent
RBUU equation for the $N^{*}$(1440) can be derived in the same way as that
of the nucleons.
The only difference between the nucleon and the $N^{*}$(1440) is the 
mass and the coupling strengths! Here the self-consistency means that we
derive both the mean field and collision term of the transport equation simultaneously
from the same effective Lagrangian. The RBUU equation for the $N^{*}$(1440)
distribution function reads 
 \begin{eqnarray}
&&\lbrace p_{\mu} [
 \partial^{\mu}_{x}-\partial^{\mu}_{x}\Sigma_{N^{*}}
^{\nu}(x) \partial_{\nu}^{p}+\partial_{x}^{\nu}\Sigma_{N^{*}}^{\mu}(x)\partial
_{\nu}^{p} ] +m^{*}_{N^{*}}\partial^{\nu}_{x}\Sigma_{N^{*}}^{S}(x)\partial_{\nu}^{p}
\rbrace \frac{f_{N^{*}}({\bf x}, {\bf p}, \tau)}{E^{*}_{N^{*}}(p)}
 \nonumber \\
 && = C^{N^{*}}(x,p).
 \end{eqnarray}
The left-hand side of Eq. (12) is the transport part and the right-hand side
is the collision term. Here we have dropped the contribution of the Fock term,
since it usually has only small effects on the mean field. 
The above equation is derived within the framework as we have done for 
the nucleon's \cite{PRC94}-\cite{PRC97} 
 \begin{eqnarray}
&&\lbrace p_{\mu} [
 \partial^{\mu}_{x}-\partial^{\mu}_{x}\Sigma
^{\nu}(x) \partial_{\nu}^{p}+\partial_{x}^{\nu}\Sigma^{\mu}(x)\partial
_{\nu}^{p} ] +m^{*}\partial^{\nu}_{x}\Sigma^{S}(x)\partial_{\nu}^{p}
\rbrace \frac{f({\bf x}, {\bf p}, \tau)}{E^{*}(p)}
 \nonumber \\
 && = C(x,p).
 \end{eqnarray}
and delta's \cite{PRC96,PLB96}
 \begin{eqnarray}
&&\lbrace p_{\mu} [
 \partial^{\mu}_{x}-\partial^{\mu}_{x}\Sigma_{\Delta}
^{\nu}(x) \partial_{\nu}^{p}+\partial_{x}^{\nu}\Sigma_{\Delta}^{\mu}(x)\partial
_{\nu}^{p} ] +m^{*}_{\Delta}\partial^{\nu}_{x}\Sigma_{\Delta}^{S}(x)\partial_{\nu}^{p}
\rbrace \frac{f_{\Delta}({\bf x}, {\bf p}, \tau)}{E^{*}_{\Delta}(p)}
 \nonumber \\
 && = C^{\Delta}(x,p).
 \end{eqnarray}
RBUU equations. Therefore, Eqs. (12), (13) and (14) stand in a consistent 
form and they are coupled together through the mean field and collision
term(i.e., in-medium scattering cross sections of different channels).  
The $\Sigma^{S}_{N^{*}}(x)$ and $\Sigma^{\mu}_{N^{*}}(x)$, which are the Hartree
terms of the scalar and vector $N^{*}$(1440) self-energies, can be written as
 \begin{eqnarray}
&&\Sigma_{N^{*}}^{S}(x)=-\frac{{\rm g}_{N^{*} N^{*}}^{\sigma}}{m_{\sigma}^{2}}
[ {\rm g}_{N N}^{\sigma}\rho_{S}(N)+{\rm g}_{N^{*} N^{*}}^{\sigma}\rho_{S}
(N^{*}) + {\rm g}_{\Delta \Delta}^{\sigma}\rho_{S}(\Delta)  ],\\
&&\Sigma_{N^{*}}^{\mu}(x)=\frac{{\rm g}_{N^{*} N^{*}}^{\omega}}{m_{\omega}^{2}}
 [{\rm g}_{N N}^{\omega}\rho_{V}^{\mu}(N) + {\rm g}_{N^{*} N^{*}}^{\omega}
 \rho_{V}^{\mu}(N^{*}) +
 {\rm g}_{\Delta \Delta}^{\omega}\rho_{V}^{\mu}(\Delta) ].
 \end{eqnarray}
The effective four momentum and effective mass of the $N^{*}$(1440) are defined as 
  \begin{eqnarray}
    &&m^{*}_{N^{*}}(x)=M_{N^{*}}+\Sigma^{S}_{N^{*}}(x) \\
    &&p^{\mu}(x)=P^{\mu} - \Sigma^{\mu}_{N^{*}}(x).
  \end{eqnarray}
If one takes into account the self-interaction of the $\sigma$, $\omega$ fields
given  in Eqs. (3) and (4), the Eqs. (15) $-$ (18) should be rewritten 
by means of  the field equations of the $\sigma$ and $\omega$ mesons within the
local density approximation 
 \begin{eqnarray}
 m_{\sigma}^{2}\sigma(x)+b({\rm g}^{\sigma}_{N N})^{3}\sigma^{2}(x)
+c({\rm g}^{\sigma}_{N N})^{4}\sigma^{3}(x)={\rm g}^{\sigma}_{N N}\rho_{S}(N)
+{\rm g}^{\sigma}_{N^{*} N^{*}}\rho_{S}(N^{*}) +{\rm g}^{\sigma}_{\Delta \Delta}\rho_{S}(\Delta), \\
 m_{\omega}^{2}\omega^{\mu}(x)+\frac{({\rm g}_{N N}^{\omega})^{2}m_{\omega}^{2}}
{Z^{2}}(\omega^{\mu}(x))^{3}
={\rm g}^{\omega}_{N N}\rho_{V}^{\mu}(N)+{\rm g}^{\omega}_{N^{*} N^{*}}\rho_{V}^{\mu}
 (N^{*}) + {\rm g}^{\omega}_{\Delta \Delta}\rho_{V}^{\mu}(\Delta),
 \end{eqnarray}
and then
  \begin{eqnarray}
    &&m^{*}_{N^{*}}(x)=M_{N^{*}} - {\rm g}^{\sigma}_{N^{*} N^{*}}\sigma(x) \\
    &&p^{\mu}(x)=P^{\mu} - {\rm g}^{\omega}_{N^{*} N^{*}} \omega^{\mu}(x).
  \end{eqnarray}
Here $\rho_{S}$(i) and $\rho_{V}^{\mu}$(i) are the scalar and vector densities
of the nucleon, $N^{*}$(1440) and delta:
 \begin{eqnarray}
&& \rho_{S}(i)=\frac{\gamma(i)}{(2 \pi)^{3}}\int d{\bf q} \frac{m^{*}_{i}}
 {\sqrt{{\bf q}^{2}+m^{*2}_{i}}} f_{i}({\bf x},{\bf q},\tau), \\
&& \rho_{V}^{\mu}(i)=\frac{\gamma(i)}{(2 \pi)^{3}}\int d{\bf q} \frac{q^{\mu}}
 {\sqrt{{\bf q}^{2}+m^{*2}_{i}}} f_{i}({\bf x},{\bf q},\tau).
 \end{eqnarray}
The abbreviations i=N, $N^{*}$, $\Delta$, and $\gamma$(i)= 4, 4, 16, correspond 
 to nucleon, $N^{*}$(1440) and delta, respectively. Eqs. (23) and (24)  
are calculated under the no-sea approximation, i.e., we drop the contribution
of negative-energy states. The effective four-momenta and effective masses of 
nucleon and delta can be defined through substituting the appropriate nucleon
and delta labels in Eqs. (21) and (22), respectively. The corresponding 
Feynman diagrams are given in Fig. 1.

  \end{sloppypar}
  \begin{center}
  \fbox{Fig. 1}
   \end{center}
  \begin{sloppypar}
The collision term can be expressed according to the transition probability,
which reads as
   \begin{eqnarray}
  C^{N^{*}}(x,p)&=&\frac{1}{2}\int \frac{d^{3}p_{2}}{(2\pi)^{3}}
 \int\frac{d^{3}p_{3}}{(2\pi)^{3}} \int\frac{d^{3}p_{4}}{(2\pi)^{3}}
 (2\pi)^{4}\delta^{(4)}(p+p_{2}-p_{3}-p_{4})\nonumber \\
 && \times W^{N^{*}}(p,\,p_{2},\,p_{3},\,p_{4})
 (F_{2}-F_{1}),
    \end{eqnarray}
where $F_{2}$, $F_{1}$ are the Uehling-Uhlenbeck factors of the gain 
($F_{2}$) and loss ($F_{1}$) terms, respectively:
    \begin{eqnarray}
&& F_{2}=[1-f_{N^{*}}({\bf x},{\bf p},\tau)][1-f_{B_{2}}({\bf x},{\bf p}_{2},\tau)]
   f_{B_{3}}({\bf x},{\bf p}_{3},\tau)f_{B_{4}}({\bf x},{\bf p}_{4},\tau), \\
&&  F_{1}=f_{N^{*}}({\bf x},{\bf p},\tau)f_{B_{2}}({\bf x},{\bf p}_{2},\tau)
   [1-f_{B_{3}}({\bf x},{\bf p}_{3},\tau)]
   [1-f_{B_{4}}({\bf x},{\bf p}_{4},\tau)] , 
    \end{eqnarray}
$B_{2}$, $B_{3}$, $B_{4}$ can be $N$, $\Delta$ and $N^{*}$(1440).
$W^{N^{*}}(p,\,p_{2},\,p_{3},\,p_{4})$  is the transition
 probability of different channels, which has the form
  \begin{equation}
 W^{N^{*}}(p,\,p_{2},\,p_{3}\,,p_{4})=
  \frac{1}{16E^{*}_{N^{*}}(p)E^{*}_{B_{2}}(p_{2})E^{*}_{B_{3}}(p_{3})
  E^{*}_{B_{4}}(p_{4})}\sum_{AB} (T_{D}\Phi_{D} - T_{E}\Phi_{E})
   +p_{3} \longleftrightarrow p_{4}.
  \end{equation}
Here $T_{D}$, $T_{E}$ are the isospin matrices and $\Phi_{D}$, $\Phi_{E}$
are the spin matrices, respectively.
 D denotes the contribution of the direct diagrams and E is that of the exchange
diagrams. $A$, $B=\sigma$, $\omega$, $\pi$ represent the contributions of 
different mesons. The exchange of $p_{3}$ and $p_{4}$ is only for the 
case of identical particles in the final state. 
 The two-body scattering reactions relevant to the $N^{*}$(1440) 
in the $N$, $\Delta$ and $N^{*}$(1440) system are follows: \\
 \indent (1) Elastic reactions: \\
\mbox{} \hspace{1.0cm} $NN^{*} \longrightarrow NN^{*}$, \hspace{1.0cm}
  $\Delta N^{*} \longrightarrow \Delta N^{*}$, \hspace{1.0cm} 
  $N^{*}N^{*} \longrightarrow N^{*} N^{*}$ .   \\
  \indent (2) Inelastic reactions:  \\
 \mbox{} \hspace{1.0cm} $N N \longleftrightarrow NN^{*}$, \hspace{1.0cm} 
  $N \Delta \longleftrightarrow N N^{*}$, \hspace{1.0cm} 
  $\Delta \Delta \longleftrightarrow N N^{*}$, \\
 \mbox{} \hspace{1.0cm} $N N^{*} \longleftrightarrow \Delta N^{*}$, \hspace{0.9cm} 
  $N N^{*} \longleftrightarrow N^{*} N^{*}$, \hspace{0.6cm} 
  $N N \longleftrightarrow \Delta N^{*}$, \\
 \mbox{} \hspace{1.0cm} $N \Delta \longleftrightarrow \Delta N^{*}$,
\hspace{1.0cm} $\Delta \Delta \longleftrightarrow \Delta N^{*}$, \hspace{1.0cm} 
  $N^{*} N^{*} \longleftrightarrow \Delta N^{*}$ , \\
 \mbox{} \hspace{1.0cm} $N N \longleftrightarrow N^{*}N^{*}$, \hspace{0.8cm} 
  $N \Delta \longleftrightarrow N^{*}N^{*}$, \hspace{0.8cm}
  $\Delta \Delta \longleftrightarrow N^{*}N^{*}$. \\
\noindent In the next section we will derive the analytical expressions
for the differential cross sections of the above reaction channels through
calculating the concrete forms of transition probability from the Born term
of the $N^{*}$(1440) self-energies. For the inelastic case    we only 
calculate the $N^{*}$(1440)-incident cross sections, its vice versa cross
sections can be obtained by means of the detailed balance \cite{Dan91}.
The corresponding Feynman diagrams of the Born term of   
the $N^{*}$(1440) self-energies contributing to the different          
reaction channels are given in Appendix A.

  \end{sloppypar} 
 \begin{center}
{\bf III. DERIVATION OF THE COLLISION TERM}
 \end{center}
  \begin{sloppypar}
Before coming to the calculation of the Born diagrams given in Appendix A, we 
firstly see
if some of them are already at hand according to our previous works on the nucleon
and delta scattering cross sections. Because $N^{*}$(1440) has the same coupling
form as that of nucleon and the only difference is the mass and the coupling strengths, the differential cross section of the $N^{*} \Delta
\rightarrow N^{*}\Delta$ reaction should be analogous to that of 
the $N \Delta \rightarrow
N \Delta$ channel given in Ref. \cite{PRC96} except for replacing the effective
  mass of
nucleon with that of the $N^{*}$(1440) and all the $N$ labels on the coupling 
strengths with $N^{*}$ labels. The same arguments apply to the $N^{*}N^{*}
\rightarrow N^{*}N^{*}$, $N^{*}\Delta \rightarrow \Delta \Delta$, $N^{*}\Delta
\rightarrow N^{*}N^{*}$ and $N^{*}N^{*} \rightarrow \Delta \Delta$ reactions.
The corresponding cross sections can be obtained from the $NN \rightarrow NN$
\cite{ZPA94}, $N \Delta \rightarrow \Delta\Delta$ \cite{PRC96},
$N \Delta \rightarrow NN$ \cite{PRC94} and $NN \rightarrow \Delta\Delta$
\cite{PRC97} scattering cross sections.
 However, for the $NN$ elastic scattering 
cross section only the contribution of $\sigma$ and $\omega$ mesons are taken
into account in Ref. \cite{ZPA94}. For completeness, in Appendix C we give the
differential cross section of $NN$ elastic scattering including the contribution
of pion. 
Using the same arguments, the differential cross section of the $NN^{*}
 \rightarrow N^{*} N^{*}$ reaction can be obtained from that of the $N^{*}N
\rightarrow NN$ reaction by the exchange of $m^{*} \leftrightarrow 
m^{*}_{N^{*}}$ and the labels $N \leftrightarrow N^{*}$ on the coupling 
strengths, and
so on for the $N^{*} \Delta \rightarrow NN$ and $N \Delta 
\rightarrow N^{*}N^{*}$ channels.
In the calculations of the $N\Delta
\rightarrow N\Delta$ \cite{PRC96} and $NN \rightarrow \Delta\Delta$ 
\cite{PRC97} cross sections we have found that the contribution of the exchange
terms is negligible. The situation should not be changed substantially when
the $N^{*}$(1440) is relevant. Therefore, we drop the exchange terms in the 
following derivation of the $N^{*} \Delta \rightarrow N \Delta$ and $N^{*}N
\rightarrow \Delta\Delta$ differential cross sections. In the other reaction
channels the exchange terms are taken into account.

  \end{sloppypar}
 \begin{sloppypar}
Now let us turn to calculate the spin and isospin matrices in Eq. (28).
Firstly, we consider the isospin factors $T_{D}$ and $T_{E}$. Here
we assume that the incident-$N^{*}$(1440) has the specific isospin and account
for the isospins of the other three particles, which is consistent with the fact that
the RBUU equations describe coupled single-particle distribution functions. In 
Eq. (12) we have averaged over isospin, i.e., $N^{*}$(0) and $N^{*}(+)$. Since $N^{*}$(1440) has
the same isospin couplings as the nucleon, we can write down $T_{D}$ and 
$T_{E}$ directly based on our previous works. The isospin factors for  the
$N^{*}N \rightarrow N^{*}N$, $N^{*}N \rightarrow NN$ and $N^{*}N^{*} 
\rightarrow NN$ reactions are given in Table II and III. For the $N^{*}N
\rightarrow N\Delta$, $N^{*}N \rightarrow \Delta N^{*}$ and $N^{*}N^{*}
\rightarrow N\Delta$ reactions only the pion contributes to the cross 
sections due to the charge conservation, and we have 
$T_{D}^{\pi}=T_{E}^{\pi}=4$. For the $N^{*}N \rightarrow
\Delta\Delta$ reaction we only consider the direct term and $T_{D}^{\pi}
=8/3$. For the $N^{*} \Delta \rightarrow N \Delta$ reaction the isospin factors
corresponding to Feynman diagram (j1) in Appendix A
are given in Table IV. In diagram (j3)
only the pion enters, we have $T_{D}^{\pi}=8/3$. The concrete expressions
for the spin matrices $\Phi_{D}$ and $\Phi_{E}$ are given in Appendix B.

 \end{sloppypar}
 \begin{center}
 \fbox{Table II} \hspace{2cm} \fbox{Table III} \hspace{2cm} \fbox{Table IV}
 \end{center}
 \begin{sloppypar}
By means of the relation between the transition probability $W^{N^{*}}(p, p_{2},
p_{3}, p_{4})$ and the differential cross section \cite{Gro80}, Eq. (25)
can be rewritten as 
  \begin{equation}
 C^{N^{*}}(x,p)=\frac{1}{2}\int \, \frac{d^{3}p_{2}}{(2\pi)^{3}}
 \upsilon \sigma_{N^{*}}(s,t)(F_{2}-F_{1})d\,\Omega.
  \end{equation}
Here $\upsilon$ is the M$\not\!\! o$ller velocity, $ \sigma_{N^{*}}(s,t)$ is the
differential cross section of different $N^{*}$-incident channels. By 
evaluating the Eqs. (B1)$-$(B24) and finally transforming it into a center of
mass system we have obtained the analytical expressions for $\sigma_{N^{*}}
(s,t)$. The in-medium $N^{*}$-incident elastic
and inelastic scattering cross sections can be calculated through the following
equations:
  \begin{eqnarray}
  && \sigma^{*}_{N^{*}N \rightarrow B_{3}B_{4}}= \frac{1}
 {8(1+\delta_{B_{3}B_{4}})}
  \int\, \sigma_{N^{*}N \rightarrow B_{3}B_{4}}(s,t) d \Omega, \\
  && \sigma^{*}_{N^{*}\Delta \rightarrow B_{3}B_{4}}= \frac{1}
 {32(1+\delta_{B_{3}B_{4}})}
  \int\, \sigma_{N^{*}\Delta \rightarrow B_{3}B_{4}}(s,t) d \Omega,\\
  && \sigma^{*}_{N^{*}N^{*} \rightarrow B_{3}B_{4}}= \frac{1}
 {8(1+\delta_{B_{3}B_{4}})}
  \int\, \sigma_{N^{*}N^{*} \rightarrow B_{3}B_{4}}(s,t) d \Omega, 
  \end{eqnarray}
where $B_{3}$, $B_{4}$ are $N$, $\Delta$ and $N^{*}$(1440). The explicit 
expressions of $\sigma_{N^{*}N \rightarrow B_{3} B_{4}}$(s,t),
$\sigma_{N^{*}\Delta \rightarrow B_{3} B_{4}}$(s,t),
and  $\sigma_{N^{*}N^{*} \rightarrow B_{3} B_{4}}$(s,t) are given in Appendix D.

 \end{sloppypar}
 \begin{center}
{\bf IV. THE CENTROID $N^{*}$(1440) MASS, COUPLING STRENGTHS AND FORM FACTORS}
  \end{center}
  \begin{sloppypar}
In quantum field theory all baryons are treated
as elementary particles as we have done in the above derivation.
But the delta and $N^{*}$(1440)
are physically decay particles. It has been pointed out that 
the wide decay widths of resonances  have strong
influence on the resonance-relevant cross sections and should be taken into
account \cite{Wol92,Dan91}, which  can be realized by introducing a distribution
function of Breit-Wigner form in the collision term 
\cite{PRC97}. However, an exact treatment of Breit-Wigner
function will cause difficulty in the derivation of in-medium cross sections
when more than one resonance is relevant. Alternatively, we introduce a 
centroid $N^{*}$(1440) mass $\langle M_{N^{*}} \rangle$ in the same way as 
we did for the delta \cite{PRC94,PRC97}, which can take into account the decay
width of resonance effectively. $\langle M_{N^{*}} \rangle $ is defined as
  \begin{equation}
  \langle M_{N^{*}} \rangle = \frac{\int_{M_{N}+m_{\pi}}^{\sqrt{S}-M_{N}}
 f(M_{N^{*}})M_{N^{*}} d \, M_{N^{*}} }
 {\int_{M_{N}+m_{\pi}}^{\sqrt{S}-M_{N}} f(M_{N^{*}}) d\, M_{N^{*}} },
  \end{equation}
$f(M_{N^{*}})$ is the Breit-Wigner function in the case of $M_{N^{*}}$ not
far away from $M_{0}$
  \begin{equation}
 f(M_{N^{*}}) = \frac{1}{2\pi} \frac{\Gamma(q)}{(M_{N^{*}}-M_{0})^{2}
 +\frac{1}{4}\Gamma^{2}(q) },
  \end{equation}
here $M_{0}=1440$ MeV and $\Gamma(q)$ is the momentum-dependent decay width of
the $N^{*}$(1440) \cite{PhD}
 \begin{equation}
 \Gamma(q)=\Gamma_{0} \frac{M_{0}}{M_{N^{*}}}(q/q_{0})^{3} 
 \frac{1.2}{1+0.2(\frac{q}{q_{0}})^{2}}, 
 \end{equation}
where
 \begin{equation}
 q^{2}=\frac{[M^{2}_{N^{*}} - (M_{N}+m_{\pi})^{2} ][M^{2}_{N^{*}} -
 (M_{N} -m_{\pi})^{2} ] } {4M^{2}_{N^{*}}} ,
 \end{equation}
$q_{0}$ is related to the case of $M_{N^{*}}=M_{0}$ and $\Gamma_{0}=200$ MeV.
 The effect of the decay
width of $N^{*}$(1440) is taken into account through replacing $M_{N^{*}}$
in Eq. (21) with $ \langle M_{N^{*}} \rangle $. The in-medium 
$N^{*} + N \rightarrow N+ N$ and $N^{*} + N \rightarrow N^{*} + N$ cross
sections can then be calculated by means of the equations
 \begin{eqnarray}
 &&\sigma^{*}_{N^{*}N \rightarrow NN} = \frac{1}{16N} \int \sigma_{N^{*}N
 \rightarrow NN} (s,t) d\, \Omega, \\
 &&\sigma^{*}_{N^{*}N \rightarrow N^{*}N} = \frac{1}{8} \int \sigma_{N^{*}N
 \rightarrow N^{*}N }(s,t) d\, \Omega,
 \end{eqnarray}
here $N$ is the normalization factor stemming from the decay width of the
$N^{*}$(1440) \cite{Wol92,PRC97,Dan91}
 \begin{equation}
 N= \int_{(M_{N}+m_{\pi})^{2}}^{(\sqrt{S}-M_{N})^{2}}
  F(M^{2}_{N^{*}}) d\, M^{2}_{N^{*}}
 \end{equation}
and $F(M^{2}_{N^{*}})$ is the Breit-Wigner function
 \begin{equation}
F(M^{2}_{N^{*}})=\frac{M_{0}}{\pi} \frac{\Gamma(q)}{(M^{2}_{N^{*}}-M^{2}_{0})
 ^{2} +M^{2}_{0}\Gamma^{2}(q) }.
 \end{equation}
The in-medium $N^{*}$(1440) production cross section can be obtained from 
Eq. (37) through detailed balance \cite{Dan91}
 \begin{equation}
\sigma^{*}_{NN \rightarrow NN^{*}} = \frac{1}{8} \int \frac{p^{2}_{NN^{*}}}
 {p^{2}_{NN}} \sigma_{N^{*}N \rightarrow NN} (s,t) d\, \Omega , 
 \end{equation}
where $p_{NN}$, $p_{NN^{*}}$ denote the c. m. three momentum of the {\em NN}
and $NN^{*}$ states, respectively. Eqs. (37), (38) and (41) will be used in our 
numerical calculations.

  \end{sloppypar}
 \begin{center}
 \fbox{Table V}
 \end{center}
  \begin{sloppypar}
Now let us specify the coupling strengths. For the coupling strength of 
${\rm g}_{NN}^{\pi}$, we take the most commonly used value $f^{2}_{\pi}
 /4\pi=0.08$. The coupling strengths of ${\rm g}_{NN}^{\sigma}$ and
${\rm g}_{NN}^{\omega}$ are determined by fitting the known ground-state
properties for infinite nuclear matter. Several sets of parameters with
nonlinear self-interaction of scalar and vector field and the corresponding
saturation properties are presented in Table V. 
For the coupling strengths of nucleon-$N^{*}$(1440) coupling we follow the 
arguments of Ref. \cite{Hub94}. The following relation is expected to be
valid
 \begin{equation}
\frac{{\rm g}_{NN^{*}}^{\pi}}{{\rm g}_{NN}^{\pi}}=
\frac{{\rm g}_{NN^{*}}^{\sigma}}{{\rm g}_{NN}^{\sigma}}=
\frac{{\rm g}_{NN^{*}}^{\omega}}{{\rm g}_{NN}^{\omega}} .
 \end{equation}
${\rm g}_{NN^{*}}^{\pi}$ is determined from the width of pion decay of the
$N^{*}$(1440)-resonance
 \begin{equation}
\frac{{\rm g}_{NN^{*}}^{\pi}}{{\rm g}_{NN}^{\pi}}=0.351 \nonumber .
 \end{equation}
 \end{sloppypar}
 \begin{sloppypar}
 For the $N^{*}N^{*}$ coupling strengths, unfortunately, there is no any 
information directly from experiment available. The same situation takes
place for the $\Delta\Delta$ coupling strengths. Based on the quark model
and mass splitting arguments several different choices for the delta coupling
strengths have been discussed in Refs. \cite{Mos74,Wal87}, which will cause
strong influence on the nuclear equation of state in relativistic mean field
calculations \cite{Wal87}. If the SU(6) symmetry is exact for baryons, 
one should use the
universal coupling strengths, that is,
 \begin{equation}
\alpha(\Delta)=\frac{{\rm g}_{\Delta\Delta}^{\omega}}
{{\rm g}_{NN}^{\omega}}=1, \hspace{2cm} 
\beta(\Delta)=\frac{{\rm g}_{\Delta\Delta}^{\sigma}}
{{\rm g}_{NN}^{\sigma}}=1, 
 \end{equation}
here we have defined the dimensionless coupling strengths $\alpha(\Delta)$ and 
$\beta(\Delta)$. However, the mass splitting of the multiplets show that the
SU(6) symmetry is not exactly fulfilled. Then, one may assume that the coupling 
strengths have a splitting similar to the mass splitting of delta and nucleon
 \begin{equation}
 \alpha(\Delta)=\beta(\Delta)=\frac{M_{\Delta}}{M_{N}} \approx 1.3.
 \end{equation}
Another choice 
 \begin{equation}
 \alpha(\Delta)=1, \hspace{2cm} \beta(\Delta) \approx 1.3
 \end{equation}
is based on the argument that the $\omega$ meson has a real quark-antiquark
structure while the structure of the hypothetical $\sigma$ is not quite clear. 
It is worth to mention that recent calculations with the QCD sum rule method
yield $\alpha(\Delta) \approx 0.5$ while no prediction for $\beta(\Delta)$
\cite{Jin95}.

 \end{sloppypar}
 \begin{sloppypar}
In numerical calculations, we assume that the above arguments apply to the
$N^{*}N^{*}$ coupling strengths. We mainly consider the following three cases:
 \begin{equation}
\alpha(N^{*})=\frac{{\rm g}_{N^{*}N^{*}}^{\omega}}
{{\rm g}_{NN}^{\omega}}=1, \hspace{2cm} 
\beta(N^{*})=\frac{{\rm g}_{N^{*}N^{*}}^{\sigma}}
{{\rm g}_{NN}^{\sigma}}=1, 
 \end{equation}
 \begin{equation}
 \alpha(N^{*})=\beta(N^{*})=\frac{M_{N^{*}}}{M_{N}} \approx 1.5.
 \end{equation}
and
 \begin{equation}
 \alpha(N^{*})=1, \hspace{2cm} \beta(N^{*}) \approx 1.5
 \end{equation}
The influence of different choices of $\alpha(\Delta) ( \alpha(N^{*}))$ and
$\beta(\Delta) (\beta(N^{*}))$ on the predicted optical potential (the real part
of self-energy) and in-medium cross sections (the imaginary part of 
self-energy) will be checked. For simplicity, an universal coupling strength
of ${\rm g}_{\Delta\Delta}^{\pi}={\rm g}_{N^{*}N^{*}}^{\pi}={\rm g}_{NN}^{\pi}$
is always assumed.

 \end{sloppypar}
 \begin{sloppypar}
To take account of the effects stemming from the finite size of hadrons and a
 part of the short range correlation, a phenomenological form factor is
 introduced at each vertex. For the nucleon-nucleon-meson vertex we take the
 commonly used form
  \begin{equation}
 F_{NNA}(t)=\frac{\Lambda_{A}^{2}}{\Lambda_{A}^{2}-t}.
  \end{equation}
 For the nucleon-$N^{*}$(1440)-meson  vertex we adopt the mixed version 
 introduced in Ref. \cite{PRC94}
  \begin{equation}
 F_{NN^{*}A}(t,\,\langle M_{N^{*}}\rangle)=\frac{\Lambda^{*2}_{A}}{\Lambda
^{*2}_{A}-t}
 \left[ \frac{\Gamma^{2}(\langle q \rangle)/4 }{ (\langle M_{N^{*}} \rangle -M
_{0})^{2}
  +\Gamma_{0}^{2}/4 } \right] ^{1/4},
  \end{equation}
 here $\Gamma_{0}=200$ MeV and  $\Gamma (\langle q \rangle )$ is obtained from
Eqs. (35) and (36)  with the M$_{N^{*}}$ replaced by the centroid mass
$\langle M_{N^{*}} \rangle$.
Here we distinguish the form factor $\Lambda^{*}_{A}$ 
for the nucleon-$N^{*}$(1440)-meson vertex to the $\Lambda_{A}$ for the 
nucleon-nucleon-meson vertex. S.~Huber and J.~Aichelin claimed that 
$\Lambda^{*}_{A}$ is about 40\% of $\Lambda_{A}$ \cite{Hub94}. We adopt this
argument in the following calculations.
 The form factor of the $N^{*}$(1440)-$N^{*}$(1440)-meson vertex is taken
  the same as
that of corresponding nucleon-nucleon-meson vertex. The cut-off masses
 $\Lambda_{\sigma}$=1200 MeV,
$\Lambda_{\omega}$=808 MeV and  $\Lambda_{\pi}$=500 MeV fixed in Refs. 
\cite{PRC94,ZPA94,PRC97}
will be used,  which are obtained by fitting  the experimental data of nucleon
mean free path
and the free {\em NN} scattering cross section. According to the above
argument, $\Lambda^{*}_{\sigma}$=480 Mev, $\Lambda^{*}_{\omega}$=323 MeV.
But we still take $\Lambda^{*}_{\pi}=\Lambda_{\pi}$=500 MeV, since this value
is already comparable to the $\Lambda^{*}_{\pi}$=400 MeV used in 
Ref. \cite{Hub94}.

 \end{sloppypar}
 \begin{center}
{\bf V. NUMERICAL RESULTS}
  \end{center}
 \begin{center}
 \fbox{Fig. 2}
 \end{center}
 \begin{sloppypar}
In this work, the numerical calculations are performed in symmetric 
nuclear matter
at zero-temperature. The distribution functions in Eqs. (23) and (24) are
replaced by the corresponding $\theta$ functions.
In Fig.~2 we display the real part of nucleon optical potential calculated
with the parameters given in Table V. The computation are performed
at $\rho=\rho_{0}$ where there is no contribution from the delta and 
$N^{*}$(1440) as expected. The experimental data from 
phenomenological
optical model analysis \cite{Ham90} is also presented.
The nonlinear self-interaction
of  vector meson is known to be important for obtaining the proper density
dependence of  vector field at high density. But the real part of nucleon
optical potential calculated with the fourth, fifth and sixth set of parameters 
in Table V exhibits an unreasonable momentum-dependence mainly due to the large
$\omega$ coupling strength ${\rm g}_{NN}^{\omega}$.  
 Because we will use the same coupling
strengths to calculate both the mean field and in-medium scattering cross
sections, this kind of unrealistic momentum-dependence exhibited by the
very rapid increase of nucleon optical potential with the increase of
energy will cause strange behavior of in-medium cross sections where $\sigma$
and $\omega$ coupling strengths enter evidently. Therefore, in the following
calculations we will take only the nonlinear self-interaction of scalar
field and mainly use the second set of parameters in Table V, which can also
reproduce the results of the G-matrix theory \cite{Bro90} quite well 
\cite{PRC96}.

 \end{sloppypar}
 \begin{center}
 \fbox{Fig. 3} \hspace{2cm} \fbox{Fig. 4}
 \end{center}
 \begin{sloppypar}
Fig.~3 and Fig.~4 depict the momentum dependence of the real part of $\Delta$
and $N^{*}$(1440) optical potential. Different sets of the $\Delta$ and
$N^{*}$(1440) coupling strengths are used. If the universal coupling strengths 
are assumed, the behavior as well as the well depths of the $\Delta$ and
$N^{*}$(1440) optical potential are quite similar to the nucleon optical
potential. But the slopes of the curves are a little smaller because of the
larger delta and $N^{*}$(1440) mass. If one takes $\beta(\Delta)=1.3$
($\beta(N^{*})=1.5$) but still remain $\alpha(\Delta)=1$ ($\alpha(N^{*})=1$),
a very attractive $\Delta$ ($N^{*}(1440)$) optical potential will be obtained
compared to the nucleon optical potential. Some estimations for the $\Delta$
optical potential were made in Refs. \cite{Bog82,Dan80}. The well depth
of the delta-nucleus effective potential turns out to be $-120$ MeV
\cite{Bog82} and $-150$ MeV \cite{Dan80}, respectively. The calculations
with $\alpha(\Delta)=1, \beta(\Delta)=1.3$ approach to this estimation. However,
one should keep in mind that no experiments for the $\Delta$ and 
$N^{*}$(1440) potential are reported up to now. The above arguments should be 
viewed with some cautions. If the coupling strengths of $\alpha(\Delta)=
\beta(\Delta)=1.3$ ($\alpha(N^{*})=\beta(N^{*})=1.5$) are used, the calculated
$\Delta$ ($N^{*}(1440)$) optical potential becomes a little more attractive at lower
energy and repulsive at higher energy compared to the case of universal
coupling strengths. But no significant difference are found due to the 
cancellation effects of the scalar and vector field.

 \end{sloppypar}
 \begin{sloppypar}
 Fig.~5 display the density dependence of the nucleon, delta and $N^{*}$(1440)
optical potential. The calculations are performed in the limit of 
zero-momentum ( $ \mid {\bf k} \mid \rightarrow 0$ ). Thus, the optical potentials are 
essentially the summation of the scalar and vector potential of the respective
baryons. Since the parameter set 2 in Table V is used as the nucleon coupling
strengths, the contributions of the delta and $N^{*}$(1440) to the baryon density
are negligible mainly due to the large effective mass $m^{*} /M_{N}=0.83$
\cite{LiAcc,CTP96,Nak84}. In other words, we don't have density isomers on
the nuclear equation of state with the present used parameters. The situation,
however, will change substantially if another set of nucleon coupling strengths
with smaller effective mass ($m^{*} /M_{N} \sim 0.6$) is used \cite{Wal87,LiAcc}.
But smaller effective mass will usually cause larger $\omega$ coupling strength
 (${\rm g}_{NN}^{\omega}$). The momentum dependence of the nucleon optical
potential will then become very steep as indicated in Fig.~2. It should be 
mentioned that finite nuclei calculations prefer a smaller effective mass
since it will give stronger spin-orbit force \cite{Sug94,Rei86,Ruf88,Gam90}.

 \end{sloppypar}
 \begin{sloppypar}
The influence of different choices of the $\Delta$ coupling strengths as well
as the $N^{*}$(1440) coupling strengths on the predicted optical potentials
is checked in Fig. 5. In the case of universal coupling strengths, the  
$\Delta$ and $N^{*}$(1440) optical potential are the same as the nucleon
optical potential. Larger scalar-delta (-$N^{*}$(1440)) coupling strength
will give a deeper $\Delta$ ($N^{*}(1440)$) effective-potential well depth. If one
uses $\alpha(\Delta)=1$, $\beta(\Delta)=1.3$ ($\alpha(N^{*})=1$, $\beta
(N^{*})=1.5$) as the $\Delta$ ($N^{*}(1440)$) coupling strengths, the $\Delta$
($N^{*}$(1440)) potential becomes so attractive that it is still a large negative
number at $\rho=3\rho_{0}$. It is currently of urgent important to have
some experimental information on the $\Delta$ and $N^{*}$(1440) coupling
strengths.

 \end{sloppypar}
 \begin{center}
 \fbox{Fig. 5} \hspace{2cm} \fbox{Fig. 6}
 \end{center}
 \begin{sloppypar}
Now we turn to discuss the imaginary part of self-energy, i.e., the two-body
scattering cross sections. Firstly, in Fig.~6 we compare our theoretical
predications 
of free $pp \rightarrow pp^{*}(1440)$ cross section to the available 
experimental data \cite{CERN}. The results of the one-boson-exchange model
computed by Huber and Aichelin \cite{Hub94} are also presented in this figure
as dashed line. Our results are  consistent with that of Ref. \cite{Hub94}.
Both of them are in good agreement with the experimental data. Here we should
point out that our calculations are almost parameter free. We do not fit any
parameters to the predicted cross section. Only the argument of $\Lambda^{*}
_{\sigma} / \Lambda_{\sigma} = \Lambda^{*}_{\omega} / \Lambda_{\omega}$=40\%
is taken from Ref. \cite{Hub94}. If $\Lambda^{*}_{\sigma}=\Lambda_{\sigma}$
and $\Lambda^{*}_{\omega}=\Lambda_{\omega}$ are adopted, the cross section will
be three times larger than the empirical value at higher energy as indicated
by the dotted line in the figure.

 \end{sloppypar}
 \begin{center}
 \fbox{Fig. 7} \hspace{2cm} \fbox{Fig. 8}
 \end{center}
 \begin{sloppypar}
Fig.~7 displays the in-medium $N^{*}$(1440) production cross section at normal 
density, where dotted line represents the contribution of direct term and dashed
 line is that of exchanged term. The calculations are performed with the
second set of parameters in Table V as the nucleon coupling strengths and
the universal coupling-strength assumption for the $\Delta$ and $N^{*}$(1440)
coupling strengths. 
In contrast to the in-medium $\Delta$
production cross section \cite{PRC94}, here the contribution of exchange term
is negligible. Fig. 8 shows the contribution of direct term and exchange term
to the in-medium $N^{*}+ N \rightarrow N^{*}+ N$ cross section. It can be seen 
that at lower energy the exchange term plays an evident cancellation effect.
With the energy increase, it decreases quickly and can be neglected at higher
energy.

 \end{sloppypar}
 \begin{center}
 \fbox{Fig. 9} \hspace{2cm} \fbox{Fig. 10}
 \end{center}
 \begin{sloppypar}
Fig.~9 and Fig.~10 depict the contributions of different mesons to the 
in-medium cross sections. The other conditions are the same as in Fig.~7. 
One can find that at lower energy the $\sigma$
term is very large. The same situation takes place for the $\omega$ term at higher
energy. However, the $\sigma + \omega$ mixed term has an opposite sign with that
of individual $\sigma$ and $\omega$ terms. There exist large cancellation
phenomena in the contributions of $\sigma$ and $\omega$ mesons. Consequently,
 pion contributes most at lower energy. At higher energy the $\pi$ meson
contributes nearly 1/4 of in-medium $N + N \rightarrow N + N^{*}$ cross
section and 1/3 of in-medium $N^{*} + N \rightarrow N^{*} + N$ cross section.
At very high energies (S=15-20 GeV$^{2}$), the main contribution of the
$N^{*} + N \rightarrow N^{*} + N$ cross section comes from the $\omega$ term.
Since only the exchange term remains, the contributions of $\sigma + \pi$
and $\omega + \pi$ mixed terms are very small.

 \end{sloppypar}
 \begin{center}
 \fbox{Fig. 11} \hspace{2cm} \fbox{Fig. 12} \hspace{2cm} \fbox{Fig. 13}
 \end{center}
 \begin{sloppypar}
Fig.~11 displays the in-medium $N^{*}$(1440) production cross sections at 
different densities and energies. The different sets of the $N^{*}$(1440)
coupling strengths are employed. For the nucleon coupling strengths we 
always use the parameter set2 in Table V. It is shown from Fig.~11 that
the $\sigma^{*}_{NN \rightarrow NN^{*}}$ increases with the increase of
density. When  the universal coupling strengths are used, only a mild 
dependence on the density is exhibited. The density dependence, however, will
become evident if one uses a larger scalar-$N^{*}$(1440) coupling strength.
The choice of the $\alpha(N^{*})$ has no influence on the predicted cross 
sections. The reason is as follows: firstly, as one can see in Appendix D,
${\rm g}_{N^{*}N^{*}}^{\omega}$ does not enter the expressions of the $\sigma
_{NN \rightarrow NN^{*}}^{*}$ explicitly; secondly, we always calculate
the in-medium total energy of two particle system (small s) from the incident
two particles, i.e., two nucleons in the case of the $\sigma^{*}_{NN \rightarrow
 NN^{*}}$. The situation will change if
one considers the $\sigma_{N^{*}N \rightarrow NN}^{*}$, where the influence
of $\alpha(N^{*})$ will enter in the calculations of in-medium total energy
(small s) from free total energy (capital S), and then affects the in-medium
cross section.

 \end{sloppypar}
 \begin{sloppypar}
Fig.~12 depicts the in-medium $N^{*}$(1440) absorption cross section. Other
conditions are the same as in Fig.~11. The cross sections drop 
very rapidly when the energy exceeds the threshold. That means that the
absorption process are most important at energy close to the threshold as in
the case of $\Delta$ absorption. When $\alpha(N^{*})=1$, $\beta(N^{*})=1.5$
is used as the $N^{*}$(1440) coupling strengths, the $\sigma_{N^{*}N \rightarrow
 NN}^{*}$ exhibits an evident density dependence. It decreases with the increase
 of density. In other two cases, i.e., $\alpha(N^{*})=\beta(N^{*})=1$ and
 $\alpha(N^{*})=\beta(N^{*})=1.5$, the dependence of the $\sigma_{N^{*}N
\rightarrow NN}^{*}$ on the density becomes weaker and less explicit.

 \end{sloppypar}
 \begin{sloppypar}
 In Fig.~13 we show the in-medium $N^{*}N \rightarrow N^{*}N$ cross section at
different densities and energies. As can be found from the figure, the cross
sections now become very sensitive to the $\alpha(N^{*})$ and $\beta(N^{*})$
used because ${\rm g}_{N^{*}N^{*}}^{\sigma}$ and ${\rm g}_{N^{*}N^{*}}^{\omega}$
 enter the expressions of the $\sigma_{N^{*}N \rightarrow N^{*}N}^{*}$
explicitly (see Appendix D). Generally speaking, the density dependence of the
cross section is not very evident when $\alpha(N^{*})=\beta(N^{*})=1$ and
$\alpha(N^{*})=\beta(N^{*})=1.5$ are used, mainly due to the strong cancellation
 effects from the $\sigma + \omega$ mixed term (see Fig.~10). A strong density
dependence appears when the set of $\alpha(N^{*})=1$, $\beta(N^{*})=1.5$ is 
used as the $N^{*}$(1440) coupling strengths. The in-medium cross section
decreases with the increase of density at lower energy and increases at higher
energy. As the energy changes, the cross section firstly decreases and then 
increases, especially in the case of $\alpha(N^{*})$=1.5. It is mainly caused 
by the contribution of the $\omega$ term. As can be seen from Fig. 10, the
$\omega$ term approaches a saturation with the increase of energy while all
other terms (especially, the important cancellation term of the $\sigma + \omega
 $ mixed term) decrease. 
The Cugnon's parameterization for free {\em NN}
elastic cross section, which is commonly used in the transport models for the 
$N^{*}N$ elastic cross section, is also plotted in Fig.~13. One can find an 
evident difference between  the
in-medium $N^{*} + N \rightarrow N^{*} + N$ cross section and the Cugnon's
parameterization. It is therefore important to take the in-medium cross sections
into account in the study of heavy-ion collisions.

 \end{sloppypar}
 \begin{center}
{\bf VI. SUMMARY AND OUTLOOK}
  \end{center}
  \begin{sloppypar}
Starting from the effective Lagrangian describing baryons interacting through 
 mesons, using the closed time-path Green's function technique and adopting
 the semi-classical, quasi-particle and Born approximations
we have developed a RBUU-type transport equation for the $N^{*}$(1440)
distribution function. The equation is derived within the same framework
which was successfully applied to the
nucleon's \cite{PRC94,ZPA94} and delta's \cite{PRC96} and thus we obtained
a set of self-consistent equations for the $N$, $\Delta$ and $N^{*}$(1440)
system. Three equations
are coupled through the self-energy terms and collision terms and should be
solved simultaneously in a numerical simulation of heavy-ion collisions. 
Both the mean field and collision term of the $N^{*}$(1440)'s RBUU equation are 
derived from the same effective Lagrangian and given explicitly, so the medium
effects on the two-body scattering cross sections are addressed automatically  
and can be studied 
self-consistently. Therefore, this approach
  provides a promising way to reach a covariant
description of the $N^{*}$(1440) in relativistic heavy-ion collisions. 

Based on this approach, we have studied both the real part and the imaginary
part of the $N^{*}$(1440) self-energy, i.e., the relativistic optical potential
and the in-medium two-body scattering cross sections. Since there is no 
information about the $N^{*}N^{*}$ coupling strengths available, several
different choices for $\alpha(N^{*})={\rm g}_{N^{*}N^{*}}^{\omega} / {\rm g}
_{NN}^{\omega}$ and $\beta(N^{*}) = {\rm g}_{N^{*}N^{*}}^{\sigma} / {\rm g}
_{NN}^{\sigma}$ are investigated. The results turn out to be sensitive to the
$\alpha(N^{*})$ and $\beta(N^{*})$ used. A very attractive $N^{*}$(1440) 
optical potential will be obtained if $\alpha(N^{*})=1$ and $\beta(N^{*})=1.5$
are used as the $N^{*}N^{*}$ coupling strengths. In the case of $\alpha(N^{*})
=\beta(N^{*})=1$ the $N^{*}$(1440) optical potential is similar to the nucleon
optical potential. When $\alpha(N^{*})=\beta(N^{*})=1.5$, the $N^{*}$(1440)
optical potential is a little more attractive than the nucleon optical potential
at lower energy/density and more repulsive at higher energy/density. The same
arguments for the $N^{*}N^{*}$ coupling strengths are applied in the study
of medium effects on the 
$N + N \rightarrow N + N^{*}$, $N^{*} + N \rightarrow N + N$ and
$N^{*} + N \rightarrow N^{*} + N$ scattering cross sections. Generally speaking,
only a mild density-dependence of in-medium cross sections is found in the
cases of $\alpha(N^{*})=\beta(N^{*})=1$ and $\alpha(N^{*})=\beta(N^{*})=1.5$.
The situation, however, is changed when the set of $\alpha(N^{*})=1$, $\beta
(N^{*})=1.5$ is adopted. An evident density-dependence  appeares. 
Qualitatively, the $\sigma^{*}_{NN \rightarrow NN^{*}}$ are found to increase
with the increase of density while the $\sigma^{*}_{N^{*}N \rightarrow NN}$
near the threshold energy decreases.
For the $\sigma^{*}_{N^{*}N \rightarrow N^{*}N}$, the situation is a little 
complicated. It decreases with the increase of density at lower energy and 
increases at higher energy. 
Because we have not included the screening and anti-screening
effects of the medium on the interaction in the present calculations,
the above arguments should  be viewed with caution.  Further
 investigations are needed.

In this work the numerical calculations are performed in static nuclear
matter with a spherical Fermi distribution in momentum space. In principle,
the initial condition of relativistic heavy-ion collisions is related to the
two interpenetrating nuclei. This kind of anisotropy of the momentum 
distributions have strong influence on the nuclear equation of state (i.e., 
mean field) when the collective velocity of the two interpenetrating nuclei
is large \cite{Lov81,Nei87}. It will certainly affect the theoretical
predictions of in-medium cross sections. A study of in-medium $NN$ elastic
scattering cross section in colliding nuclear matter has recently been carried
out by Sehn et al. \cite{Seh96}. It is quite interesting to address the problem
in the present transport theory, in which it can be investigated
more naturally since the single-particle distribution functions of transport
equations contain essentially the information of the initial longitudinal
momentum excess. This output can be used in the study of
heavy-ion collisions directly, both in RBUU models and in (Ur)QMD models 
\cite{URQMD}. Work on this aspect is in progress. 

As has been pointed out before, the temperature degree of freedom is not 
taken into account in the present  microscopic transport theory for
non-equilibrium system. However, relativistic heavy-ion collisions allow the
  study of
the dynamical process under extreme conditions of high temperature and density.
The temperature degree of freedom should be incorporated in order to study the
phase transition. It is realized in macroscopic theories such as two/three
fluid model, but it is still a major challenge to  RBUU-type transport theories.
Nevertheless, one can discuss the effects of temperature on the mean field
and in-medium cross sections in  static nuclear matter by means of the formula
obtained in this work. A simpler way is to replace the single-particle 
distribution functions with the Fermi-Dirac distribution functions. Then one
can study the temperature-dependent in-medium cross sections relativisticlly,
  which was never
done before. An evident influence of 
temperature on the in-medium cross sections is to be expected, with  
implications due to
  investigations on the temperature-dependent 
 imaginary part of optical potential \cite{Han97}.

It is straightforward to develop a transport equation for the $N^{*}$(1535)
resonance within the current framework. Theoretically, the main difference
between the $N^{*}$(1440) and the $N^{*}$(1535) is that the $N^{*}$(1535) has
a negative parity. However, we do not think this will cause significant 
technical problems.
The extension to the $N^{*}$(1535) will appear in forthcoming paper. 
As the colliding energy increases, it becomes important to include other
$N^{*}$ as well as $\Delta$ resonances with higher resonance-masses. Most/all
of in-medium cross sections relevant to these high-mass resonances are 
experimentally unavalaible, and very little theoretical work on this aspect
has been done. On the other hand, they are urgently needed in  realistic
transport models extended to describe ultra-relativistic heavy-ion collisions.
These in-medium cross sections can be studied in the present
microscopic transport theory. Work on this direction is continuing.

The proposed enhancement of strange particle production in heavy-ion
collisions may be a very promising experimental signal in the search of
quark-gluon plasma \cite{Koc86}, connected with 
 the possibility of existence of stable 
or meta-stable multistrange objects \cite{Sch93,Greiner} (which will have very
important consequence on the
equation of state of neutron stars \cite{Gle92}) has stimulated substantial
  further activity. Experimentally, the search for
strange composites--strange clusters (MEMOs), strange droplets of quark matter
 (strangelets) is under investigation by a number of groups at the AGS and the
SPS (e.g. E882, E814, E813/836, P864, NA52). Theoretical work about this topic
 receives  attention currently. 
With the effective Lagrangian proposed by Schaffer et al. \cite{Sch93,Sch94}
the self-consistent RBUU approach is a promising model to be generalized to
include the hyperon degree of freedom on the octet of spin 1/2- and decouplet
of spin 3/2-baryons such as $\Sigma$, $\Xi$, $\Lambda$ et al. \cite{Greiner}.
Then, problems relevant to strangeness can be studied in a relativistic 
microscopic transport theory.

It should be pointed out that in this work all mesons ($\sigma$, $\omega$
and $\pi$) are treated as virtual mesons. 
Strictly speaking, this picture is valid only under the
assumption that the mesons remain in equilibrium during transport \cite{Dav91}.
In a reasonable physical scenario, it should be possible to describe the
creation and destruction of real as well as virtual mesons and not just one or
the other. It is practical to firstly treat pions explicitly, considering that
the pion is a physical observed meson, while the remaining other mesons, 
 such as the $\sigma$
and the $\omega$ are still treated as virtual mesons \cite{Pre}. It is  
important  to develop transport equations for other physical mesons,  
such as $K$, $K^{*}$, $\rho$, $\eta$, $\phi$, $f_{2}$ ..., in the meson 
multiplets 
within the present framework.

It was argued that chiral symmetry might be restored or partial restored in
the hot and dense matter, characterized by vanishing (dropping) effective
nucleon and meson masses \cite{Bro91,Bir94}. 
Experimentally, this could be verified by measuring the dileptons produced
from heavy-ion collisions \cite{Li95}. Other experimental observations have
also been cited as signals for the modification of hadron properties in 
nuclear matter, consistent with partial restoration of chiral symmetry. 
Unfortunately, almost none of them provides unambiguous evidence, since there
are {\em convential} mechanisms, which can generate similar effects. Although
it has recently become one of the most exciting topics in nuclear physics
to seek evidence of chiral symmetry restoration in heavy-ion collisions, 
the practical relativistic dynamical equations for describing production
{\em with chiral symmetry} are not available yet. Derivations of such
transport equations with chiral symmetry were carried out by several groups
based on the chiral Lagrangian of the Nambo-Jano-Lasinio model
\cite{Zha92,Aba95,Gre96,Cse95,Nuth}. 
 These attempts, however, are still at an early stage
and a complete numerical realization is not available as of yet. It might be
numerically more practical to solve transport equations developed from an
effective chiral Lagrangian with baryonic and mesonic degrees of freedom
\cite{Pap97,Sub}. The Green's function techniques of non-equilibrium system
used in this work can be directly applied to such chiral Lagrangians
to develop a chiral transport theory, in which the requirement of chiral
symmetry from QCD can be realized.   

\end{sloppypar}
 \begin{center}
{\bf ACKNOWLEDGMENTS}
 \end{center}
 \begin{sloppypar}
 We thank S.~A.~Bass and C.~Ernst for fruitful discussions.
 G.~Mao and Z.~Li are  grateful to the Alexander von 
Humboldt-Stiftung for financial support and to the Institut f\"{u}r
Theoretische Physik der J.~W.~Goethe Universit\"{a}t for their hospitality.
This work was supported by DFG-Graduiertenkolleg Theoretische \& Experimentelle
Schwerionenphysik, GSI, BMBF, DFG, and A.v.Humboldt-Stiftung.

 \end{sloppypar}
 \setcounter{equation}{0}
\renewcommand{\theequation}{A\arabic{equation}}
 \begin{center}
{\bf APPENDIX A}
 \end{center}
 \begin{sloppypar}
 Here we present the Feynman diagrams of  the Born term of the $N^{*}$(1440)
self-energies contributing to the reaction channels discussed in Sects. II and
 III. Dashed lines denote mesons, double lines denote deltas, and solid and
bold-solid lines represent the nucleon and $N^{*}$(1440), respectively:

 \end{sloppypar}
 \begin{figure}[htbp]
 \indent (a) $N^{*} N \longrightarrow N^{*} N$  \\
 \vskip -4.5cm
\hskip -2.2cm \psfig{file=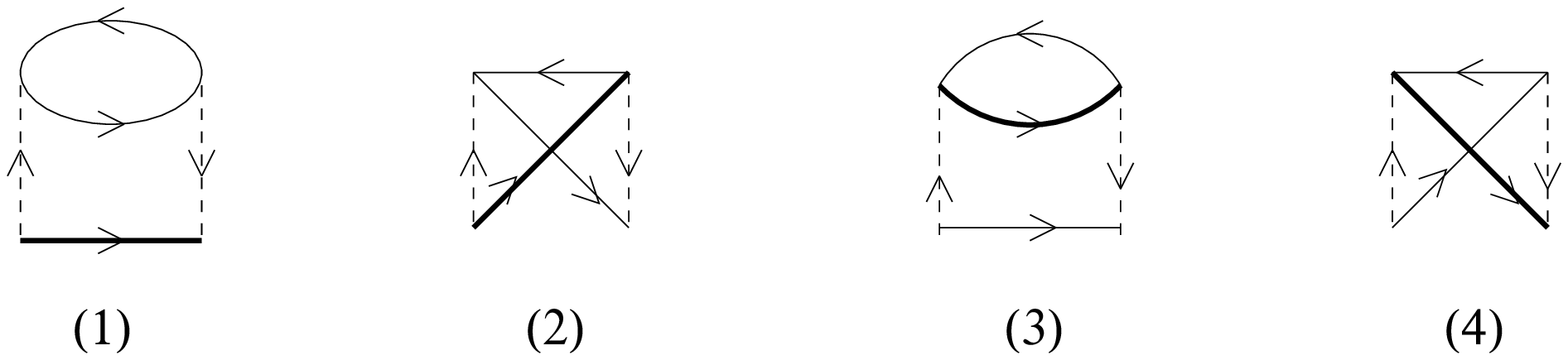,width=11cm,height=7cm,angle=-90}
\end{figure}
 \begin{figure}[htbp]
 \indent (b) $N^{*} \Delta \longrightarrow N^{*} \Delta$ \\
 \vskip -4.5cm
\hskip -2.2cm \psfig{file=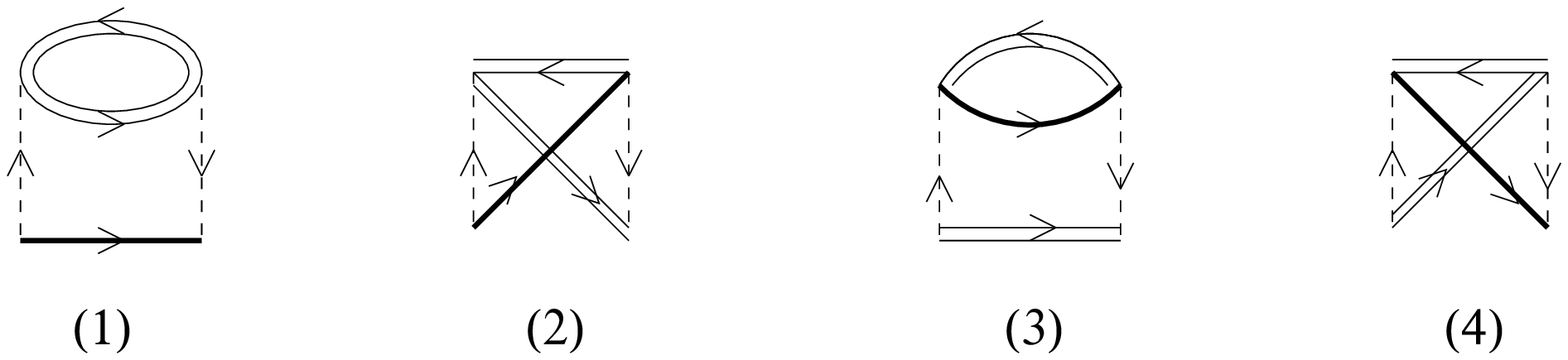,width=11cm,height=7cm,angle=-90}
\end{figure} 
 \begin{figure}[htbp]
 \indent  (c) $N^{*} N^{*} \longrightarrow N^{*} N^{*}$  \\
 \vskip -4.5cm
\hskip -2.0cm \psfig{file=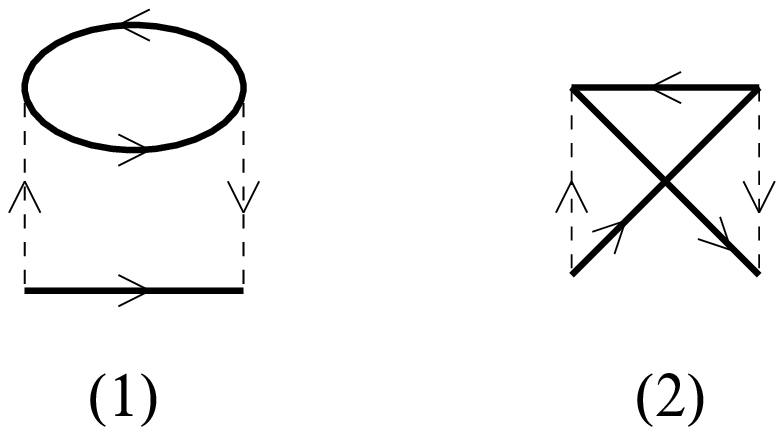,width=5.5cm,height=7cm,angle=-90}
\end{figure}
 \begin{figure}[htbp]
 \indent  (d) $N^{*} N \longrightarrow N N$  \\
 \vskip -4.5cm
\hskip -2.0cm \psfig{file=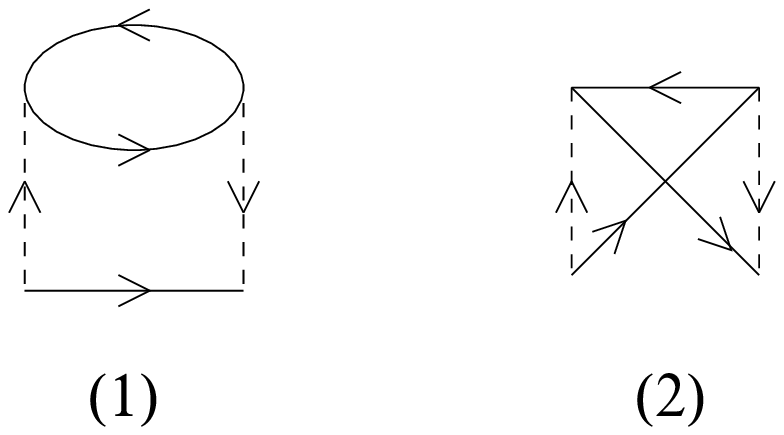,width=5.5cm,height=7cm,angle=-90}
\end{figure}
 \begin{figure}[htbp]
 \indent  (e) $N^{*} N \longrightarrow N \Delta$  \\
 \vskip -4.5cm
\hskip -2.2cm \psfig{file=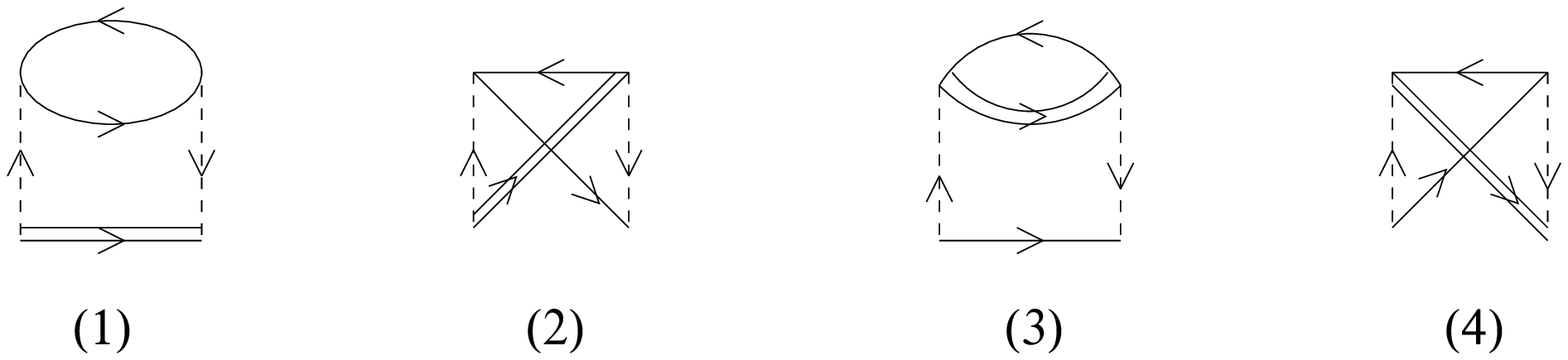,width=11cm,height=7cm,angle=-90}
\end{figure}


 \begin{figure}[htbp]
 \indent  (f) $N^{*} N \longrightarrow \Delta \Delta$  \\
 \vskip -4.5cm
\hskip -2.0cm \psfig{file=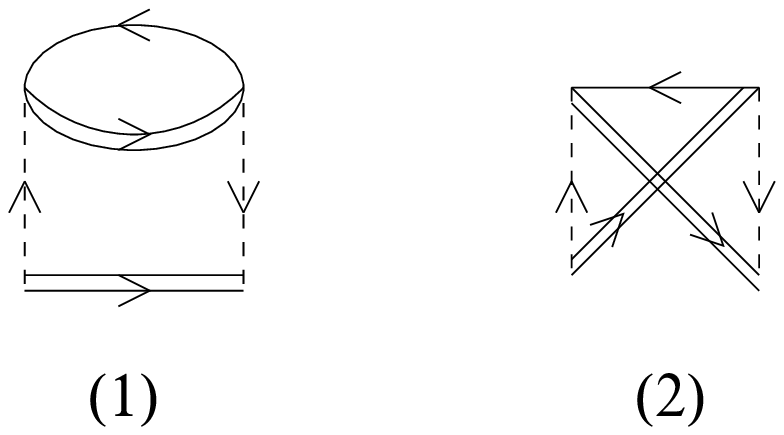,width=5.5cm,height=7cm,angle=-90}
\end{figure}
 \begin{figure}[htbp]
 \indent  (g) $N^{*} N \longrightarrow \Delta N^{*} $  \\
 \vskip -4.5cm
\hskip -2.2cm \psfig{file=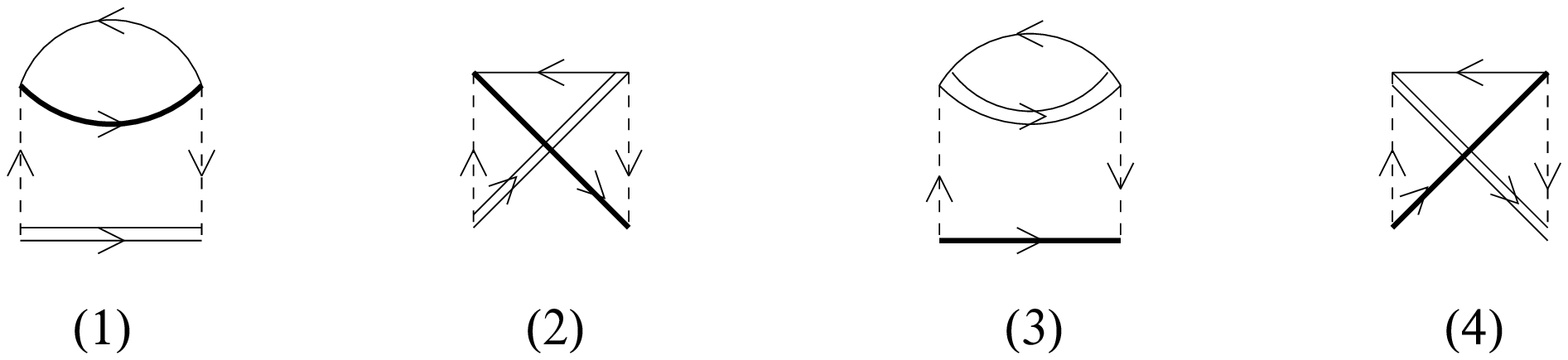,width=11cm,height=7cm,angle=-90}
\end{figure}
 \begin{figure}[htbp]
 \indent  (h) $N^{*} N \longrightarrow N^{*} N^{*}$  \\
 \vskip -4.5cm
\hskip -2.0cm \psfig{file=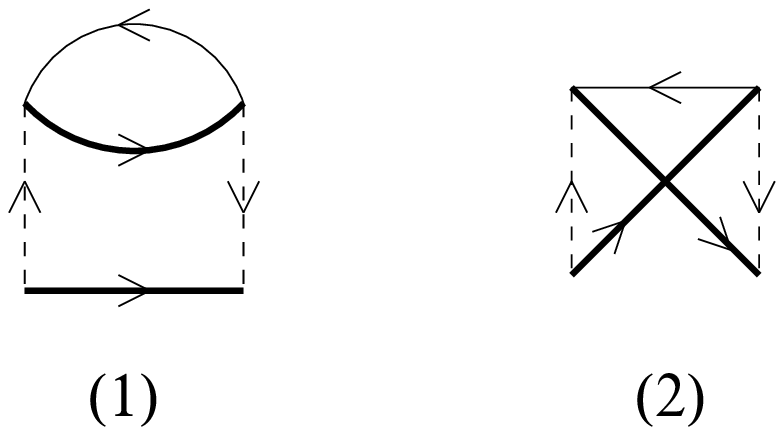,width=5.5cm,height=7cm,angle=-90}
\end{figure}
 \begin{figure}[htbp]
 \indent  (i) $N^{*} \Delta \longrightarrow N N$  \\
 \vskip -4.5cm
\hskip -2.0cm \psfig{file=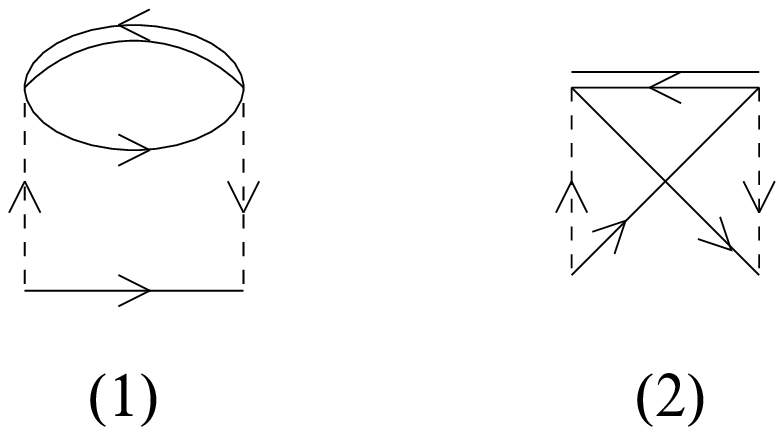,width=5.5cm,height=7cm,angle=-90}
\end{figure}
 \begin{figure}[htbp]
 \indent  (j) $N^{*} \Delta \longrightarrow N \Delta$  \\
 \vskip -4.5cm
\hskip -2.2cm \psfig{file=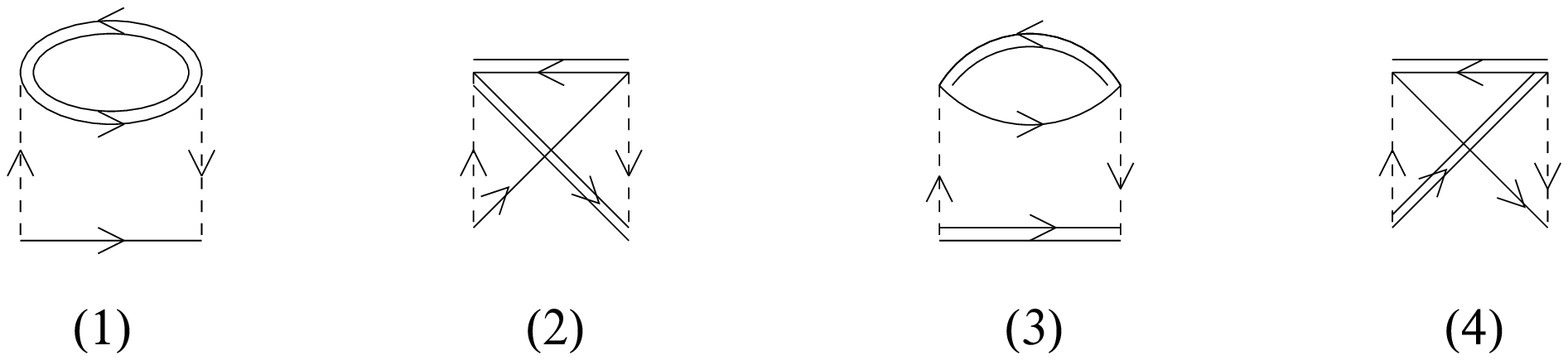,width=11cm,height=7cm,angle=-90}
\end{figure}


 \begin{figure}[htbp]
 \indent  (k) $N^{*} \Delta \longrightarrow \Delta \Delta$  \\
 \vskip -4.5cm
\hskip -2.0cm \psfig{file=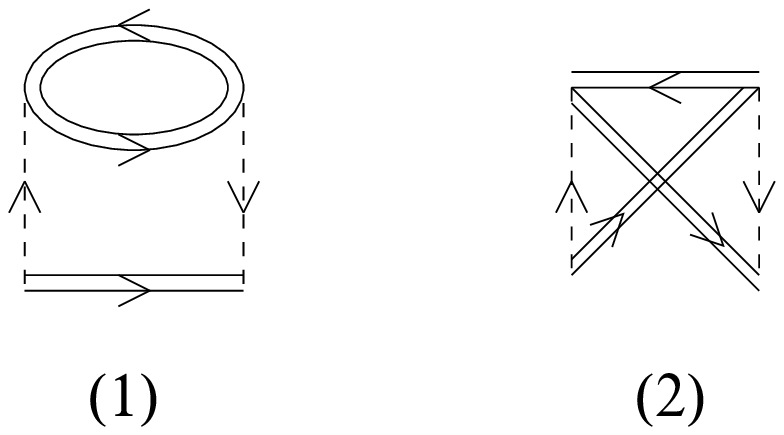,width=5.5cm,height=7cm,angle=-90}
\end{figure}
 \begin{figure}[htbp]
 \indent  (l) $N^{*} \Delta \longrightarrow N^{*} N^{*}$  \\
 \vskip -4.5cm
\hskip -2.0cm \psfig{file=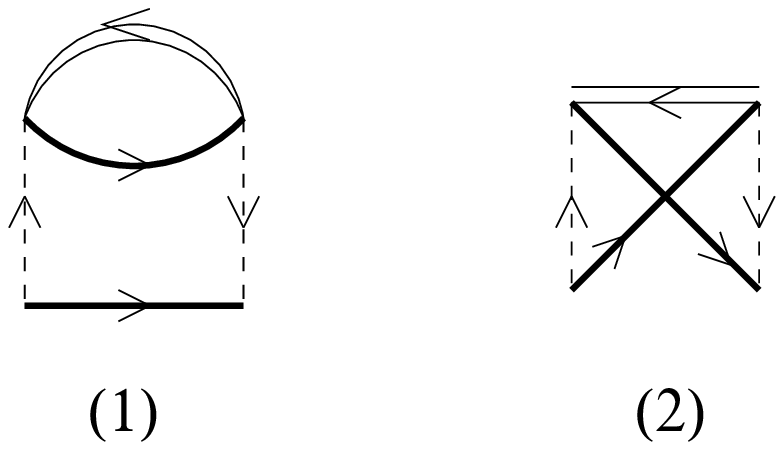,width=5.5cm,height=7cm,angle=-90}
\end{figure}
 \begin{figure}[htbp]
 \indent  (m) $N^{*} N^{*} \longrightarrow N N$  \\
 \vskip -4.5cm
\hskip -2.0cm \psfig{file=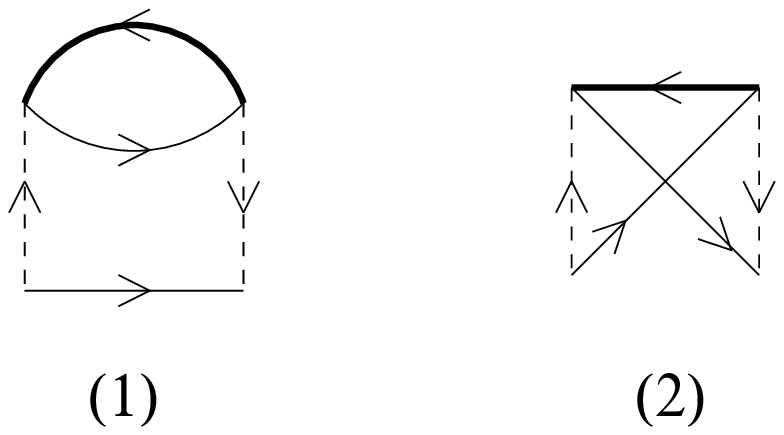,width=5.5cm,height=7cm,angle=-90}
\end{figure}
 \begin{figure}[htbp]
 \indent  (n) $N^{*} N^{*} \longrightarrow N \Delta$  \\
 \vskip -4.5cm
\hskip -2.2cm \psfig{file=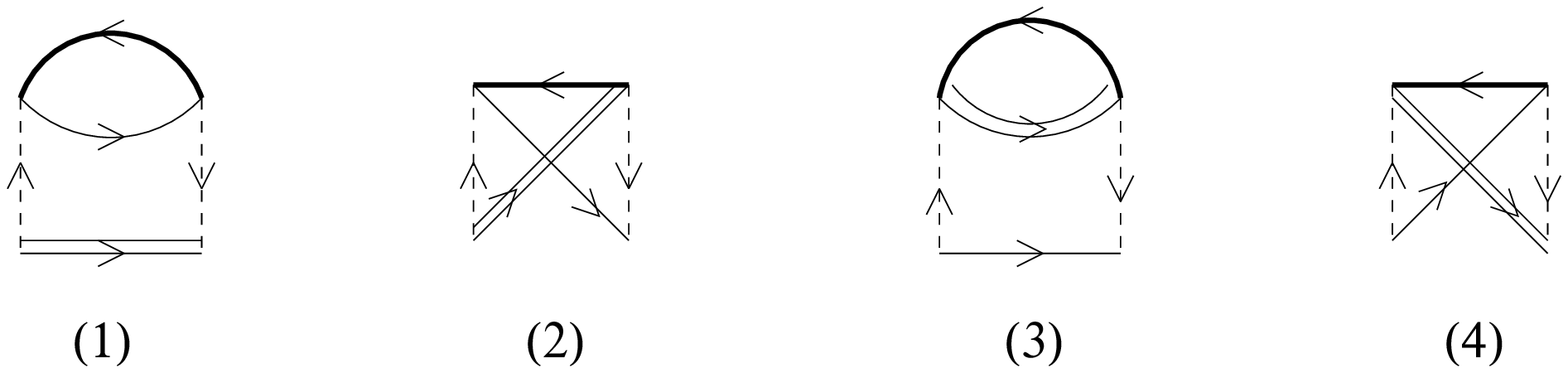,width=11cm,height=7cm,angle=-90}
\end{figure}
 \begin{figure}[htbp]
 \indent  (o) $N^{*} N^{*} \longrightarrow \Delta \Delta$  \\
 \vskip -4.5cm
\hskip -2.0cm \psfig{file=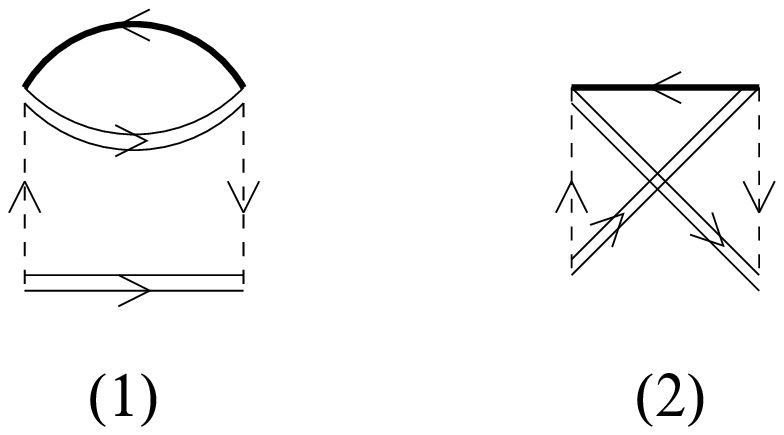,width=5.5cm,height=7cm,angle=-90}
\end{figure}
\newpage
 \setcounter{equation}{0}
\renewcommand{\theequation}{B\arabic{equation}}
 \begin{center}
{\bf APPENDIX B}
 \end{center}
 \begin{sloppypar}
In this appendix we give the concrete expressions for the spin matrices
$\Phi_{D}$ and $\Phi_{E}$, the subscripts denote the terms contributed from 
the different Feynman diagrams given in appendix A:
  \begin{eqnarray}
 \Phi_{a1}&=& tr \lbrace {\rm g}_{N^{*}N^{*}}^{A}\gamma_{A}
      (\not\! p_{3}+m^{*}_{N^{*}}) {\rm g}_{N^{*}N^{*}}^{B}\gamma_{B}
     tr[ {\rm g}_{NN}^{A}\gamma_{A}(\not\! p_{4}+m^{*})
      {\rm g}_{NN}^{B}\gamma_{B}(\not\! p_{2}+m^{*}) ] 
      (\not\! p +m^{*}_{N^{*}}) \nonumber \\                   
      && D_{A}^{\mu}D_{B}^{\nu} \rbrace
     \frac{1}{(p-p_{3})^{2}-m_{A}^{2}}
      \frac{1}{(p-p_{3})^{2}-m_{B}^{2}} , \\
 \Phi_{a2}&=& tr \lbrace {\rm g}_{NN^{*}}^{A}\gamma_{A}
      (\not\! p_{4}+m^{*}) {\rm g}_{NN}^{B}\gamma_{B}
      (\not\! p_{2}+m^{*}){\rm g}_{NN^{*}}^{A}\gamma_{A}
      (\not\! p_{3}+m^{*}_{N^{*}}){\rm g}_{N^{*}N^{*}}^{B}\gamma_{B}
      (\not\! p +m^{*}_{N^{*}}) \nonumber \\                 
      && D_{A}^{\mu}D_{B}^{\nu} \rbrace
     \frac{1}{(p-p_{4})^{2}-m_{A}^{2}}
      \frac{1}{(p-p_{3})^{2}-m_{B}^{2}} , \\
 \Phi_{a3}&=& tr \lbrace {\rm g}_{NN^{*}}^{A}\gamma_{A}
      (\not\! p_{3}+m^{*}) {\rm g}_{NN^{*}}^{B}\gamma_{B}
     tr[ {\rm g}_{NN^{*}}^{A}\gamma_{A}(\not\! p_{4}+m^{*}_{N^{*}})
      {\rm g}_{NN^{*}}^{B}\gamma_{B}(\not\! p_{2}+m^{*}) ] 
      (\not\! p +m^{*}_{N^{*}}) \nonumber \\                   
      && D_{A}^{\mu}D_{B}^{\nu} \rbrace
     \frac{1}{(p-p_{3})^{2}-m_{A}^{2}}
      \frac{1}{(p-p_{3})^{2}-m_{B}^{2}} , \\
 \Phi_{a4}&=& tr \lbrace {\rm g}_{N^{*}N^{*}}^{A}\gamma_{A}
      (\not\! p_{4}+m^{*}_{N^{*}}) {\rm g}_{NN^{*}}^{B}\gamma_{B}
      (\not\! p_{2}+m^{*}){\rm g}_{NN}^{A}\gamma_{A}
      (\not\! p_{3}+m^{*}){\rm g}_{NN^{*}}^{B}\gamma_{B}
      (\not\! p +m^{*}_{N^{*}}) \nonumber \\                   
      && D_{A}^{\mu}D_{B}^{\nu} \rbrace
     \frac{1}{(p-p_{4})^{2}-m_{A}^{2}}
      \frac{1}{(p-p_{3})^{2}-m_{B}^{2}} , \\
 \Phi_{d1}&=& tr \lbrace {\rm g}_{NN^{*}}^{A}\gamma_{A}
      (\not\! p_{3}+m^{*}) {\rm g}_{NN^{*}}^{B}\gamma_{B}
     tr[ {\rm g}_{NN}^{A}\gamma_{A}(\not\! p_{4}+m^{*})
      {\rm g}_{NN}^{B}\gamma_{B}(\not\! p_{2}+m^{*}) ] 
      (\not\! p +m^{*}_{N^{*}}) \nonumber \\                 
      && D_{A}^{\mu}D_{B}^{\nu} \rbrace
     \frac{1}{(p-p_{3})^{2}-m_{A}^{2}}
      \frac{1}{(p-p_{3})^{2}-m_{B}^{2}} , \\
 \Phi_{d2}&=& tr \lbrace {\rm g}_{NN^{*}}^{A}\gamma_{A}
      (\not\! p_{4}+m^{*}) {\rm g}_{NN}^{B}\gamma_{B}
      (\not\! p_{2}+m^{*}){\rm g}_{NN}^{A}\gamma_{A}
      (\not\! p_{3}+m^{*}){\rm g}_{NN^{*}}^{B}\gamma_{B}
      (\not\! p +m^{*}_{N^{*}}) \nonumber \\                 
      && D_{A}^{\mu}D_{B}^{\nu} \rbrace
     \frac{1}{(p-p_{4})^{2}-m_{A}^{2}}
      \frac{1}{(p-p_{3})^{2}-m_{B}^{2}} , \\
 \Phi_{e1}&=& -tr \lbrace {\rm g}_{\Delta N^{*}}^{\pi}
      (\not\! p_{3}+m^{*}_{\Delta}) (p-p_{3})^{\nu}
      (p-p_{3})^{\mu} D_{\nu \mu}(p_{3}) {\rm g}_{\Delta N^{*}}^{\pi}
     tr[ {\rm g}_{NN}^{\pi}(\not\! p-\not\! p_{3})\gamma_{5}        
      (\not\! p_{2} + m^{*}) \nonumber \\                              
      &&{\rm g}_{NN}^{\pi}(\not\! p - \not\! p_{3})
     \gamma_{5}  (\not\! p_{4}+m^{*}) ]
      (\not\! p +m^{*}_{N^{*}}) \rbrace
     \frac{1}{(p-p_{3})^{2}-m_{\pi}^{2}}
      \frac{1}{(p-p_{3})^{2}-m_{\pi}^{2}} , \\
 \Phi_{e2}&=& - tr \lbrace {\rm g}_{NN^{*}}^{\pi}(\not\! p - \not\! p_{4})
      \gamma_{5}(\not\! p_{4}+m^{*}) {\rm g}_{NN}^{\pi}(\not\! p- \not\!
      p_{3})\gamma_{5}
      (\not\! p_{2}+m^{*}){\rm g}_{\Delta N}^{\pi}
      (\not\! p_{3}+m_{\Delta}^{*})
      (p - p_{4})^{\rho}\nonumber \\                          
     && (p - p_{3})^{\mu}D_{\rho \mu}(p_{3})
      {\rm g}_{\Delta N^{*}}^{\pi}
      (\not\! p +m^{*}_{N^{*}}) \rbrace
     \frac{1}{(p-p_{4})^{2}-m_{\pi}^{2}}
      \frac{1}{(p-p_{3})^{2}-m_{\pi}^{2}} , \\
 \Phi_{e3}&=& - tr \lbrace {\rm g}_{NN^{*}}^{\pi}(\not\! p - \not\! p_{3})
      \gamma_{5}(\not\! p_{3}+m^{*}) {\rm g}_{NN^{*}}^{\pi}(\not\! p- \not\!
      p_{3})\gamma_{5}
      tr [ {\rm g}_{\Delta N}^{\pi}                
      (\not\! p_{2}+m^{*}){\rm g}_{\Delta N}^{\pi} 
      (p_{4} + m_{\Delta}^{*}) \nonumber \\                           
     && (p - p_{3})^{\rho}(p - p_{3})^{\sigma}
      D_{\rho \sigma}(p_{4}) ]
      (\not\! p +m^{*}_{N^{*}}) \rbrace
     \frac{1}{(p-p_{3})^{2}-m_{\pi}^{2}}
      \frac{1}{(p-p_{3})^{2}-m_{\pi}^{2}} , \\
 \Phi_{e4}&=& -tr \lbrace {\rm g}_{\Delta N^{*}}^{\pi}
      (\not\! p_{4}+m^{*}_{\Delta}) (p-p_{4})^{\nu}
      (p-p_{3})^{\sigma} D_{\nu \sigma}(p_{4}) {\rm g}_{\Delta N}^{\pi}
      (\not\! p_{2} + m^{*}){\rm g}_{NN}^{\pi}             
       (\not\! p - \not\! p_{4}) \nonumber \\                      
      &&  \gamma_{5}  (\not\! p_{3}+m^{*}) 
     {\rm g}_{N N^{*}}^{\pi}(\not\! p - \not\! p_{3}) \gamma_{5} 
      (\not\! p +m^{*}_{N^{*}}) \rbrace
     \frac{1}{(p-p_{4})^{2}-m_{\pi}^{2}}
      \frac{1}{(p-p_{3})^{2}-m_{\pi}^{2}} , \\
 \Phi_{f1}&=& tr \lbrace {\rm g}_{\Delta N^{*}}^{\pi}
      (\not\! p_{3}+m^{*}_{\Delta}) (p-p_{3})^{\nu}
      (p-p_{3})^{\mu} D_{\nu \mu}(p_{3}) {\rm g}_{\Delta N^{*}}^{\pi}
     tr[ {\rm g}_{\Delta N}^{\pi}        
      (\not\! p_{2} + m^{*}) {\rm g}_{\Delta N}^{\pi}
      (\not\! p_{4} + m^{*}_{\Delta}) \nonumber \\                              
      &&(p -  p_{3})^{\rho} (p-p_{3})^{\sigma}
      D_{\rho \sigma}(p_{4}) ]              
      (\not\! p +m^{*}_{N^{*}}) \rbrace
     \frac{1}{(p-p_{3})^{2}-m_{\pi}^{2}}
      \frac{1}{(p-p_{3})^{2}-m_{\pi}^{2}} , \\
 \Phi_{{\rm g}1}&=& -tr \lbrace {\rm g}_{\Delta N^{*}}^{\pi}
      (\not\! p_{3}+m^{*}_{\Delta}) (p-p_{3})^{\nu}
      (p-p_{3})^{\mu} D_{\nu \mu}(p_{3}) {\rm g}_{\Delta N^{*}}^{\pi}
     tr[ {\rm g}_{NN^{*}}^{\pi}(\not\! p-\not\! p_{3})\gamma_{5}        
      (\not\! p_{2} + m^{*}) \nonumber \\                              
      &&{\rm g}_{NN^{*}}^{\pi}(\not\! p - \not\! p_{3})
     \gamma_{5}  (\not\! p_{4}+m^{*}_{N^{*}}) ]
      (\not\! p +m^{*}_{N^{*}}) \rbrace
     \frac{1}{(p-p_{3})^{2}-m_{\pi}^{2}}
      \frac{1}{(p-p_{3})^{2}-m_{\pi}^{2}} , \\
 \Phi_{{\rm g}2}&=& - tr \lbrace {\rm g}_{N^{*}N^{*}}^{\pi}(\not\! p - \not\! p_{4})
 \gamma_{5}(\not\! p_{4}+m^{*}_{N^{*}}) {\rm g}_{NN^{*}}^{\pi}(\not\! p- \not\!
      p_{3})\gamma_{5}
      (\not\! p_{2}+m^{*}){\rm g}_{\Delta N}^{\pi}
      (\not\! p_{3}+m_{\Delta}^{*})
      (p - p_{4})^{\rho}\nonumber \\                          
     && (p - p_{3})^{\mu}D_{\rho \mu}(p_{3})
      {\rm g}_{\Delta N^{*}}^{\pi}
      (\not\! p +m^{*}_{N^{*}}) \rbrace
     \frac{1}{(p-p_{4})^{2}-m_{\pi}^{2}}
      \frac{1}{(p-p_{3})^{2}-m_{\pi}^{2}} , \\
 \Phi_{{\rm g}3}&=& - tr \lbrace {\rm g}_{N^{*}N^{*}}^{\pi}(\not\! p - \not\! p_{3})
      \gamma_{5}(\not\! p_{3}+m^{*}_{N^{*}}) {\rm g}_{N^{*}N^{*}}^{\pi}(\not\! p- \not\!
      p_{3})\gamma_{5}
      tr [ {\rm g}_{\Delta N}^{\pi}                
      (\not\! p_{2}+m^{*}){\rm g}_{\Delta N}^{\pi} 
      (p_{4} + m_{\Delta}^{*}) \nonumber \\                           
     && (p - p_{3})^{\rho}(p - p_{3})^{\sigma}
      D_{\rho \sigma}(p_{4}) ]
      (\not\! p +m^{*}_{N^{*}}) \rbrace
     \frac{1}{(p-p_{3})^{2}-m_{\pi}^{2}}
      \frac{1}{(p-p_{3})^{2}-m_{\pi}^{2}} , \\
 \Phi_{{\rm g}4}&=& -tr \lbrace {\rm g}_{\Delta N^{*}}^{\pi}
      (\not\! p_{4}+m^{*}_{\Delta}) (p-p_{4})^{\nu}
      (p-p_{3})^{\sigma} D_{\nu \sigma}(p_{4}) {\rm g}_{\Delta N}^{\pi}
      (\not\! p_{2} + m^{*}){\rm g}_{NN^{*}}^{\pi}             
       (\not\! p - \not\! p_{4}) \nonumber \\                      
      &&  \gamma_{5}  (\not\! p_{3}+m^{*}_{N^{*}}) 
     {\rm g}_{N^{*} N^{*}}^{\pi}(\not\! p - \not\! p_{3}) \gamma_{5} 
      (\not\! p +m^{*}_{N^{*}}) \rbrace
     \frac{1}{(p-p_{4})^{2}-m_{\pi}^{2}}
      \frac{1}{(p-p_{3})^{2}-m_{\pi}^{2}} , \\
 \Phi_{j1}&=& tr \lbrace {\rm g}_{NN^{*}}^{A}\gamma_{A}
      (\not\! p_{3}+m^{*}) {\rm g}_{NN^{*}}^{B}\gamma_{B}
     tr[ {\rm g}_{\Delta \Delta}^{B}\gamma_{B}(\not\! p_{2}+m^{*}_{\Delta})
    D_{\sigma \rho}(p_{2}){\rm g}_{\Delta \Delta}^{A}\gamma_{A} 
    (\not\! p_{4} +m^{*}_{\Delta})D^{\rho \sigma}(p_{4}) ] \nonumber \\ 
      && (\not\! p + m^{*}_{N^{*}}) D_{A}^{\mu}D_{B}^{\nu} \rbrace
     \frac{1}{(p-p_{3})^{2}-m_{A}^{2}}
      \frac{1}{(p-p_{3})^{2}-m_{B}^{2}} , \\
 \Phi_{j3}&=& tr \lbrace {\rm g}_{\Delta N^{*}}^{\pi}
      (\not\! p_{3}+m^{*}_{\Delta}) (p-p_{3})^{\nu}
      (p-p_{3})^{\mu} D_{\nu \mu}(p_{3}) {\rm g}_{\Delta N^{*}}^{\pi}
     tr[ {\rm g}_{\Delta N}^{\pi}(\not\! p_{2} + m^{*}_{\Delta})          
      (p - p_{3})^{\sigma} \nonumber \\                              
      &&(p - p_{3})^{\rho} D_{\sigma \rho}(p_{2}) {\rm g}_{\Delta N}^{\pi}
      (\not\! p_{4}+m^{*}) ]
      (\not\! p +m^{*}_{N^{*}}) \rbrace
     \frac{1}{(p-p_{3})^{2}-m_{\pi}^{2}}
      \frac{1}{(p-p_{3})^{2}-m_{\pi}^{2}} , \\
 \Phi_{m1}&=& tr \lbrace {\rm g}_{NN^{*}}^{A}\gamma_{A}
      (\not\! p_{3}+m^{*}) {\rm g}_{NN^{*}}^{B}\gamma_{B}
     tr[ {\rm g}_{NN^{*}}^{B}\gamma_{B}(\not\! p_{2}+m^{*}_{N^{*}})
      {\rm g}_{NN^{*}}^{A}\gamma_{A}(\not\! p_{4}+m^{*}) ] 
      (\not\! p +m^{*}_{N^{*}}) \nonumber \\                 
      && D_{A}^{\mu}D_{B}^{\nu} \rbrace
     \frac{1}{(p-p_{3})^{2}-m_{A}^{2}}
      \frac{1}{(p-p_{3})^{2}-m_{B}^{2}} , \\
 \Phi_{m2}&=& tr \lbrace {\rm g}_{NN^{*}}^{A}\gamma_{A}
      (\not\! p_{4}+m^{*}) {\rm g}_{NN^{*}}^{B}\gamma_{B}
      (\not\! p_{2}+m^{*}_{N^{*}}){\rm g}_{NN^{*}}^{A}\gamma_{A}
      (\not\! p_{3}+m^{*}){\rm g}_{NN^{*}}^{B}\gamma_{B}
      (\not\! p +m^{*}_{N^{*}}) \nonumber \\                 
      && D_{A}^{\mu}D_{B}^{\nu} \rbrace
     \frac{1}{(p-p_{4})^{2}-m_{A}^{2}}
      \frac{1}{(p-p_{3})^{2}-m_{B}^{2}} , \\
 \Phi_{n1}&=& -tr \lbrace {\rm g}_{\Delta N^{*}}^{\pi}
      (\not\! p_{3}+m^{*}_{\Delta}) (p-p_{3})^{\nu}
      (p-p_{3})^{\mu} D_{\nu \mu}(p_{3}) {\rm g}_{\Delta N^{*}}^{\pi}
     tr[ {\rm g}_{NN^{*}}^{\pi}(\not\! p-\not\! p_{3})\gamma_{5}        
      (\not\! p_{2} + m^{*}_{N^{*}}) \nonumber \\                              
      &&{\rm g}_{NN^{*}}^{\pi}(\not\! p - \not\! p_{3})
     \gamma_{5}  (\not\! p_{4}+m^{*}) ]
      (\not\! p +m^{*}_{N^{*}}) \rbrace
     \frac{1}{(p-p_{3})^{2}-m_{\pi}^{2}}
      \frac{1}{(p-p_{3})^{2}-m_{\pi}^{2}} , \\
 \Phi_{n2}&=& - tr \lbrace {\rm g}_{NN^{*}}^{\pi}(\not\! p - \not\! p_{4})
 \gamma_{5}(\not\! p_{4}+m^{*}) {\rm g}_{NN^{*}}^{\pi}(\not\! p- \not\!
      p_{3})\gamma_{5}
      (\not\! p_{2}+m^{*}_{N^{*}}){\rm g}_{\Delta N^{*}}^{\pi}
      (\not\! p_{3}+m_{\Delta}^{*})
      (p - p_{4})^{\rho}\nonumber \\                          
     && (p - p_{3})^{\mu}D_{\rho \mu}(p_{3})
      {\rm g}_{\Delta N^{*}}^{\pi}
      (\not\! p +m^{*}_{N^{*}}) \rbrace
     \frac{1}{(p-p_{4})^{2}-m_{\pi}^{2}}
      \frac{1}{(p-p_{3})^{2}-m_{\pi}^{2}} , \\
 \Phi_{n3}&=& - tr \lbrace {\rm g}_{NN^{*}}^{\pi}(\not\! p - \not\! p_{3})
      \gamma_{5}(\not\! p_{3}+m^{*}) {\rm g}_{NN^{*}}^{\pi}(\not\! p- \not\!
      p_{3})\gamma_{5}
      tr [ {\rm g}_{\Delta N^{*}}^{\pi}                
      (\not\! p_{2}+m^{*}_{N^{*}}){\rm g}_{\Delta N^{*}}^{\pi} 
      (\not\! p_{4} + m_{\Delta}^{*}) \nonumber \\                           
     && (p - p_{3})^{\rho}(p - p_{3})^{\sigma}
      D_{\rho \sigma}(p_{4}) ]
      (\not\! p +m^{*}_{N^{*}}) \rbrace
     \frac{1}{(p-p_{3})^{2}-m_{\pi}^{2}}
      \frac{1}{(p-p_{3})^{2}-m_{\pi}^{2}} , \\
 \Phi_{n4}&=& -tr \lbrace {\rm g}_{\Delta N^{*}}^{\pi}
      (\not\! p_{4}+m^{*}_{\Delta}) (p-p_{4})^{\nu}
      (p-p_{3})^{\sigma} D_{\nu \sigma}(p_{4}) {\rm g}_{\Delta N^{*}}^{\pi}
      (\not\! p_{2} + m^{*}_{N^{*}}){\rm g}_{NN^{*}}^{\pi}             
       (\not\! p - \not\! p_{4}) \nonumber \\                      
      &&  \gamma_{5}  (\not\! p_{3}+m^{*}) 
     {\rm g}_{N N^{*}}^{\pi}(\not\! p - \not\! p_{3}) \gamma_{5} 
      (\not\! p +m^{*}_{N^{*}}) \rbrace
     \frac{1}{(p-p_{4})^{2}-m_{\pi}^{2}}
      \frac{1}{(p-p_{3})^{2}-m_{\pi}^{2}} , 
  \end{eqnarray}
where
  \begin{equation}
 D_{\mu \nu}(p)={\rm g}_{\mu \nu} - \frac{1}{3}\gamma_{\mu}\gamma_{\nu}
  -\frac{1}{3m_{\Delta}^{*}}(\gamma_{\mu}p_{\nu}-\gamma_{\nu}p_{\mu})
  -\frac{2}{3m_{\Delta}^{*2}}p_{\mu}p_{\nu},
  \end{equation}
and  $m^{*}$, $m^{*}_{\Delta}$ and $m^{*}_{N^{*}}$ are the effective masses of
 the nucleon, delta and $N^{*}$(1440)
 which are defined in Sect. II.                                          

 \end{sloppypar}
 \setcounter{equation}{0}
\renewcommand{\theequation}{C\arabic{equation}}
 \begin{center}
{\bf APPENDIX C}
 \end{center}
 \begin{sloppypar}
In this appendix we present the analytical expressions for
differential cross section of in-medium $NN$ elastic scattering including 
the contribution of pion. The in-medium $N^{*}N^{*}$ elastic differential
cross section can be obtained by replacing  $m^{*}$ with $m^{*}_{N^{*}}$
and ${\rm g}_{NN}^{A}$ with ${\rm g}_{N^{*}N^{*}}^{A}$.
  \begin{equation}
  \sigma_{NN \rightarrow NN}(s,t) = \frac{1}{(2 \pi)^{2} s} \lbrack D(s,t)
  + E(s,t) + (s, t \longleftrightarrow u) \rbrack, 
  \end{equation}
 \begin{eqnarray}
  D(s,t)&=& \frac{({\rm g}_{NN}^{\sigma})^{4}}{2(t-m_{\sigma}^{2})^{2}}
 (t- 4 m^{*2})^{2} + \frac{({\rm g}_{NN}^{\omega})^{4}}{(t-m_{\omega}^{2})^{2}}
 (2 s^{2} + 2st +t^{2} -8m^{*2}s +8m^{*4}) \nonumber \\
  &&  + \frac{24({\rm g}_{NN}^{\pi})^{4}}{(t- m_{\pi}^{2})^{2}} m^{*4}t^{2}
 - \frac{4({\rm g}_{NN}^{\sigma}{\rm g}_{NN}^{\omega})^{2}}
{(t - m_{\sigma}^{2})(t-m_{\omega}^{2})} (2s + t -4m^{*2})m^{*2}, \\
  E(s,t)&=& -\frac{({\rm g}_{NN}^{\sigma})^{4}}{8(t-m_{\sigma}^{2})
 (u-m_{\sigma}^{2})} \lbrack t (t+s) + 4m^{*2}(s-t) \rbrack
  + \frac{({\rm g}_{NN}^{\omega})^{4}}{2(t-m_{\omega}^{2})(u-m_{\omega}^{2})}
 (s- 2m^{*2}) \nonumber \\ && \times (s-6m^{*2})  
    - \frac{6({\rm g}_{NN}^{\pi})^{4}}{(t- m_{\pi}^{2})(u-m_{\pi}^{2})}
  (4m^{*2}-s -t ) m^{*4}t \nonumber \\
  && + ({\rm g}_{NN}^{\sigma}{\rm g}_{NN}^{\omega})^{2}
 \lbrack \frac{t^{2} - 4m^{*2}s -10m^{*2}t +24m^{*4}}{4(t-m_{\sigma}^{2})
(u-m_{\omega}^{2})} + \frac{(t+s)^{2} - 2m^{*2}s + 2m^{*2}t }{4 (t-m_{\omega}
^{2})(u-m_{\sigma}^{2})}  \rbrack \nonumber \\
  && + ({\rm g}_{NN}^{\sigma}{\rm g}_{NN}^{\pi})^{2}
 \lbrack \frac{3m^{*2}(4m^{*2}-s-t)(4m^{*2}-t)}{2(t-m_{\sigma}^{2})
(u-m_{\pi}^{2})} + \frac{3t(t+s)m^{*2}}{2 (t-m_{\pi}
^{2})(u-m_{\sigma}^{2})}  \rbrack \nonumber \\
  && + ({\rm g}_{NN}^{\omega}{\rm g}_{NN}^{\pi})^{2}
 \lbrack \frac{3m^{*2}(t+s-4m^{*2})(t+s-2m^{*2})}{(t-m_{\omega}^{2})
(u-m_{\pi}^{2})} + \frac{3m^{*2}(t^{2}-2m^{*2}t)}{ (t-m_{\pi}
^{2})(u-m_{\omega}^{2})}  \rbrack, 
  \end{eqnarray}
where the function D represents the contribution of the direct term and E is the
exchange term and
  \begin{eqnarray}
  &&s=(p+p_{2})^{2}= \lbrack E^{*}(p)+E^{*}(p_{2}) \rbrack ^{2}
   - ( {\bf p} + {\bf p}_{2})^{2}, \\
  &&t=(p-p_{3})^{2}=\frac{1}{2}(s - 4m^{*2}) ( \cos \theta -1 ), \\
  &&u=(p-p_{4})^{2}=4m^{*2}-s -t,
  \end{eqnarray}
  $\theta$ is the scattering angle in c.m. system  and
  \begin{eqnarray}
  &&\mid {\bf p} \mid =\mid {\bf p}_{3} \mid = \frac{1}{2}\sqrt{s-4m^{*2}}, \\
  && E^{*}(p)=\sqrt{{\bf p}^{2}+m^{*2}}, \\
  && E^{*}(p_{2})=\sqrt{{\bf p}_{2}^{2}+m^{*2} }.
  \end{eqnarray}

 \end{sloppypar}
 \setcounter{equation}{0}
\renewcommand{\theequation}{D\arabic{equation}}
 \begin{center}
{\bf APPENDIX D}
 \end{center}
 \begin{sloppypar}
 Here we present the analytical expressions of in-medium differential 
cross sections for different channels.  \\
 (a) Differential cross section of in-medium $N^{*}N \longrightarrow N^{*}N$
   scattering:
  \begin{equation}
\sigma_{N^{*}N \rightarrow N^{*}N}(s,t) = \frac{1}{(2 \pi)^{2} s} \lbrack D(s,t)
  + E(s,t) \rbrack, 
  \end{equation}
 \begin{eqnarray}
D(s,t) & = & \frac{({\rm g}_{NN}^{\sigma})^{2}({\rm g}_{N^{*}N^{*}}^{\sigma})
 ^{2}}{2(u-m_{\sigma}^{2})^{2}}(4m^{*2}-u)(4m^{*2}_{N^{*}}-u)
 + \frac{({\rm g}_{NN^{*}}^{\sigma})^{4}}{2(t-m_{\sigma}^{2})^{2}}
 \lbrack (m^{*}_{N^{*}} + m^{*})^{2} -t \rbrack ^{2} \nonumber \\
 && + \frac{({\rm g}_{NN}^{\omega})^{2}({\rm g}_{N^{*}N^{*}}^{\omega})^{2}}
 {(u-m_{\omega}^{2})^{2}} \lbrack 2(m^{*2}+m^{*2}_{N^{*}})^{2}
 +s(s-4m^{*2}-4m^{*2}_{N^{*}}) + (s+u)^{2} \rbrack \nonumber \\
 && + \frac{({\rm g}_{NN^{*}}^{\omega})^{4}}{(t-m_{\omega}^{2})^{2}}
 \lbrack (m^{*2}_{N^{*}}+m^{*2})^{2}+(s+t)^{2}+4m^{*2}(m^{*2}_{N^{*}}-s)
 +s(s-4m^{*2}_{N^{*}}) \nonumber \\
&& -2(t+2m^{*}m^{*}_{N^{*}})(m^{*}_{N^{*}}-m^{*})^{2} \rbrack 
 +\frac{24({\rm g}_{NN}^{\pi})^{2}({\rm g}_{N^{*}N^{*}}^{\pi})^{2}}
 {(u-m_{\pi}^{2})^{2}}m^{*2}m^{*2}_{N^{*}}u^{2} \nonumber \\
&& + \frac{3({\rm g}_{NN^{*}}^{\pi})^{4}}{2(t-m_{\pi}^{2})^{2}}
 \lbrack (m^{*}_{N^{*}}-m^{*})^{2}-t \rbrack ^{2}(m^{*}_{N^{*}}+m^{*})^{4} 
 \nonumber \\
&&+ \frac{4{\rm g}_{NN}^{\sigma}{\rm g}_{N^{*}N^{*}}^{\sigma}{\rm g}_{NN}
^{\omega}{\rm g}_{N^{*}N^{*}}^{\omega}}{(u-m_{\sigma}^{2})(u-m_{\omega}^{2})}
m^{*}m^{*}_{N^{*}}(2m^{*2}+2m^{*2}_{N^{*}}-2s-u) \nonumber \\
&& + \frac{2({\rm g}_{NN^{*}}^{\sigma})^{2}({\rm g}_{NN^{*}}^{\omega})^{2} }
{(t-m_{\sigma}^{2})(t-m_{\omega}^{2}) } [(m^{*2}+m^{*2}_{N^{*}})
 (2m^{*2}+2m^{*2}_{N^{*}} -s -t) \nonumber \\
&&+2m^{*}m^{*}_{N^{*}}(m^{*}_{N^{*}}-m^{*})^{2}-2m^{*}
m^{*}_{N^{*}}s \rbrack , \\
E(s,t)&=& \frac{{\rm g}_{NN}^{\sigma}{\rm g}_{N^{*}N^{*}}^{\sigma}
({\rm g}_{NN^{*}}^{\sigma})^{2}}{4(t-m_{\sigma}^{2})(u-m_{\sigma}^{2})}
 \lbrack 2(m^{*2}_{N^{*}}-m^{*2})^{2}+(m^{*}_{N^{*}}+m^{*})^{2}(t-s)-st
 -t^{2} \rbrack \nonumber \\
  && + \frac{{\rm g}_{NN}^{\omega}{\rm g}_{N^{*}N^{*}}^{\omega}
({\rm g}_{NN^{*}}^{\omega})^{2}}{(t-m_{\omega}^{2})(u-m_{\omega}^{2})}
 \lbrack 3(m^{*2}_{N^{*}}+m^{*2})^{2}-(m^{*}_{N^{*}}-m^{*})^{2}t
 \nonumber \\
 &&+s(s-3m^{*2}-3m^{*2}_{N^{*}}-2m^{*}m^{*}_{N^{*}}) \rbrack \nonumber \\
  && + \frac{3{\rm g}_{NN}^{\pi}{\rm g}_{N^{*}N^{*}}^{\pi}
({\rm g}_{NN^{*}}^{\pi})^{2}}{(t-m_{\pi}^{2})(u-m_{\pi}^{2})}
 m^{*}m^{*}_{N^{*}}(m^{*}+m^{*}_{N^{*}})^{2}(2m^{*2}+2m^{*2}_{N^{*}}-s-t)
 \lbrack (m^{*}_{N^{*}}-m^{*})^{2}-t \rbrack \nonumber \\
  && + \frac{({\rm g}_{NN^{*}}^{\sigma})^{2}{\rm g}_{NN}^{\omega}
{\rm g}_{N^{*}N^{*}}^{\omega}}{2(t-m_{\sigma}^{2})(u-m_{\omega}^{2})}
 \lbrack (m^{*2}_{N^{*}}+m^{*2})(3(m^{*}_{N^{*}}+m^{*})^{2}-3t-s)
 +t^{2}-2m^{*}m^{*}_{N^{*}}(s+2t)  \rbrack \nonumber \\
  && - \frac{{\rm g}_{NN}^{\sigma}{\rm g}_{N^{*}N^{*}}^{\sigma}
({\rm g}_{NN^{*}}^{\omega})^{2}}{2(u-m_{\sigma}^{2})(t-m_{\omega}^{2})}
 \lbrack 4(m^{*2}_{N^{*}}-m^{*2})^{2}+ 2m^{*}m^{*}_{N^{*}}(s-t)
 -(s+t)^{2}  \rbrack \nonumber \\
  && + \frac{3({\rm g}_{NN^{*}}^{\sigma})^{2}{\rm g}_{NN}^{\pi}
{\rm g}_{N^{*}N^{*}}^{\pi}}{(t-m_{\sigma}^{2})(u-m_{\pi}^{2})}
 m^{*}m^{*}_{N^{*}}(2m^{*2}+2m^{*2}_{N^{*}}-s-t)
 \lbrack (m^{*}_{N^{*}}+m^{*})^{2}-t \rbrack \nonumber \\
  && - \frac{3{\rm g}_{NN}^{\sigma}{\rm g}_{N^{*}N^{*}}^{\sigma}
({\rm g}_{NN^{*}}^{\pi})^{2}}{4(u-m_{\sigma}^{2})(t-m_{\pi}^{2})}
 (m^{*}+m^{*}_{N^{*}})^{2}
 \lbrack 2(m^{*2}_{N^{*}}-m^{*2})^{2}+ (m^{*}_{N^{*}}-m^{*})^{2}(t-s)
 -st-t^{2}  \rbrack \nonumber \\
  && + \frac{6({\rm g}_{NN^{*}}^{\omega})^{2}{\rm g}_{NN}^{\pi}
{\rm g}_{N^{*}N^{*}}^{\pi}}{(t-m_{\omega}^{2})(u-m_{\pi}^{2})}
 m^{*}m^{*}_{N^{*}}(2m^{*2}-2m^{*}m^{*}_{N^{*}}+2m^{*2}_{N^{*}}-s-t)
 \nonumber \\
 && \times (2m^{*2}+2m^{*2}_{N^{*}} -s -t) \nonumber \\
  && + \frac{3{\rm g}_{NN}^{\omega}{\rm g}_{N^{*}N^{*}}^{\omega}
({\rm g}_{NN^{*}}^{\pi})^{2}}{2(u-m_{\omega}^{2})(t-m_{\pi}^{2})}
 (m^{*}+m^{*}_{N^{*}})^{2}
 \lbrack 3(m^{*2}_{N^{*}}+m^{*2})(m^{*}_{N^{*}}-m^{*})^{2}
 \nonumber \\
 && - (m^{*2}_{N^{*}}+m^{*2})(s+3t) + 2m^{*}m^{*}_{N^{*}}(s+2t) +t^{2}
  \rbrack,
  \end{eqnarray}
 where
 \begin{eqnarray}
&&t=\frac{1}{2}(2m^{*2}_{N^{*}}+2m^{*2}-s)
  +\frac{1}{2s}(m_{N^{*}}^{*2}-m^{*2})^{2}
+2\mid {\bf p} \mid \mid {\bf p}_{3} \mid
 \cos \theta, \\
&&u=2m_{N^{*}}^{*2}+2m^{*2}-s-t,\\
&& \mid {\bf p} \mid =\mid {\bf p}_{3} \mid=\frac{1}{2 \sqrt{s}}\sqrt{(s-m^{*2}
-m_{N^{*}}^{*2})^{2}
 -4m^{*2}m_{N^{*}}^{*2}}.
 \end{eqnarray}
the definition of s is the same as in Eq. (C4),
$\theta$ is the scattering angle in c.m. system. In numerical calculations the
following constraints
 \begin{equation}
 t \leq 0,
 \hspace{3cm}
 u\leq 0
 \end{equation}
 must be guaranteed. Therefore
 \begin{equation}
 -1 \leq \cos \theta \leq \frac{s(s-2m_{N^{*}}^{*2}-2m^{*2})-(m_{N^{*}}^{*2}
 -m^{*2})^{2}}{s(s-2m_{N^{*}}^{*2}-2m^{*2})+(m_{N^{*}}^{*2}-m^{*2})^{2}}.
 \end{equation}
 \end{sloppypar}
 (b) Differential cross section of in-medium $N^{*}N \longrightarrow NN$
   scattering:
  \begin{equation}
\sigma_{N^{*}N \rightarrow NN}(s,t) = \frac{1}{(2 \pi)^{2} s} 
  \left[ \frac{s(s-4m^{*2})}{(s-m^{*2}-m^{*2}_{N^{*}})^{2}-4m^{*2}
 m^{*2}_{N^{*}} } \right] ^{1/2} 
  \lbrack D(s,t)+ E(s,t) + (s, t \longleftrightarrow u) \rbrack, 
  \end{equation}
 \begin{eqnarray}
 D(s,t) & = & \frac{({\rm g}_{NN}^{\sigma})^{2}({\rm g}_{NN^{*}}^{\sigma})^{2}}
 {2(t-m_{\sigma}^{2})^{2} }(4m^{*2}-t) \lbrack (m^{*}+m^{*}_{N^{*}})^{2}-t
 \rbrack + \frac{({\rm g}_{NN}^{\omega})^{2}({\rm g}_{NN^{*}}^{\omega})^{2} }
 {(t-m_{\omega}^{2})^{2} } \nonumber \\
 && \times \lbrack 2m^{*2}(m^{*}+m^{*}_{N^{*}})^{2} - t(m^{*}_{N^{*}}-m^{*})^{2}
 +(s+t)^{2}+s(s-6m^{*2}-2m^{*2}_{N^{*}}) \rbrack \nonumber \\
 && - \frac{6({\rm g}_{NN}^{\pi})^{2}({\rm g}_{NN^{*}}^{\pi})^{2} }
{(t-m_{\pi}^{2})^{2} }m^{*2}t (m^{*}+m^{*}_{N^{*}})^{2} \lbrack (m^{*}_{N^{*}}
 -m^{*})^{2} -t \rbrack \nonumber \\
 &&+ \frac{2{\rm g}_{NN}^{\sigma}{\rm g}_{NN^{*}}^{\sigma}{\rm g}_{NN}^{\omega}
 {\rm g}_{NN^{*}}^{\omega} }{(t-m_{\sigma}^{2})(t-m_{\omega}^{2}) }
 m^{*}(m^{*}+m^{*}_{N^{*}})(3m^{*2}+m^{*2}_{N^{*}}-2s-t), \\
E(s,t) & = &\frac{({\rm g}_{NN}^{\sigma})^{2}({\rm g}_{NN^{*}}^{\sigma})^{2} }
 {8 (t-m_{\sigma}^{2})(u-m_{\sigma}^{2}) } \lbrack m^{*}(m^{*}+m^{*}_{N^{*}})
 ( (m^{*}_{N^{*}}-m^{*})^{2} -2s ) 
  + t(3m^{*2}+m^{*2}_{N^{*}}-s-t) \rbrack  \nonumber \\                      
 && + \frac{({\rm g}_{NN}^{\omega})^{2}
 ({\rm g}_{NN^{*}}^{\omega})^{2} }{2(t-m_{\omega}^{2})(u-m_{\omega}^{2}) } 
 (4m^{*2} +m^{*}m^{*}_{N^{*}} + m^{*2}_{N^{*}}-s) 
 (m^{*2}+m^{*}m^{*}_{N^{*}}-s)  \nonumber \\                    
 && + \frac{3({\rm g}_{NN}^{\pi})^{2}
 ({\rm g}_{NN^{*}}^{\pi})^{2} }{ 2(t-m_{\pi}^{2} )(u-m_{\pi}^{2} ) }
 m^{*2}(m^{*}+m^{*}_{N^{*}})^{2} \lbrack m^{*}(m^{*}_{N^{*}}+m^{*})
 (m^{*}_{N^{*}}-m^{*})^{2} \nonumber \\
 && +t(s+t-m^{*2}_{N^{*}}-3m^{*2} ) \rbrack  + \frac{{\rm g}_{NN}^{\sigma}
 {\rm g}_{NN^{*}}^{\sigma}{\rm g}_{NN}^{\omega}{\rm g}_{NN^{*}}^{\omega}}
 {4(t-m_{\sigma}^{2})(u-m_{\omega}^{2})} \lbrack m^{*}(m^{*}+m^{*}_{N^{*}})
 \nonumber \\
 && \times (7m^{*2}-2s +4m^{*}m^{*}_{N^{*}} + m^{*2}_{N^{*}} ) + t(t
 -6m^{*2}-3m^{*}m^{*}_{N^{*}}-m^{*2}_{N^{*}} ) \rbrack  \nonumber \\
 &&- \frac{{\rm g}_{NN}^{\sigma}{\rm g}_{NN^{*}}^{\sigma}{\rm g}_{NN}^{\omega}
 {\rm g}_{NN^{*}}^{\omega} }{4 (u-m_{\sigma}^{2})(t-m_{\omega}^{2}) }
 \lbrack 2m^{*}(m^{*}_{N^{*}}+m^{*})(m^{*}_{N^{*}}-m^{*})^{2}
 +s(2m^{*2}+m^{*2}_{N^{*}} -m^{*}m^{*}_{N^{*}} ) \nonumber \\
 && + t(m^{*2}_{N^{*}} - 3m^{*}m^{*}_{N^{*}} ) -(s+t)^{2} \rbrack 
 + \frac{3{\rm g}_{NN}^{\sigma}{\rm g}_{NN^{*}}^{\sigma}{\rm g}_{NN}^{\pi}
 {\rm g}_{NN^{*}}^{\pi} }{4 (t-m_{\sigma}^{2})(u-m_{\pi}^{2}) }
 m^{*}(m^{*}+m^{*}_{N^{*}}) \nonumber \\
 && \times \lbrack m^{*2}(5m^{*2}+ 7m^{*}m^{*}_{N^{*}} + 3m^{*2}_{N^{*}}
 -2s -5t ) + m^{*}m^{*}_{N^{*}} (m^{*2}_{N^{*}} -2s-2t) \nonumber \\ &&+t(t+s-
 m^{*2}_{N^{*}} ) ] 
 - \frac{3{\rm g}_{NN}^{\sigma}{\rm g}_{NN^{*}}^{\sigma}{\rm g}_{NN}^{\pi}
 {\rm g}_{NN^{*}}^{\pi} }{4 (u-m_{\sigma}^{2})(t-m_{\pi}^{2}) }
 m^{*}(m^{*}+m^{*}_{N^{*}}) \nonumber \\ && \times
 \lbrack (m^{*}_{N^{*}} - m^{*})^{2}
 (m^{*2}+m^{*}m^{*}_{N^{*}}+t) -st -t^{2} \rbrack 
 + \frac{3{\rm g}_{NN}^{\omega}{\rm g}_{NN^{*}}^{\omega}{\rm g}_{NN}^{\pi}
 {\rm g}_{NN^{*}}^{\pi} }{2 (t-m_{\omega}^{2})(u-m_{\pi}^{2}) }
 m^{*}(m^{*}+m^{*}_{N^{*}}) \nonumber \\ &&  \times 
 (2m^{*2}+2m^{*}m^{*}_{N^{*}}-s-t) 
  (2m^{*2}-m^{*}m^{*}_{N^{*}}+m^{*2}_{N^{*}} -s -t) \nonumber \\
  && + \frac{3{\rm g}_{NN}^{\omega}{\rm g}_{NN^{*}}^{\omega}{\rm g}_{NN}^{\pi}
 {\rm g}_{NN^{*}}^{\pi} }{2 (u-m_{\omega}^{2})(t-m_{\pi}^{2}) }
 m^{*}(m^{*}+m^{*}_{N^{*}}) (m^{*2}+m^{*}m^{*}_{N^{*}}-t)
 \lbrack ( m^{*}_{N^{*}} - m^{*} )^{2} -t \rbrack.
 \end{eqnarray}
Here
 \begin{eqnarray}
&&t=\frac{1}{2}(3m^{*2} + m^{*2}_{N^{*}}-s)
+2\mid {\bf p} \mid \mid {\bf p}_{3} \mid
 \cos \theta, \\
&&u=m_{N^{*}}^{*2}+3m^{*2}-s-t,\\
&& \mid {\bf p} \mid =\frac{1}{2 \sqrt{s}}\sqrt{(s-m^{*2}
-m_{N^{*}}^{*2})^{2}
 -4m^{*2}m_{N^{*}}^{*2}}, \\
&& \mid {\bf p}_{3} \mid = \frac{1}{2} \sqrt{s-4m^{*2} }.
 \end{eqnarray}
 (c) Differential cross section of in-medium $N^{*}N \longrightarrow N\Delta$
   scattering:
  \begin{equation}
\sigma_{N^{*}N \rightarrow N\Delta}(s,t) = \frac{1}{(2 \pi)^{2} s} 
  \left[ \frac{(s-m^{*2}-m^{*2}_{\Delta})^{2}-4m^{*2}m^{*2}_{\Delta}}
  {(s-m^{*2}-m^{*2}_{N^{*}})^{2}-4m^{*2}
 m^{*2}_{N^{*}} } \right] ^{1/2} 
  \lbrack D(s,t)+ E(s,t) \rbrack, 
  \end{equation}
 \begin{eqnarray}
D(s,t) & = & -\frac{2({\rm g}_{NN}^{\pi})^{2}({\rm g}_{\Delta N^{*}}^{\pi})
 ^{2} }{3m^{*2}_{\Delta}(u-m_{\pi}^{2})^{2} }
 m^{*2}u \lbrack (m^{*}_{\Delta}+m^{*}_{N^{*}})^{2}-u \rbrack ^{2}
 \lbrack (m^{*}_{N^{*}}-m^{*}_{\Delta} )^{2}-u \rbrack \nonumber \\
 &  & +\frac{({\rm g}_{NN^{*}}^{\pi})^{2}({\rm g}_{\Delta N}^{\pi})
 ^{2} }{6m^{*2}_{\Delta}(t-m_{\pi}^{2})^{2} }
 (m^{*}+m^{*}_{N^{*}})^{2} \lbrack (m^{*}_{\Delta} + m^{*} )^{2} -t \rbrack
 ^{2} \lbrack (m^{*}_{\Delta} -m^{*})^{2} -t \rbrack \nonumber \\           
  && \lbrack (m^{*}_{N^{*}} -m^{*})^{2} -t \rbrack , \\
E(s,t) & = & \frac{{\rm g}_{NN}^{\pi}{\rm g}_{NN^{*}}^{\pi}{\rm g}_{\Delta N}
 ^{\pi}{\rm g}_{\Delta N^{*}}^{\pi} }{ 3m^{*2}_{\Delta} (t-m_{\pi}^{2}) 
 (u-m_{\pi}^{2}) } m^{*}(m^{*}+m^{*}_{N^{*}}) \sum^{9}_{i=1} E_{i} , 
 \end{eqnarray}
  \begin{eqnarray}
 E_{1} & = & m^{*5}_{\Delta}m^{*}_{N^{*}}
 \lbrack m^{*}_{\Delta}m^{*}_{N^{*}} + 3m^{*2}_{N^{*}} -s +t \rbrack 
 \nonumber \\
 && +m^{*4}_{\Delta} (m^{*4}_{N^{*}}-m^{*2}_{N^{*}}s + m^{*2}_{N^{*}} t
 -2st +t^{2} ), \\
 E_{2} & = & m^{*3}_{\Delta}m^{*}_{N^{*}} (3m^{*4}_{N^{*}} -4m^{*2}_{N^{*}}s
 -2m^{*2}_{N^{*}}t +s^{2} -2st -2t^{2} )  \nonumber \\     
 && + m^{*2}_{\Delta}t (2m^{*4}_{N^{*}}-4m^{*2}_{N^{*}}s-m^{*2}_{N^{*}}t
+2s^{2}-2t^{2} ) , \\
 E_{3} & = & m^{*}_{\Delta}m^{*}_{N^{*}}t \lbrack -2m^{*2}_{N^{*}}s
 -m^{*2}_{N^{*}}t + (2s+t)(s+t) \rbrack \nonumber \\
 && +m^{*}_{N^{*}}t^{2} (m^{*}t -m^{*}_{N^{*}}s -m^{*}_{N^{*}}t )
 +t^{2}(s+t)^{2}, \\
 E_{4} & = & m^{*2}m^{*2}_{\Delta} (2m^{*}_{\Delta}m^{*3}_{N^{*}} +6m^{*}
 _{\Delta}m^{*}_{N^{*}}t -2m^{*4}_{N^{*}}+4m^{*2}_{N^{*}}s +6m^{*2}_{N^{*}}t
 -s^{2}-2st +7t^{2} ) \nonumber \\
 && + m^{*2}m^{*}_{\Delta}m^{*}_{N^{*}} (2m^{*2}_{N^{*}}s +4m^{*2}_{N^{*}}t
 -s^{2}-8st-5t^{2} ) , \\
 E_{5} & = & 2m^{*2}m^{*2}_{N^{*}} t(s+2t) -2m^{*2}t(s+t)(s+3t)
 +m^{*}m^{*3}_{\Delta}(4m^{*}_{\Delta}m^{*3}_{N^{*}} -2m^{*}_{\Delta}
 m^{*}_{N^{*}}s \nonumber \\
 && + 2m^{*}_{\Delta}m^{*}_{N^{*}}t -4m^{*4}_{N^{*}} +4m^{*2}_{N^{*}}s
 -s^{2}+t^{2} ) , \\
 E_{6} & = & 2m^{*}m^{*}_{\Delta}m^{*2}_{N^{*}}(s+t)(t-m^{*}_{\Delta}
 m^{*}_{N^{*}}) +m^{*}m^{*2}_{\Delta}m^{*}_{N^{*}}(s+t)(s-3t) \nonumber \\
 && -m^{*}m^{*}_{\Delta}t(s+t)^{2} +m^{*}m^{*}_{N^{*}}st(s+2t) , \\
 E_{7} & = & 2m^{*6}(2m^{*2} +2m^{*}m^{*}_{\Delta} -2m^{*}m^{*}_{N^{*}}
 +2m^{*2}_{\Delta}-m^{*}_{\Delta}m^{*}_{N^{*}} +m^{*2}_{N^{*}}-2s-6t) 
 \nonumber \\
 && +4m^{*5}(m^{*}_{N^{*}} -m^{*}_{\Delta}) (s+2t+m^{*}_{\Delta}m^{*}_{N^{*}}), \\
 E_{8} & = & m^{*4}(-m^{*4}_{\Delta} -3m^{*3}_{\Delta}m^{*}_{N^{*}}
 -5m^{*2}_{\Delta}m^{*2}_{N^{*}} -8m^{*2}_{\Delta}t -3m^{*}_{\Delta}
 m^{*3}_{N^{*}} +3m^{*}_{\Delta}m^{*}_{N^{*}}s \nonumber \\                   
 &&+7m^{*}_{\Delta}m^{*}_{N^{*}}t
  -m^{*2}_{N^{*}}s -5m^{*2}_{N^{*}}t +s^{2}+10st +13t^{2} ), \\
 E_{9} & = & m^{*3} \lbrack 4m^{*2}_{\Delta}m^{*2}_{N^{*}} (m^{*}_{N^{*}} 
 -m^{*}_{\Delta} ) +2m^{*3}_{\Delta}(s-t) +2m^{*}_{\Delta}m^{*}_{N^{*}}
 (4m^{*}_{\Delta}t -m^{*}_{N^{*}}s -3m^{*}_{N^{*}}t ) \nonumber \\
 && + ( m^{*}_{\Delta} -m^{*}_{N^{*}} )(s+t)(s+5t) \rbrack 
 +2m^{*2}m^{*4}_{\Delta} (s-2m^{*2}_{N^{*}} ) . 
  \end{eqnarray}
Here
 \begin{eqnarray}
t&=&\frac{1}{2}(2m^{*2} + m^{*2}_{\Delta} + m^{*2}_{N^{*}}-s)
 + \frac{1}{2s} (m^{*2}_{N^{*}} -m^{*2} )(m^{*2}_{\Delta} -m^{*2})
 \nonumber \\
 && +2\mid {\bf p} \mid \mid {\bf p}_{3} \mid
 \cos \theta, \\
u&=&2m^{*2} + m^{*2}_{\Delta} +m^{*2}_{N^{*}}-s-t,\\
 \mid {\bf p} \mid &=&\frac{1}{2 \sqrt{s}}\sqrt{(s-m^{*2}
-m_{N^{*}}^{*2})^{2}
 -4m^{*2}m_{N^{*}}^{*2}}, \\
 \mid {\bf p}_{3} \mid & =&\frac{1}{2 \sqrt{s}}\sqrt{(s-m^{*2}
-m_{\Delta}^{*2})^{2}
 -4m^{*2}m_{\Delta}^{*2}}. 
  \end{eqnarray}
(d) Differential cross section of in-medium $N^{*}N\longrightarrow \Delta\Delta$
   scattering:
  \begin{equation}
\sigma_{N^{*}N \rightarrow \Delta\Delta}(s,t) = \frac{1}{(2 \pi)^{2} s} 
  \left[  \frac{s(s-4m^{*2}_{\Delta})}
  {(s-m^{*2}-m^{*2}_{N^{*}})^{2}-4m^{*2}
 m^{*2}_{N^{*}} } \right] ^{1/2} 
  \lbrack D(s,t)+ (s, t \longleftrightarrow u) \rbrack, 
  \end{equation}
 \begin{eqnarray}
 D(s,t) & = & \frac{({\rm g}_{\Delta N}^{\pi})^{2} ({\rm g}_{\Delta N^{*}}
 ^{\pi})^{2} }{ 54m^{*4}_{\Delta}(t-m_{\pi}^{2} )^{2} }
 \lbrack (m^{*}_{\Delta} +m^{*})^{2} -t \rbrack ^{2} \lbrack (m^{*}_{\Delta}
 -m^{*} )^{2} -t \rbrack  \nonumber \\
 && \lbrack (m^{*}_{N^{*}} +m^{*}_{\Delta} )^{2} -t \rbrack ^{2} 
 \lbrack (m^{*}_{N^{*}} -m^{*}_{\Delta} )^{2} -t \rbrack ,
 \end{eqnarray}
where
 \begin{eqnarray}
&&t=\frac{1}{2}(2m^{*2}_{\Delta} + m^{*2} + m^{*2}_{N^{*}}-s)
+2\mid {\bf p} \mid \mid {\bf p}_{3} \mid
 \cos \theta, \\
&&u=2m^{*2}_{\Delta} + m^{*2} +m^{*2}_{N^{*}} -s-t,\\
&& \mid {\bf p} \mid =\frac{1}{2 \sqrt{s}}\sqrt{(s-m^{*2}
-m_{N^{*}}^{*2})^{2}
 -4m^{*2}m_{N^{*}}^{*2}}, \\
&& \mid {\bf p}_{3} \mid = \frac{1}{2} \sqrt{s-4m^{*2}_{\Delta} }.
 \end{eqnarray}
 (e) Differential cross section of in-medium $N^{*}N \longrightarrow \Delta 
   N^{*} $ scattering:
  \begin{equation}
\sigma_{N^{*}N \rightarrow \Delta N^{*}}(s,t) = \frac{1}{(2 \pi)^{2} s} 
  \left[ \frac{(s-m^{*2}_{\Delta}-m^{*2}_{N^{*}})^{2}-4m^{*2}_{\Delta}m^{*2}_{N^{*}}}
  {(s-m^{*2}-m^{*2}_{N^{*}})^{2}-4m^{*2}
 m^{*2}_{N^{*}} } \right] ^{1/2} 
  \lbrack D(s,t)+ E(s,t) \rbrack, 
  \end{equation}
 \begin{eqnarray}
 D(s,t) & = & \frac{({\rm g}_{NN^{*}}^{\pi})^{2}({\rm g}_{\Delta N^{*}}^{\pi})
 ^{2} }{6m^{*2}_{\Delta} (u-m_{\pi}^{2})^{2} } (m^{*} +m^{*}_{N^{*}} )^{2}   
  \lbrack (m^{*}_{N^{*}} -m^{*}) ^{2} -u \rbrack \lbrack (m^{*}_{N^{*}}
 +m^{*}_{\Delta} )^{2} -u \rbrack ^{2} \nonumber \\
 && \times \lbrack (m^{*}_{N^{*}} -m^{*}_{\Delta}
 )^{2} -u \rbrack 
  - \frac{2({\rm g}_{N^{*}N^{*}}^{\pi} )^{2} ({\rm g}_{\Delta N}^{\pi})^{2} }
 {3m^{*2}_{\Delta} (t-m_{\pi}^{2} )^{2} } m^{*2}_{N^{*}} t
 \lbrack (m^{*}_{\Delta} +m^{*})^{2} -t \rbrack ^{2} \nonumber \\
  && \times \lbrack (m^{*}_{\Delta} -m^{*})^{2} -t \rbrack , \\
 E(s,t) & = & \frac{{\rm g}_{NN^{*}}^{\pi}{\rm g}_{N^{*}N^{*}}^{\pi}
 {\rm g}_{\Delta N}^{\pi}{\rm g}_{\Delta N^{*}}^{\pi} }
 {3m^{*2}_{\Delta} (t-m_{\pi}^{2}) (u-m_{\pi}^{2}) } m^{*}_{N^{*}}
 (m^{*} +m^{*}_{N^{*}} ) \sum_{i=1}^{8} E_{i}, 
 \end{eqnarray}
 \begin{eqnarray}
 E_{1} & = & -m^{*2}t^{2} (4s+3t) +m^{*}m^{*4}_{\Delta}m^{*}_{N^{*}}
 (m^{*}_{\Delta}m^{*}_{N^{*}} +m^{*2}_{N^{*}} -s+t ) \nonumber \\
 && + m^{*}m^{*2}_{\Delta}t (-2m^{*}_{\Delta}s +m^{*}_{\Delta}t +2m^{*3}_{N^{*}}
 -2m^{*}_{N^{*}}s ) , \\       
 E_{2} & = & m^{*}m^{*}_{\Delta}t (m^{*4}_{N^{*}} -2m^{*2}_{N^{*}}s
 +m^{*2}_{N^{*}}t +s^{2} -t^{2} ) +m^{*}m^{*}_{N^{*}}t^{2}(m^{*2}_{N^{*}}-s-t)
 \nonumber \\
 && + m^{*5}_{\Delta}m^{*}_{N^{*}} (m^{*}_{\Delta}m^{*}_{N^{*}} +m^{*2}_{N^{*}}
 -s +t ) , \\
 E_{3} & = & m^{*4}_{\Delta} (t^{2} -2st) +m^{*2}_{\Delta}t (-2m^{*}_{\Delta}
 m^{*}_{N^{*}}t +2m^{*4}_{N^{*}} -4m^{*2}_{N^{*}}s +m^{*2}_{N^{*}}t +2s^{2}
 -2t^{2} ) \nonumber \\
 && +m^{*}_{\Delta}m^{*}_{N^{*}}t^{2} (-m^{*2}_{N^{*}} +s +t) , \\
 E_{4} & = & m^{*2}_{N^{*}}t^{2} (m^{*2}_{N^{*}} -2s -2t) +t^{2}(s+t)^{2}, \\
 E_{5} & = & m^{*5} (m^{*2}m^{*}_{N^{*}} +m^{*}m^{*}_{\Delta}m^{*}_{N^{*}}
 -m^{*}t -2m^{*2}_{\Delta}m^{*}_{N^{*}} +m^{*}_{\Delta}m^{*2}_{N^{*}}
  \nonumber \\
 && -m^{*}_{\Delta}t +m^{*3}_{N^{*}} -m^{*}_{N^{*}}s -3m^{*}_{N^{*}}t ) , \\
 E_{6} & = & m^{*4} (-2m^{*3}_{\Delta}m^{*}_{N^{*}} +m^{*2}_{\Delta}m^{*2}
 _{N^{*}} +m^{*}_{\Delta}m^{*3}_{N^{*}} -m^{*}_{\Delta}m^{*}_{N^{*}}s 
 - m^{*}_{\Delta}m^{*}_{N^{*}}t \nonumber \\ 
 && -2m^{*2}_{N^{*}}t +2st +3t^{2} ) , \\
 E_{7} & = & m^{*3} (m^{*4}_{\Delta}m^{*}_{N^{*}} -2m^{*3}_{\Delta}m^{*2}
 _{N^{*}} +2m^{*3}_{\Delta}t -2m^{*2}_{\Delta}m^{*3}_{N^{*}} +2m^{*2}_{\Delta}
 m^{*}_{N^{*}}s +2m^{*2}_{\Delta}m^{*}_{N^{*}}t \nonumber \\
 && -2m^{*}_{\Delta}m^{*2}_{N^{*}}t +2m^{*}_{\Delta}t^{2} -2m^{*3}_{N^{*}}t
 + 2m^{*}_{N^{*}}st +3m^{*}_{N^{*}}t^{2} ) , \\
E_{8} & = & m^{*2} (m^{*5}_{\Delta}m^{*}_{N^{*}} -2m^{*4}_{\Delta}m^{*2}_{N^{*}}
 +2m^{*4}_{\Delta}t -2m^{*3}_{\Delta}m^{*3}_{N^{*}} +2m^{*3}_{\Delta}
 m^{*}_{N^{*}}s -2m^{*2}_{\Delta}st \nonumber \\ && +2m^{*2}_{\Delta}t^{2}
  -m^{*}_{\Delta}m^{*}_{N^{*}}t^{2} -m^{*4}_{N^{*}}t +2m^{*2}_{N^{*}}st
 + 4m^{*2}_{N^{*}}t^{2} -s^{2}t \rbrack .
 \end{eqnarray}
Here
 \begin{eqnarray}
t&=&\frac{1}{2}(2m^{*2}_{N^{*}} + m^{*2} + m^{*2}_{\Delta}-s)
 - \frac{1}{2s} (m^{*2}_{N^{*}} -m^{*2} )(m^{*2}_{N^{*}} -m^{*2}_{\Delta})
 \nonumber \\
 && +2\mid {\bf p} \mid \mid {\bf p}_{3} \mid
 \cos \theta, \\
u&=&2m^{*2}_{N^{*}} + m^{*2} + m^{*2}_{\Delta} -s-t,\\
 \mid {\bf p} \mid &=&\frac{1}{2 \sqrt{s}}\sqrt{(s-m^{*2}
-m_{N^{*}}^{*2})^{2}
 -4m^{*2}m_{N^{*}}^{*2}}, \\
 \mid {\bf p}_{3} \mid & =&\frac{1}{2 \sqrt{s}}\sqrt{(s-m^{*2}_{\Delta}
-m_{N^{*}}^{*2})^{2}
 -4m^{*2}_{\Delta}m_{N^{*}}^{*2}}. 
  \end{eqnarray}
 (f) Differential cross section of in-medium $N^{*}\Delta \longrightarrow N 
   \Delta $ scattering:
  \begin{equation}
\sigma_{N^{*}\Delta \rightarrow N\Delta }(s,t) = \frac{1}{(2 \pi)^{2} s} 
  \left[ \frac{(s-m^{*2}-m^{*2}_{\Delta})^{2}-4m^{*2}m^{*2}_{\Delta}}
  {(s-m^{*2}_{\Delta}-m^{*2}_{N^{*}})^{2}-4m^{*2}_{\Delta}
 m^{*2}_{N^{*}} } \right] ^{1/2} 
  \lbrack D(s,t)+ E(s,t)  \rbrack, 
  \end{equation}
 \begin{eqnarray}
 D(s,t) & = & \frac{({\rm g}_{N N^{*}}^{\sigma})^{2}({\rm g}_{\Delta \Delta}
 ^{\sigma})^{2} }{ 9m^{*4}_{\Delta}(u-m_{\sigma}^{2})^{2} }
 (4m^{*2}_{\Delta} - u)(18m^{*4}_{\Delta} -6m^{*2}_{\Delta}u +u^{2} )
 \lbrack (m^{*}_{N^{*}} +m^{*} )^{2} -u \rbrack \nonumber \\
  &  & -\frac{2({\rm g}_{N N^{*}}^{\omega})^{2}({\rm g}_{\Delta \Delta}
 ^{\omega})^{2} }{ 9m^{*4}_{\Delta}(u-m_{\omega}^{2})^{2} }
 \lbrace (2m^{*2}_{\Delta} -u)^{2} \lbrack (m^{*}_{N^{*}} -m^{*} )^{2}u
 -2(m^{*}m^{*}_{N^{*}} +m^{*2}_{\Delta} )^{2} \nonumber \\
 && +2s (2m^{*2}_{\Delta} +m^{*2}_{N^{*}} +m^{*2} -s -u) -u^{2} \rbrack
 +2m^{*2}m^{*4}_{\Delta}(14s +5u -2m^{*2} -10m^{*2}_{N^{*}} ) \nonumber \\
 && -4m^{*}m^{*4}_{\Delta}m^{*}_{N^{*}} (14m^{*2}_{\Delta} +u) 
 +2m^{*4}_{\Delta} (28m^{*2}_{\Delta}s -14m^{*4}_{\Delta} -2m^{*4}_{N^{*}}
 +14m^{*2}_{N^{*}}s   
 +5m^{*2}_{N^{*}}u \nonumber \\ && -14s^{2} -14su -3u^{2} ) \rbrace 
 + \frac{4{\rm g}_{N N^{*}}^{\sigma}{\rm g}_{N N^{*}}^{\omega}{\rm g}
 _{\Delta \Delta}^{\sigma}{\rm g}_{\Delta \Delta}^{\omega} }
 {9m^{*3}_{\Delta} (u-m_{\sigma}^{2} )( u-m_{\omega}^{2} ) }
 (m^{*2} +2m^{*2}_{\Delta} +m^{*2}_{N^{*}} -2s -u) \nonumber \\
 && \times (m^{*}+m^{*}_{N^{*}} ) ( 18m^{*4}_{\Delta} -6m^{*2}_{\Delta} u
 +u^{2} ) - \frac{5({\rm g}_{N N^{*}}^{\pi} )^{2} ({\rm g}_{\Delta \Delta}
 ^{\pi} )^{2} }{ 3m^{*2}_{\Delta} (u-m_{\pi}^{2} )^{2} }
 u (m^{*}+m^{*}_{N^{*}} )^{2} \nonumber \\
 && \times (10m^{*4}_{\Delta} -2m^{*2}_{\Delta}u +u^{2} ) \lbrack 
 (m^{*}_{N^{*}} -m^{*} )^{2} -u \rbrack
 + \frac{({\rm g }_{\Delta N}^{\pi} )^{2}({\rm g}_{\Delta N^{*}}^{\pi})^{2} }
 {54 m^{*4}_{\Delta} (t-m_{\pi}^{2} )^{2} } \lbrack (m^{*}_{\Delta} 
 +m^{*} )^{2} -t \rbrack ^{2}  \nonumber \\
 && \times \lbrack (m^{*}_{\Delta} -m^{*} )^{2} -t \rbrack 
 \lbrack (m^{*}_{N^{*}} +m^{*}_{\Delta} )^{2} -t \rbrack ^{2}
 \lbrack (m^{*}_{N^{*}} -m^{*}_{\Delta} )^{2} -t \rbrack ,
 \end{eqnarray}
Here
 \begin{eqnarray}
t&=&\frac{1}{2}(2m^{*2}_{\Delta} + m^{*2} + m^{*2}_{N^{*}}-s)
 - \frac{1}{2s} (m^{*2}_{N^{*}} -m^{*2}_{\Delta} )(m^{*2}_{\Delta} -m^{*2})
 \nonumber \\
 && +2\mid {\bf p} \mid \mid {\bf p}_{3} \mid
 \cos \theta, \\
u&=&2m^{*2}_{\Delta} + m^{*2} + m^{*2}_{N^{*}} -s-t,\\
 \mid {\bf p} \mid &=&\frac{1}{2 \sqrt{s}}\sqrt{(s-m^{*2}_{\Delta}
-m_{N^{*}}^{*2})^{2}
 -4m^{*2}_{\Delta}m_{N^{*}}^{*2}}, \\
 \mid {\bf p}_{3} \mid & =&\frac{1}{2 \sqrt{s}}\sqrt{(s-m^{*2}
-m_{\Delta}^{*2})^{2}
 -4m^{*2}m_{\Delta}^{*2}}. 
  \end{eqnarray}
 (g) Differential cross section of in-medium $N^{*}N^{*} \longrightarrow N 
   N $ scattering:
  \begin{equation}
\sigma_{N^{*}N^{*} \rightarrow N N }(s,t) = \frac{1}{(2 \pi)^{2} s} 
  \left[ \frac{s-4m^{*2}}
 {s-4m^{*2}_{N^{*}} } \right] ^{1/2} 
  \lbrack D(s,t)+ E(s,t) + (s, t \longleftrightarrow u) \rbrack, 
  \end{equation}
 \begin{eqnarray}
 D(s,t) & = & \frac{ ({\rm g}_{N N^{*}}^{\sigma} )^{4} }
 { 2(t-m_{\sigma}^{2} )^{2} } 
 \lbrack (m^{*}_{N^{*}} +m^{*} )^{2} -t \rbrack ^{2} 
   + \frac{ ({\rm g}_{N N^{*}}^{\omega} )^{4} }
 { (t-m_{\omega}^{2} )^{2} }  \lbrack (m^{*2} +m^{*2}_{N^{*}} )^{2} \nonumber \\
 && -2(m^{*}_{N^{*}} -m^{*} )^{2} (t+2m^{*}m^{*}_{N^{*}} ) +2s(s+t -2m^{*2}
 -2m^{*2}_{N^{*}} ) +4m^{*2}m^{*2}_{N^{*}} +t^{2} \rbrack \nonumber \\
 && +\frac{3({\rm g}_{N N^{*}}^{\pi} )^{4} }{ 2(t-m_{\pi}^{2} )^{2} }
 (m^{*} +m^{*}_{N^{*}} )^{4} \lbrack (m^{*}_{N^{*}} -m^{*} )^{2} -t \rbrack 
 ^{2} \nonumber \\
 && + \frac{2 ({\rm g}_{N N^{*}}^{\sigma})^{2} ({\rm g}_{N N^{*}}^{\omega})
 ^{2} }{ (t-m_{\sigma}^{2}) (t-m_{\omega}^{2} ) } \lbrack (m^{*}_{N^{*}}
 +m^{*} )^{2} (2m^{*}m^{*}_{N^{*}} -s ) -2m^{*}m^{*}_{N^{*}} t \rbrack , \\
 E(s,t) & = & - \frac{({\rm g}_{N N^{*}}^{\sigma})^{4} }{ 8(t-m_{\sigma}^{2})
 (u-m_{\sigma}^{2} ) }
 \lbrack (m^{*2}_{N^{*}} -m^{*2} )^{2} +(m^{*}_{N^{*}} +m^{*})^{2}s
 +t(s+t-2m^{*2}-2m^{*2}_{N^{*}} ) \rbrack \nonumber \\
  &  & + \frac{({\rm g}_{N N^{*}}^{\omega})^{4} }{ 2(t-m_{\omega}^{2})
 (u-m_{\omega}^{2} ) }
 (12m^{*2}m^{*2}_{N^{*}} -3m^{*2}s -2m^{*}m^{*}_{N^{*}}s -3m^{*2}_{N^{*}}s
 +s^{2} ) \nonumber \\
  &  & + \frac{3({\rm g}_{N N^{*}}^{\pi})^{4} }{ 8(t-m_{\pi}^{2})
 (u-m_{\pi}^{2} ) } (m^{*} +m^{*}_{N^{*}} )^{4} \lbrack (m^{*2}_{N^{*}}
 -m^{*2} )^{2} +(m^{*}_{N^{*}} -m^{*})^{2}s
 \nonumber \\ && +t(s+t -2m^{*2}-2m^{*2}_{N^{*}} ) \rbrack 
  + \frac{({\rm g}_{N N^{*}}^{\sigma})^{2}({\rm g}_{N N^{*}}^{\omega})^{2}}
 {4(t-m_{\sigma}^{2})(u-m_{\omega}^{2} ) }
 \lbrack (m^{*2}_{N^{*}} -m^{*2} )^{2}+(m^{*}_{N^{*}} +m^{*} )^{2} \nonumber \\
 && \times (6m^{*}m^{*}_{N^{*}} -s -2t) +t(t-2m^{*}m^{*}_{N^{*}} ) \rbrack 
  + \frac{({\rm g}_{N N^{*}}^{\sigma})^{2}({\rm g}_{N N^{*}}^{\omega})^{2}}
 {4(u-m_{\sigma}^{2})(t-m_{\omega}^{2} ) }
 \lbrack (m^{*2}_{N^{*}} -m^{*2} )^{2}\nonumber \\ &&-(m^{*}_{N^{*}} -m^{*})^{2}
 (6m^{*}m^{*}_{N^{*}} +3s +2t) +2m^{*}m^{*}_{N^{*}} (t-s) +(s+t)^{2} \rbrack
 \nonumber \\
 && + \frac{3({\rm g}_{N N^{*}}^{\sigma})^{2}({\rm g}_{N N^{*}}^{\pi})^{2}}
 {8(t-m_{\sigma}^{2})(u-m_{\pi}^{2} ) }
 (m^{*}+m^{*}_{N^{*}} )^{2} \lbrack (m^{*}_{N^{*}} +m^{*})^{2} -s-t \rbrack
 \lbrack (m^{*}_{N^{*}} +m^{*} )^{2} -t \rbrack \nonumber \\
 && + \frac{3({\rm g}_{N N^{*}}^{\sigma})^{2}({\rm g}_{N N^{*}}^{\pi})^{2}}
 {8(u-m_{\sigma}^{2})(t-m_{\pi}^{2} ) }
 (m^{*}+m^{*}_{N^{*}} )^{2} \lbrack (m^{*}_{N^{*}} -m^{*})^{2} -s-t \rbrack
 \lbrack (m^{*}_{N^{*}} -m^{*} )^{2} -t \rbrack \nonumber \\
 && + \frac{3({\rm g}_{N N^{*}}^{\omega})^{2}({\rm g}_{N N^{*}}^{\pi})^{2}}
 {4(t-m_{\omega}^{2})(u-m_{\pi}^{2} ) }
 (m^{*} +m^{*}_{N^{*}} )^{2} \lbrack (m^{*2}_{N^{*}} +m^{*2} )^{2}
 -(m^{*}_{N^{*}} +m^{*} )^{2}(s+2t) \nonumber \\ && 
  +2m^{*}m^{*}_{N^{*}}(m^{*2}+m^{*2}_{N^{*}}
 -s +t) +(s+t)^{2} \rbrack 
  + \frac{3({\rm g}_{N N^{*}}^{\omega})^{2}({\rm g}_{N N^{*}}^{\pi})^{2}}
 {4(u-m_{\omega}^{2})(t-m_{\pi}^{2} ) }
 (m^{*} +m^{*}_{N^{*}} )^{2} \nonumber \\ && \times  \lbrack (m^{*2}_{N^{*}} +m^{*2} )^{2}
+(m^{*}_{N^{*}} -m^{*} )^{2}(s-2t)  -2m^{*}m^{*}_{N^{*}} (m^{*2}
 + m^{*2}_{N^{*}} +t )+t^{2} \rbrack ,
 \end{eqnarray}
where
 \begin{eqnarray}
&&t=\frac{1}{2}( 2m^{*2} + 2m^{*2}_{N^{*}}-s)
+2\mid {\bf p} \mid \mid {\bf p}_{3} \mid
 \cos \theta, \\
&&u= 2m^{*2} +2m^{*2}_{N^{*}} -s-t,\\
&& \mid {\bf p} \mid =\frac{1}{2}\sqrt{s-4m^{*2}_{N^{*}} } , \\
&& \mid {\bf p}_{3} \mid = \frac{1}{2} \sqrt{s-4m^{*2}}.
 \end{eqnarray}
 (h) Differential cross section of in-medium $N^{*}N^{*} \longrightarrow N 
   \Delta $ scattering:
  \begin{equation}
\sigma_{N^{*}N^{*}  \rightarrow N\Delta }(s,t) = \frac{1}{(2 \pi)^{2} s} 
  \left[ \frac{(s-m^{*2}-m^{*2}_{\Delta})^{2}-4m^{*2}m^{*2}_{\Delta} }
  {s(s-4m^{*2}_{N^{*}} )} \right] ^{1/2}
  \lbrack D(s,t)+ E(s,t) + (s, t \longleftrightarrow u) \rbrack, 
  \end{equation}
 \begin{eqnarray}
 D(s,t) & = & \frac{({\rm g}_{N N^{*}}^{\pi})^{2} ({\rm g}_{\Delta N^{*}}^{\pi})
 ^{2} }{6m^{*2}_{\Delta} (t-m_{\pi}^{2} )^{2} }
 (m^{*} +m^{*}_{N^{*}} )^{2} \lbrack (m^{*}_{N^{*}} -m^{*} )^{2} -t \rbrack 
 \nonumber \\
 && \lbrack (m^{*}_{N^{*}} +m^{*}_{\Delta})^{2} -t \rbrack  ^{2}
 \lbrack (m^{*}_{N^{*}} -m^{*}_{\Delta} )^{2} -t \rbrack, \\
 E(s,t) & = & \frac{ ({\rm g}_{N N^{*}} ^{\pi})^{2}({\rm g}_{\Delta N^{*}}
 ^{\pi} )^{2} }{ 12m^{*2}_{\Delta} (t-m_{\pi}^{2} ) ( u-m_{\pi}^{2} ) }
 (m^{*} +m^{*}_{N^{*}} )^{2} \sum_{i=1}^{10} E_{i}, 
 \end{eqnarray}
 \begin{eqnarray}
 E_{1} & = & m^{*2}_{\Delta} (6m^{*4}_{N^{*}}t +m^{*2}_{N^{*}}s^{2}
 -8m^{*2}_{N^{*}}st +2s^{2}t -2t^{3} ) \nonumber \\
 && + m^{*}_{\Delta}m^{*}_{N^{*}} (3m^{*4}_{N^{*}}s -m^{*2}_{N^{*}}s^{2}
 -6m^{*2}_{N^{*}}st +3s^{2}t +3st^{2} ) , \\          
 E_{2} & = & m^{*2}_{N^{*}} (m^{*6}_{N^{*}} -4m^{*4}_{N^{*}}t -m^{*2}_{N^{*}}
 s^{2} +2m^{*2}_{N^{*}}st +6m^{*2}_{N^{*}}t^{2} -4st^{2} -4t^{3} )
  \nonumber \\ && +t^{2}(s+t)^{2} , \\
 E_{3} & = & 2m^{*2}t^{2} (2m^{*2}_{N^{*}} -s -t) +m^{*}m^{*4}_{\Delta}
 m^{*}_{N^{*}}(-m^{*}_{\Delta}m^{*}_{N^{*}} +3m^{*2}_{N^{*}} -s +t)
 \nonumber \\ &&  +m^{*}m^{*3}_{\Delta}s (4m^{*2}_{N^{*}} -s +2t) , \\
 E_{4} & = & m^{*}m^{*2}_{\Delta}m^{*3}_{N^{*}} (-m^{*2}_{N^{*}} +4s +2t)
 -m^{*}m^{*2}_{\Delta}m^{*}_{N^{*}}(s^{2}+t^{2}) \nonumber \\
 && +m^{*}m^{*}_{\Delta}(-2m^{*4}_{N^{*}}s +m^{*2}_{N^{*}}s^{2}
 +4m^{*2}_{N^{*}}st -2s^{2}t -2st^{2} ) , \\        
 E_{5} & = & m^{*}m^{*}_{N^{*}}(-m^{*4}_{N^{*}}s +m^{*2}_{N^{*}}s^{2}
 +2m^{*2}_{N^{*}}st -s^{2}t -st^{2} ) \nonumber \\
 && + m^{*5}_{\Delta}m^{*}_{N^{*}} (m^{*}_{\Delta}m^{*}_{N^{*}} +5m^{*2}
 _{N^{*}} -s +t) , \\
 E_{6} & = & m^{*4}_{\Delta}(3m^{*4}_{N^{*}} -2m^{*2}_{N^{*}}s +4m^{*2}
 _{N^{*}}t -2st +t^{2} ) +m^{*2}_{\Delta}m^{*}_{N^{*}}(-m^{*}_{\Delta}
 m^{*4}_{N^{*}} -4m^{*}_{\Delta}m^{*2}_{N^{*}}s \nonumber \\
 && +2m^{*}_{\Delta}m^{*2}_{N^{*}}t +m^{*}_{\Delta}s^{2} -4m^{*}_{\Delta}st
 -m^{*}_{\Delta}t^{2} -4m^{*5}_{N^{*}} ) , \\
 E_{7} & = & m^{*5}m^{*}_{N^{*}} (m^{*2}_{\Delta} -m^{*}_{\Delta}m^{*}_{N^{*}}
 +m^{*2}_{N^{*}} -t ) +m^{*4}(m^{*4}_{\Delta} +m^{*3}_{\Delta}m^{*}_{N^{*}}
 \nonumber \\
 && + m^{*2}_{\Delta} m^{*2}_{N^{*}} -2m^{*2}_{\Delta}t +3m^{*}_{\Delta}
 m^{*3}_{N^{*}} -m^{*}_{\Delta}m^{*}_{N^{*}}t -m^{*4}_{N^{*}} +t^{2} ), \\
 E_{8} & = & m^{*3} (-m^{*4}_{\Delta}m^{*}_{N^{*}} +2m^{*3}_{\Delta}
 m^{*2}_{N^{*}} -2m^{*3}_{\Delta}s -4m^{*2}_{\Delta}m^{*3}_{N^{*}}
 +2m^{*}_{\Delta}st \nonumber \\  
 && + m^{*5}_{N^{*}} -2m^{*3}_{N^{*}}s -2m^{*3}_{N^{*}}t +2m^{*}_{N^{*}}st
 + m^{*}_{N^{*}}t^{2} ) , \\
E_{9} & = & m^{*2}m^{*4}_{\Delta} (-m^{*}_{\Delta}m^{*}_{N^{*}} -6m^{*2}_{N^{*}}
 +2s -2t ) +m^{*2}m^{*3}_{\Delta} (-8m^{*3}_{N^{*}} +4m^{*}_{N^{*}} s ) 
 \nonumber \\
 && +2m^{*2}m^{*2}_{\Delta} (2m^{*4}_{N^{*}} -m^{*2}_{N^{*}}s +2t^{2} ) , \\
 E_{10} & = & m^{*2}m^{*}_{N^{*}} ( m^{*}_{\Delta}m^{*4}_{N^{*}}
 -2m^{*}_{\Delta}m^{*2}_{N^{*}}s -2m^{*}_{\Delta}m^{*2}_{N^{*}}t
 -2m^{*}_{\Delta}st +m^{*}_{\Delta}t^{2}  \nonumber \\                         
 && +2m^{*3}_{N^{*}}s -2m^{*3}_{N^{*}}t ),
 \end{eqnarray}
where
 \begin{eqnarray}
&&t=\frac{1}{2}( 2m^{*2}_{N^{*}} + m^{*2} +m^{*2}_{\Delta}-s)
+2\mid {\bf p} \mid \mid {\bf p}_{3} \mid
 \cos \theta, \\
&&u= 2m^{*2}_{N^{*}} +m^{*2} + m^{*2}_{\Delta} -s-t,\\
&& \mid {\bf p} \mid =\frac{1}{2}\sqrt{s-4m^{*2}_{N^{*}} } , \\
&& \mid {\bf p}_{3} \mid =\frac{1}{2 \sqrt{s}} \sqrt{(s-m^{*2} -m^{*2}_{\Delta})
 ^{2} -4m^{*2}m^{*2}_{\Delta} }
 \end{eqnarray}
 
 \newpage
  \vspace{3cm}
  \begin{center}
 {\bf TABLES}
 \end{center}
 \vspace{2cm}
{\bf TABLE I}: Some symbols and notation used in this paper, k$_{\mu}$ is the 
 transformed four-momentum.
     \vspace{-0.5cm}
    \begin{center}
    \tabcolsep 0.10in
  \begin{tabular}{|c|c|c|c|c|c|c|c|c|c|c|c|}
      \hline
 {\rm A} & m$_{A}$ & g$^{A}_{NN}$ & g$^{A}_{N^{*}N^{*}}$ & 
 g$^{A}_{\Delta \Delta}$ & g$^{A}_{NN^{*}}$ & $\gamma_{A} $ & $\tau_{A}$ & $ T_{A} $ &
  $\Phi_{A} (x) $ & D$_{A}^{\mu}$ & D$_{A}^{i}$ \\
 \hline
 $\sigma$ & m$_{\sigma}$ & g$^{\sigma}_{NN}$ & g$^{\sigma}_{N^{*}N^{*}}$ &
 g$_{\Delta \Delta}^{\sigma}$ & g$^{\sigma}_{NN^{*}}$ & 1 & 1 & 1 & $\sigma(x)$ &
 1 & 1   \\
 \hline
$\omega$ & m$_{\omega}$ & $-$ g$^{\omega}_{NN} $ & $-$ g$^{\omega}_{N^{*}N^{*}}$ & 
 $-$ g$_{\Delta \Delta}^{\omega}$ & $-$ g$^{\omega}_{NN^{*}}$ &  $\gamma_{\mu} $ &
 1 & 1 & $\omega^{\mu}(x)$ & $-$ g$^{\mu \nu}$ & 1 \\
 \hline
 $\pi$ & m$_{\pi}$ & g$_{NN}^{\pi}$& g$^{\pi}_{N^{*}N^{*}}$ & 
 g$_{\Delta \Delta}^{\pi}$ & g$^{\pi}_{NN^{*}} $ & 
 $\not\! k \gamma_{5} $
& \mbox{\boldmath $\tau$} &${\bf T}$
 & \mbox{\boldmath $\pi$}(x) & 1 & $\delta_{ij}$ \\
       \hline
      \end{tabular}
          \end{center}
   \vspace{1.0cm}
{\bf TABLE II}: Isospin factors for the direct term of $N^{*}N \rightarrow
N^{*}N$, $N^{*}N \rightarrow N N$ and $N^{*}N^{*} \rightarrow NN$ reactions.
    \vspace{0.5cm}
    \begin{center}
    \tabcolsep 0.15in
  \begin{tabular}{|c|c|c|c|}
      \hline
 T$_{D}^{AB}$ & $\sigma$ & $\omega$ & $\pi$  \\
 \hline
 $\sigma$  & 2  &  2  &  0  \\
 \hline
 $\omega$  &  2   &  2  &  0  \\
 \hline
  $\pi$    &  0   &  0  &  6   \\
       \hline
      \end{tabular}
          \end{center}
   \vspace{1.0cm}
{\bf TABLE III}: Isospin factors for the exchange term of $N^{*}N \rightarrow
N^{*}N$, $N^{*}N \rightarrow N N$ and $N^{*}N^{*} \rightarrow NN$ reactions.
    \vspace{0.5cm}
    \begin{center}
    \tabcolsep 0.15in
  \begin{tabular}{|c|c|c|c|}
      \hline
 T$_{E}^{AB}$ & $\sigma$ & $\omega$ & $\pi$  \\
 \hline
 $\sigma$  & 1  &  1  &  3  \\
 \hline
 $\omega$  &  1   &  1  &  3  \\
 \hline
  $\pi$    &  3   &  3  &  $-$ 3  \\
       \hline
      \end{tabular}
          \end{center}
   \vspace{1.0cm}
{\bf TABLE IV}: Isospin factors for the direct term of $N^{*}\Delta \rightarrow
N\Delta$ reaction (corresponding to Feynman diagram (j1) in Appendix A).
    \vspace{0.5cm}
    \begin{center}
    \tabcolsep 0.15in
  \begin{tabular}{|c|c|c|c|}
      \hline
 T$_{D}^{AB}$ & $\sigma$ & $\omega$ & $\pi$  \\
 \hline
 $\sigma$  & 4  &  4  &  0  \\
 \hline
 $\omega$  &  4   &  4  &  0  \\
 \hline
  $\pi$    &  0   &  0  &  15  \\
       \hline
      \end{tabular}
          \end{center}
     \vspace{1.0cm}
 {\bf TABLE V}: Mean field parameters and the corresponding
 nuclear saturation properties. The sixth set is the TM1 parameters from
Ref. \cite{Sug94}.
   \vspace{-0.5cm}
           \begin{center}
    \tabcolsep 0.10 in
\begin{tabular}{|c|c|c|c|c|c|c|c|c|c|}
      \hline
       & g$^{\sigma}_{NN}$ & g$^{\omega}_{NN}$ & b$({\rm g}^{\sigma}_{NN})^{3}$
       & c$({\rm g}^{\sigma}_{NN})^{4}$ & Z
       & E$_{\rm bin}$ & m$^{*}/M_{N}$ & K(MeV) & $\rho_{0}$ \\
      \hline
   {\rm 1} & 9.40 & 10.95 & $-$0.69 & 40.44 & $\infty$ & $-$15.57 & 0.70 & 380 & 0.145 \\
      \hline
  {\rm 2} & 6.90 & 7.54 & $-$40.49 & 383.07 & $\infty$ & $-$15.76 & 0.83 & 380 & 0.145 \\
       \hline
  {\rm 3} & 7.937 & 6.696 & 42.35 & 157.55 & $\infty$ & $-$16.00 & 0.85 & 210 & 0.153  \\
       \hline
  {\rm 4} & 12.419 & 15.063 & 12.276 & 189.008 & 1.7042  & $-$15.29 & 0.60 &
 210 & 0.185 \\  
       \hline
  {\rm 5} & 11.536  & 14.528 & $-$4.252 & 102.141 & 2.8335 & $-$15.56 & 0.60 &
 310 & 0.150 \\
       \hline
  {\rm 6} & 10.029 & 12.614 & $-$7.233 & 0.6183 & 5.9273 & $-$16.30  & 0.634 &
 281 & 0.145 \\
       \hline
      \end{tabular}
     \end{center}
 \newpage
\begin{center}
{\bf CAPTIONS}
 \end{center}
 \begin{description}
 \item[\tt Fig.1 ] Feynman diagrams contributing to the Hartree term of
the (a) nucleon, (b) delta and (c)
$N^{*}$(1440) self-energies. A dashed line denotes a meson, a double line
denotes the delta resonance, a solid and bold-solid line represent the nucleon
and the $N^{*}$(1440), respectively.
 \item[\tt Fig.2 ] The relativistic nucleon optical potential calculated at
$\rho=\rho_{0}$ as a function
of the kinetic energy. The different curves correspond to the different 
parameter sets (Table V) 
as indicated in the figure. The hatched area
shows the experimental data from Ref. \cite{Ham90}.
 \item[\tt Fig.3 ] The momentum dependence of the relativistic $\Delta$
 optical potential calculated at $\rho=\rho_{0}$. The second set of parameters
 in Table V is used as the nucleon coupling strengths. For the delta coupling
strengths several different choices are  tested and discussed in the text.
 \item[\tt Fig.4 ] The same as Fig.~3, but for the $N^{*}$(1440) optical
potential.
 \item[\tt Fig.5 ] The density dependence of the nucleon, delta and 
$N^{*}$(1440) optical potential calculated in the limit of zero-momentum. 
The parameter set2 in Table V is used as the nucleon coupling strengths.
The different choices of the delta and $N^{*}$(1440) coupling strengths are
employed in the calculations and discussed in the text.
 \item[\tt Fig.6 ] Free scattering cross section for reaction $pp \rightarrow
 pp^{*}(1440)$. The parameter set2 in Table V is used as the nucleon
coupling strengths. Solid line represents the results of this work, and dashed 
line denotes the results of Ref. \cite{Hub94}.
 The experimental data are taken from Ref. \cite{CERN}.
The unitary form factor for $NN$ and $NN^{*}$ vertex, i.e., $\Lambda^{*}_{A}
 =\Lambda_{A}$ is also tested in the calculations, which is depicted by the
 dotted line.
 \item[\tt Fig.7 ] The in-medium $NN \rightarrow NN^{*}$ cross section at 
normal density. The parameter set2 in Table V is used as the nucleon coupling
strengths. For the $\Delta$ and $N^{*}$(1440) coupling strengths we employ the
universal coupling-strength assumption.
The contributions of the direct term and exchange term are denoted by the dotted
and dashed line, respectively. The solid line gives the summation of these
two terms.
 \item[\tt Fig.8 ] The same as Fig. 7, but for an in-medium $N^{*}N \rightarrow
N^{*}N$ cross section.
 \item[\tt Fig.9 ] The contributions of different terms to the in-medium      
 $NN \rightarrow NN^{*}$ cross section. Others are the same as in Fig.~7.
 \item[\tt Fig.10 ] The same as Fig. 9, but for an in-medium $N^{*}N \rightarrow
N^{*}N$ cross section.
 \item[\tt Fig.11 ] The in-medium $NN \rightarrow NN^{*}$ cross section at
different densities and energies. The calculations are performed with parameter
set2 in Table V and different sets of $\alpha(N^{*})$ and $\beta(N^{*})$.
 \item[\tt Fig.12 ] The same as Fig.~11, but for an in-medium $N^{*} N
 \rightarrow NN$ cross section.
 \item[\tt Fig.13 ] The same as Fig.~11, but for an in-medium $N^{*} N
 \rightarrow N^{*}N$ cross section.
  \end{description}
 \newpage
 {\Large Fig. 1}
 \begin{figure}[htbp]
  \vspace{0cm}
 \hskip  1cm \psfig{file=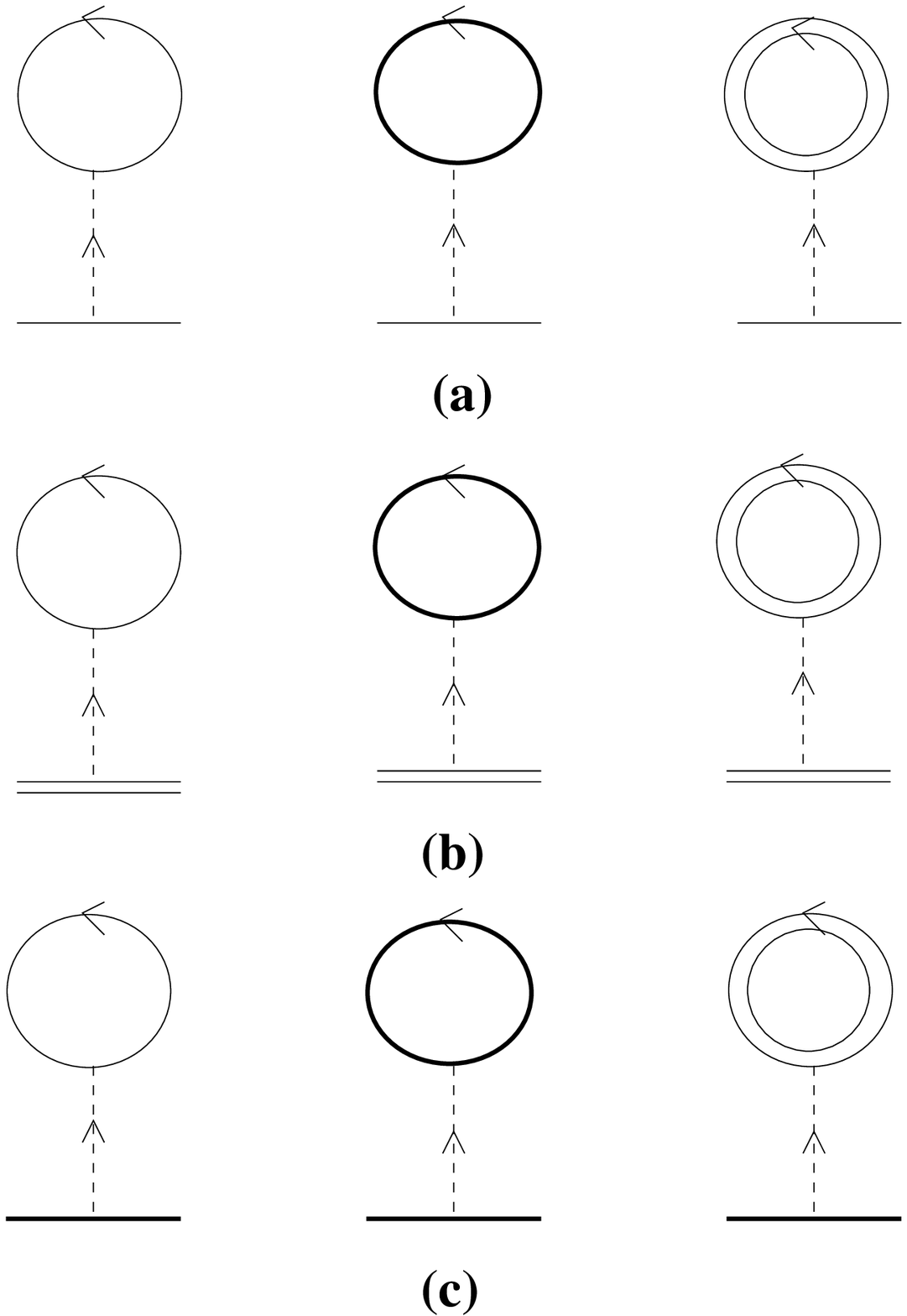,width=12.cm,height=17cm,angle=0}
\end{figure}
 \newpage
 {\Large Fig. 2}
 \begin{figure}[htbp]
  \vspace{0cm}
 \hskip  -1cm \psfig{file=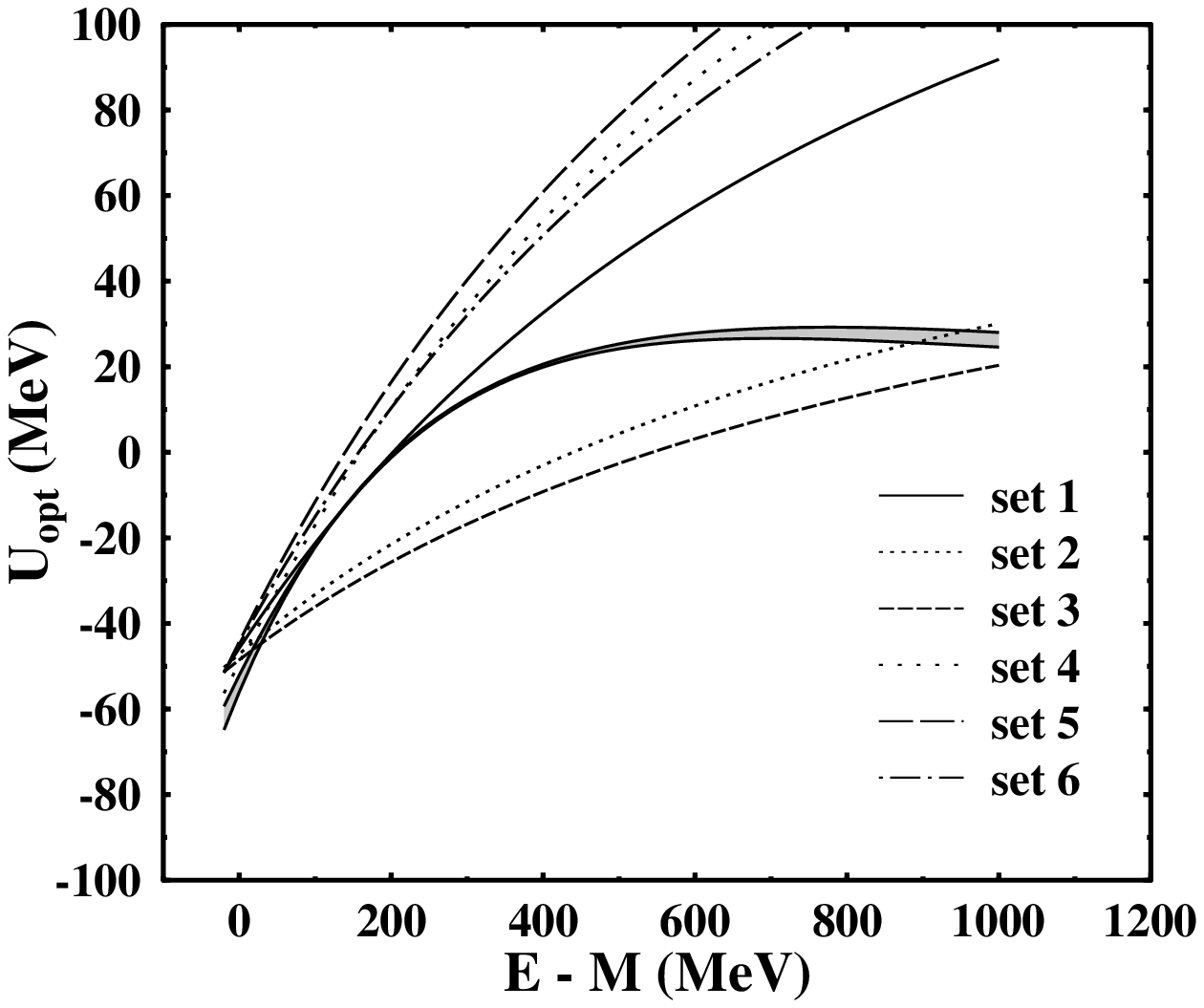,width=15.cm,height=17cm,angle=-90}
\end{figure}
 \newpage
 {\Large Fig. 3}
 \begin{figure}[htbp]
  \vspace{0cm}
 \hskip  -1cm \psfig{file=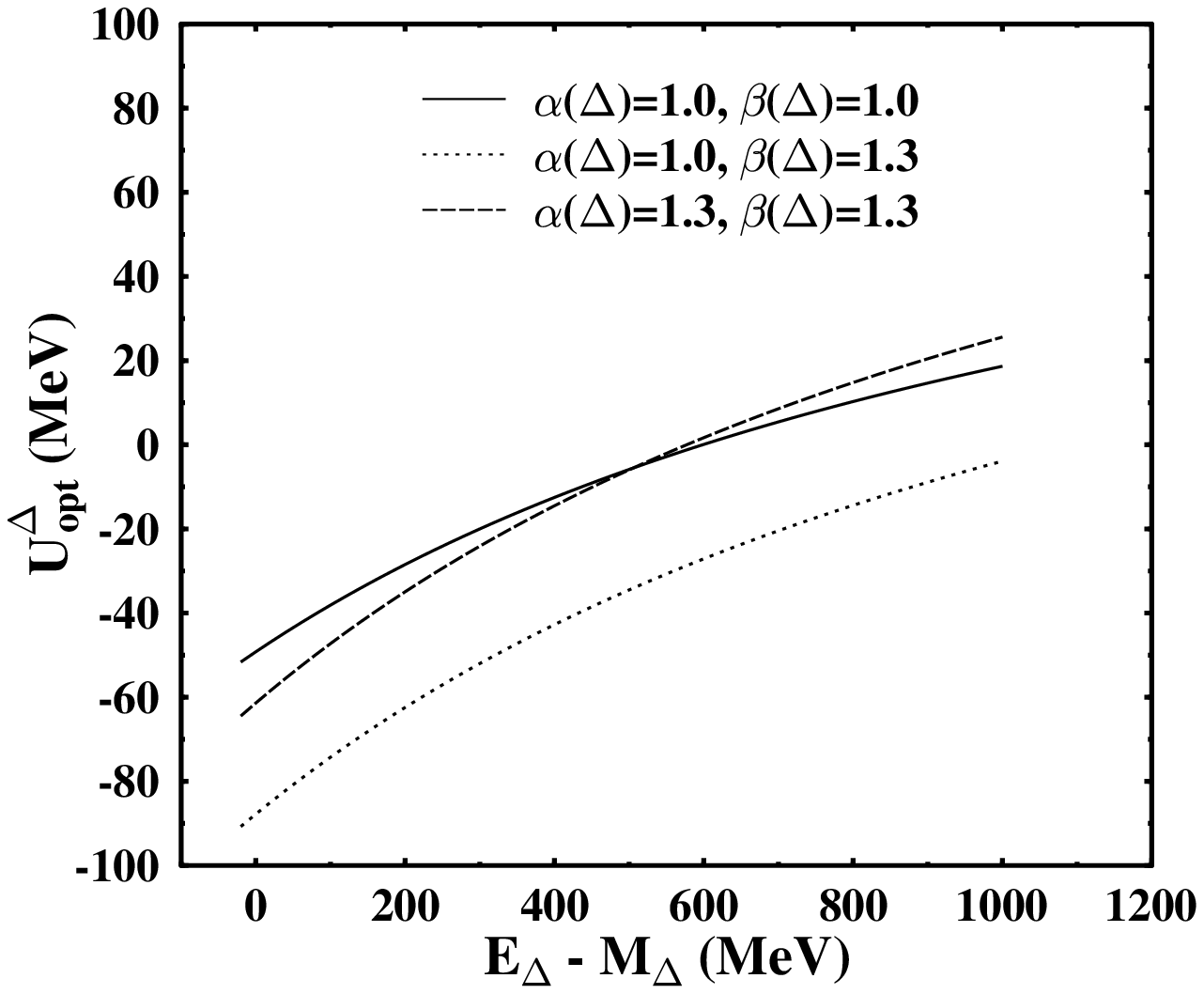,width=15.cm,height=17cm,angle=-90}
\end{figure}
 \newpage
 {\Large Fig. 4}
 \begin{figure}[htbp]
  \vspace{0cm}
 \hskip  -1cm \psfig{file=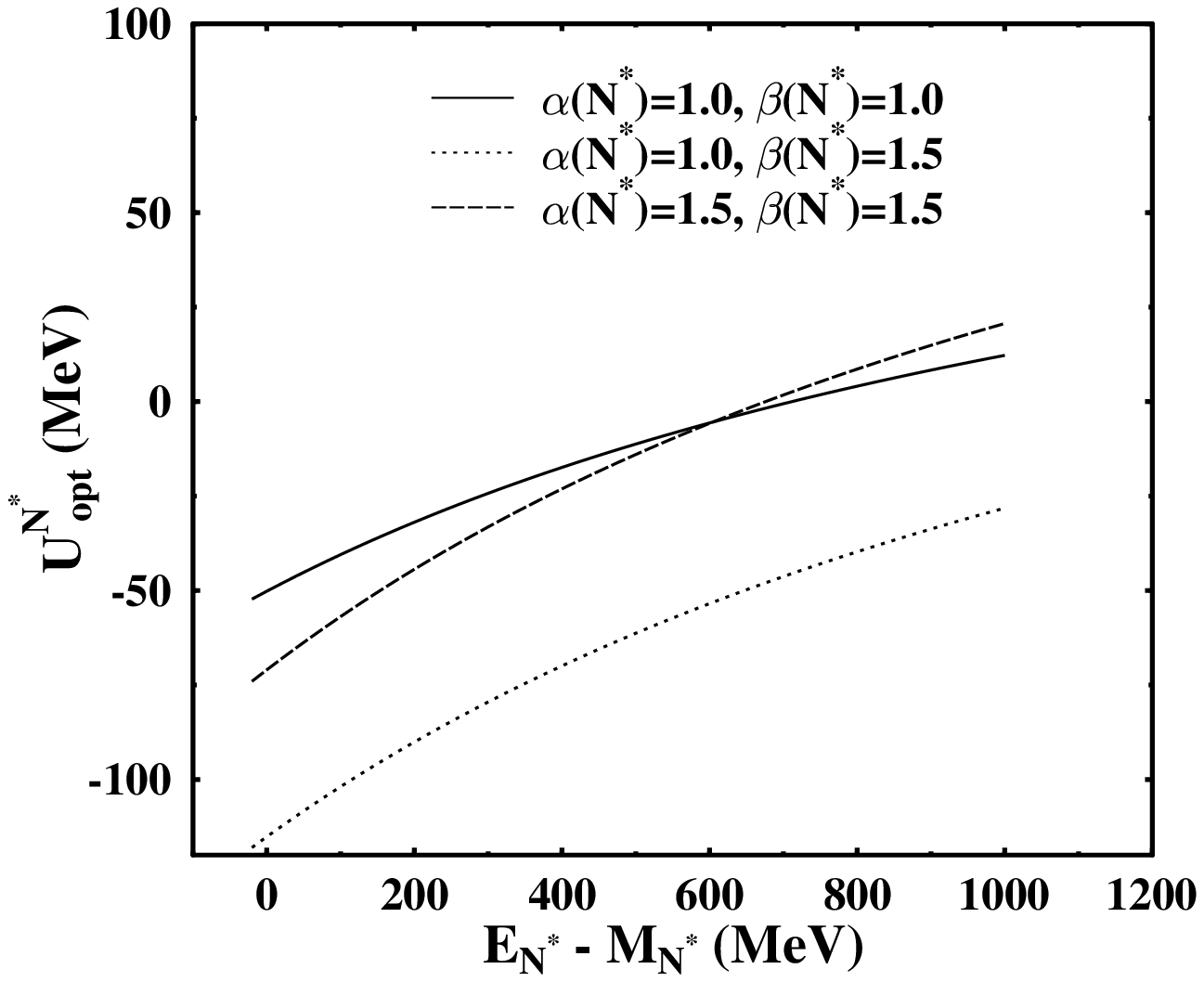,width=15.cm,height=17cm,angle=-90}
\end{figure}
 \newpage
 {\Large Fig. 5}
 \begin{figure}[htbp]
  \vspace{0cm}
 \hskip  -1cm \psfig{file=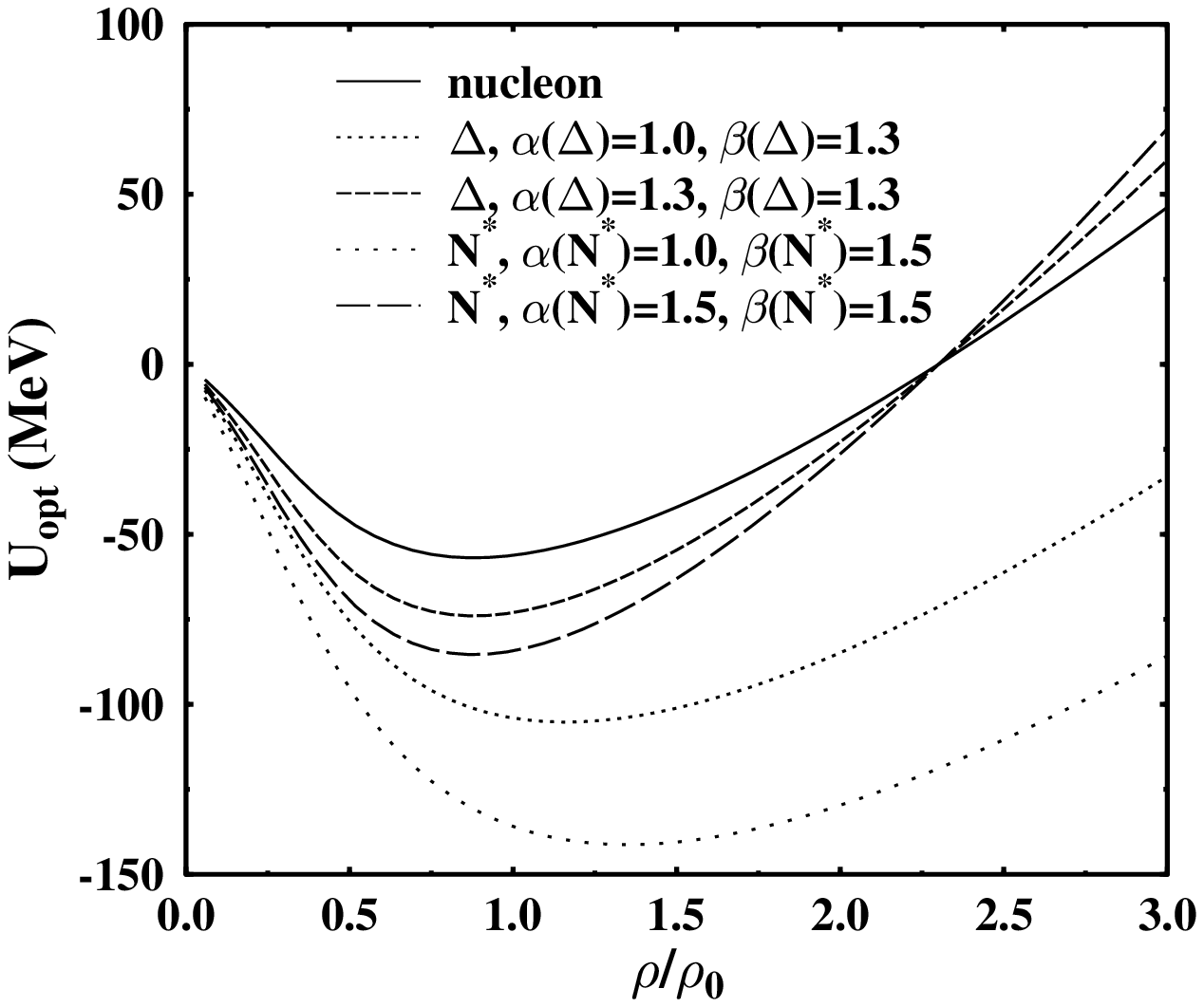,width=15.cm,height=17cm,angle=-90}
\end{figure}
 \newpage
 {\Large Fig. 6}
 \begin{figure}[htbp]
  \vspace{0cm}
 \hskip  -1cm \psfig{file=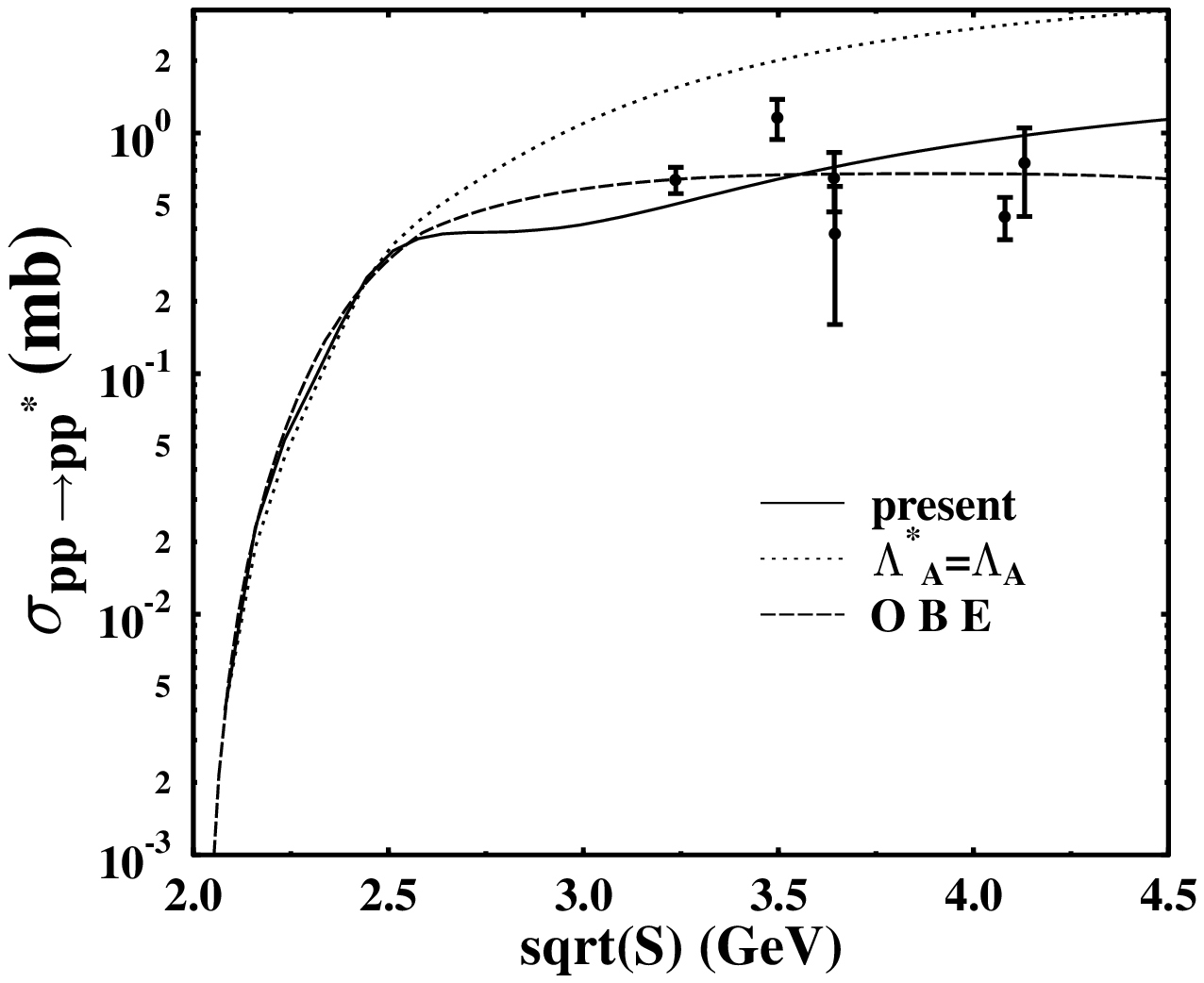,width=15.cm,height=17cm,angle=-90}
\end{figure}
 \newpage
 {\Large Fig. 7}
 \begin{figure}[htbp]
  \vspace{0cm}
 \hskip  -1cm \psfig{file=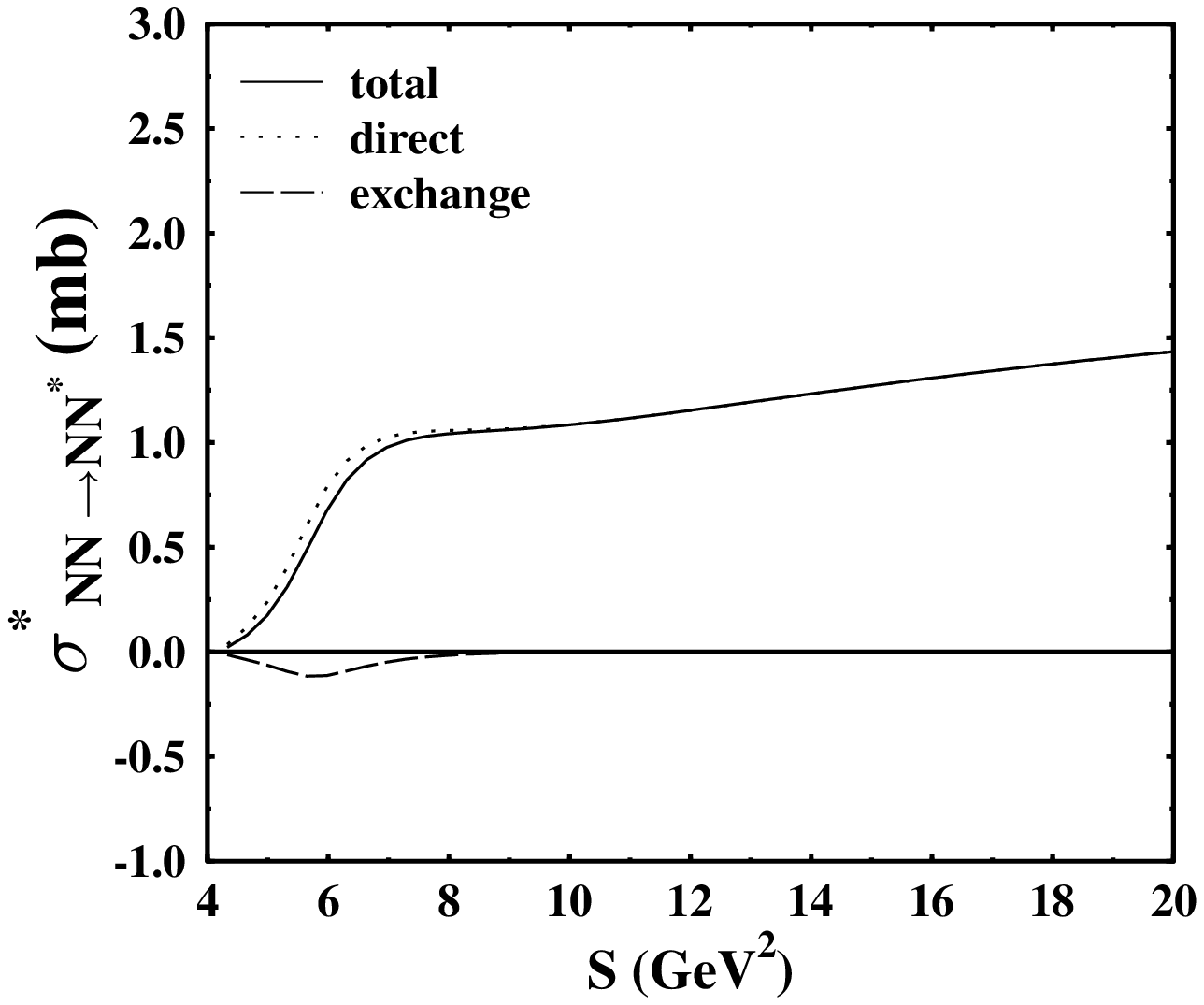,width=15.cm,height=17cm,angle=-90}
\end{figure}
 \newpage
 {\Large Fig. 8}
 \begin{figure}[htbp]
  \vspace{0cm}
 \hskip  -1cm \psfig{file=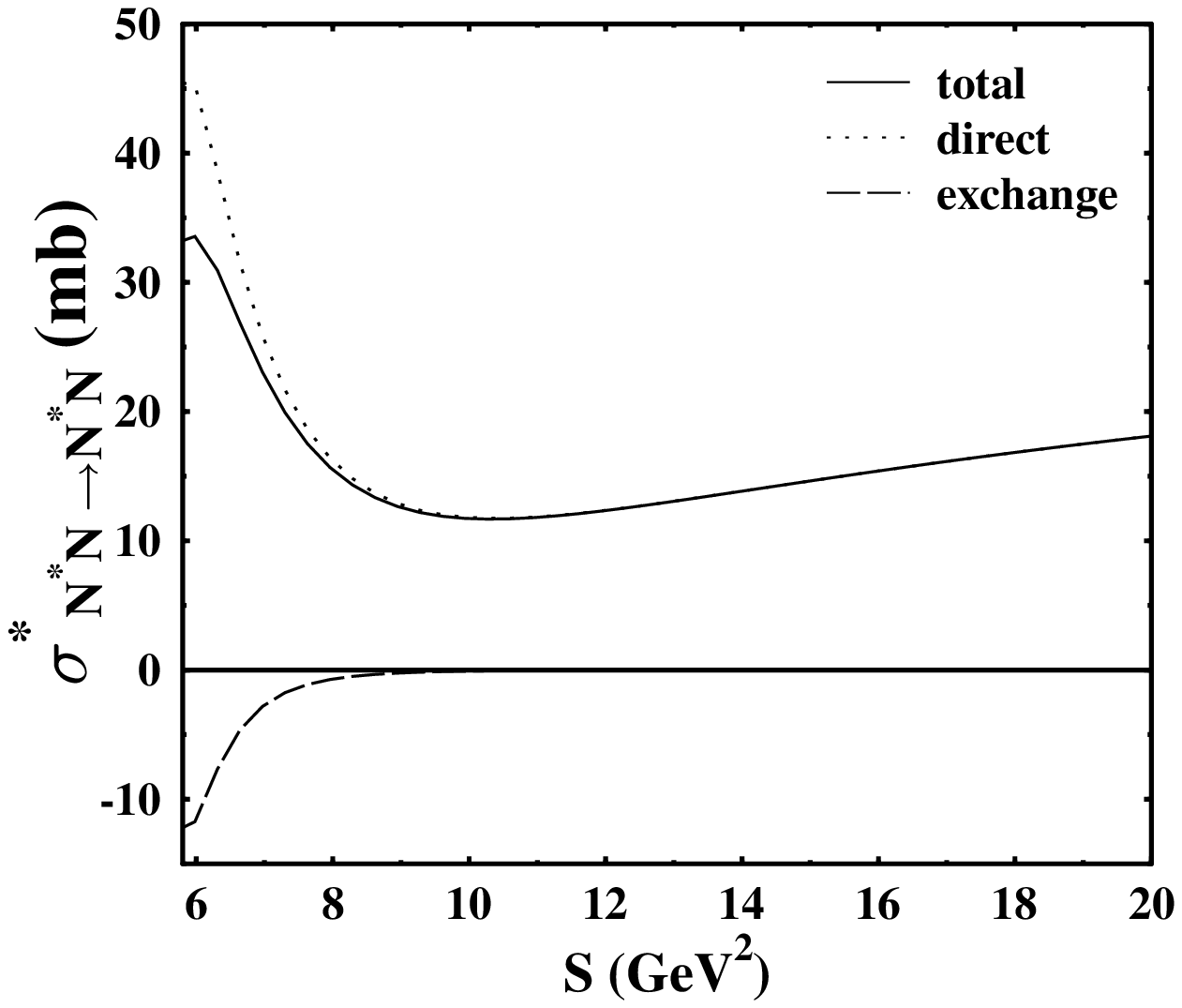,width=15.cm,height=17cm,angle=-90}
\end{figure}
 \newpage
 {\Large Fig. 9}
 \begin{figure}[htbp]
  \vspace{0cm}
 \hskip  -1cm \psfig{file=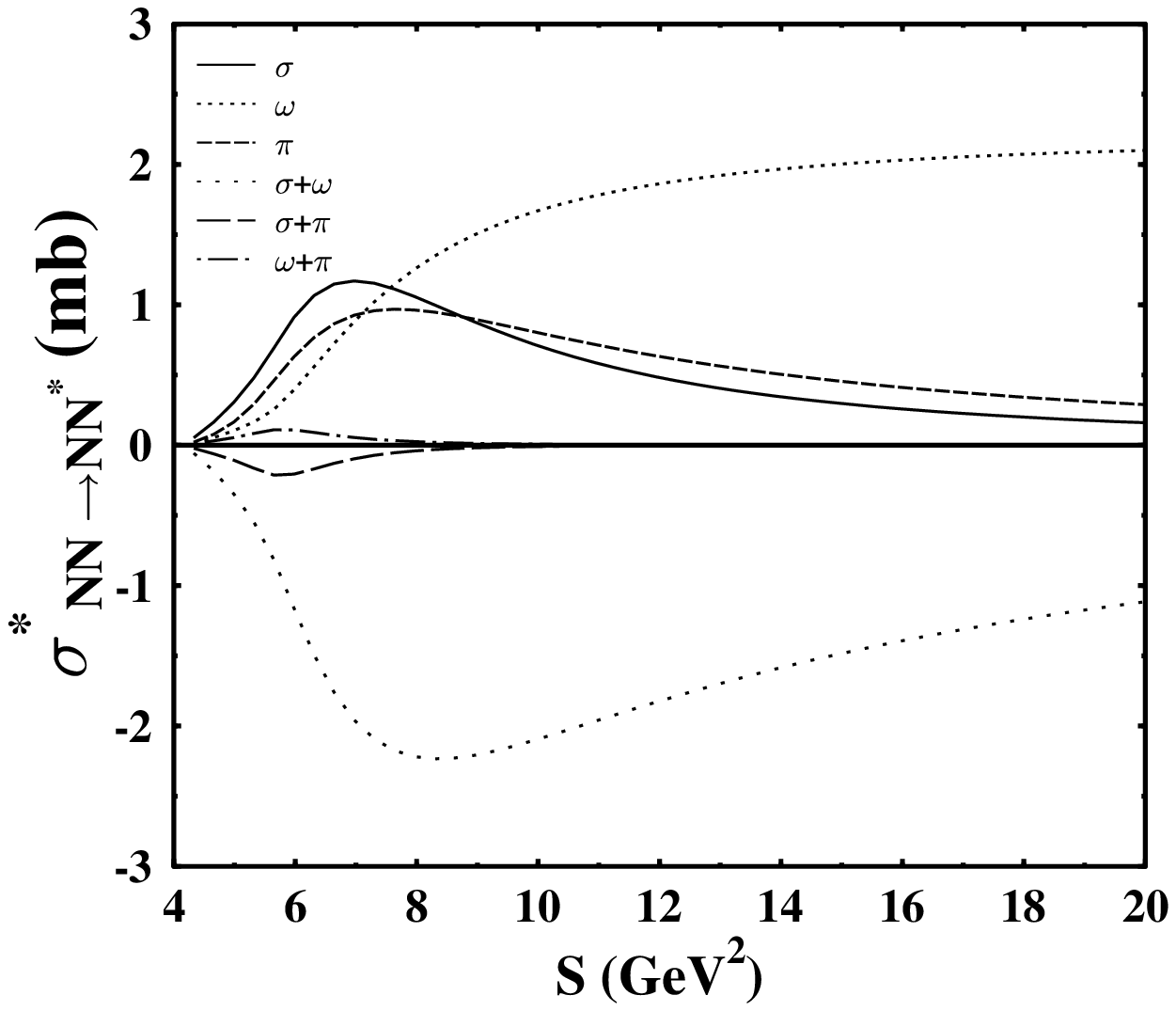,width=15.cm,height=17cm,angle=-90}
\end{figure}
 \newpage
 {\Large Fig. 10}
 \begin{figure}[htbp]
  \vspace{0cm}
 \hskip  -1cm \psfig{file=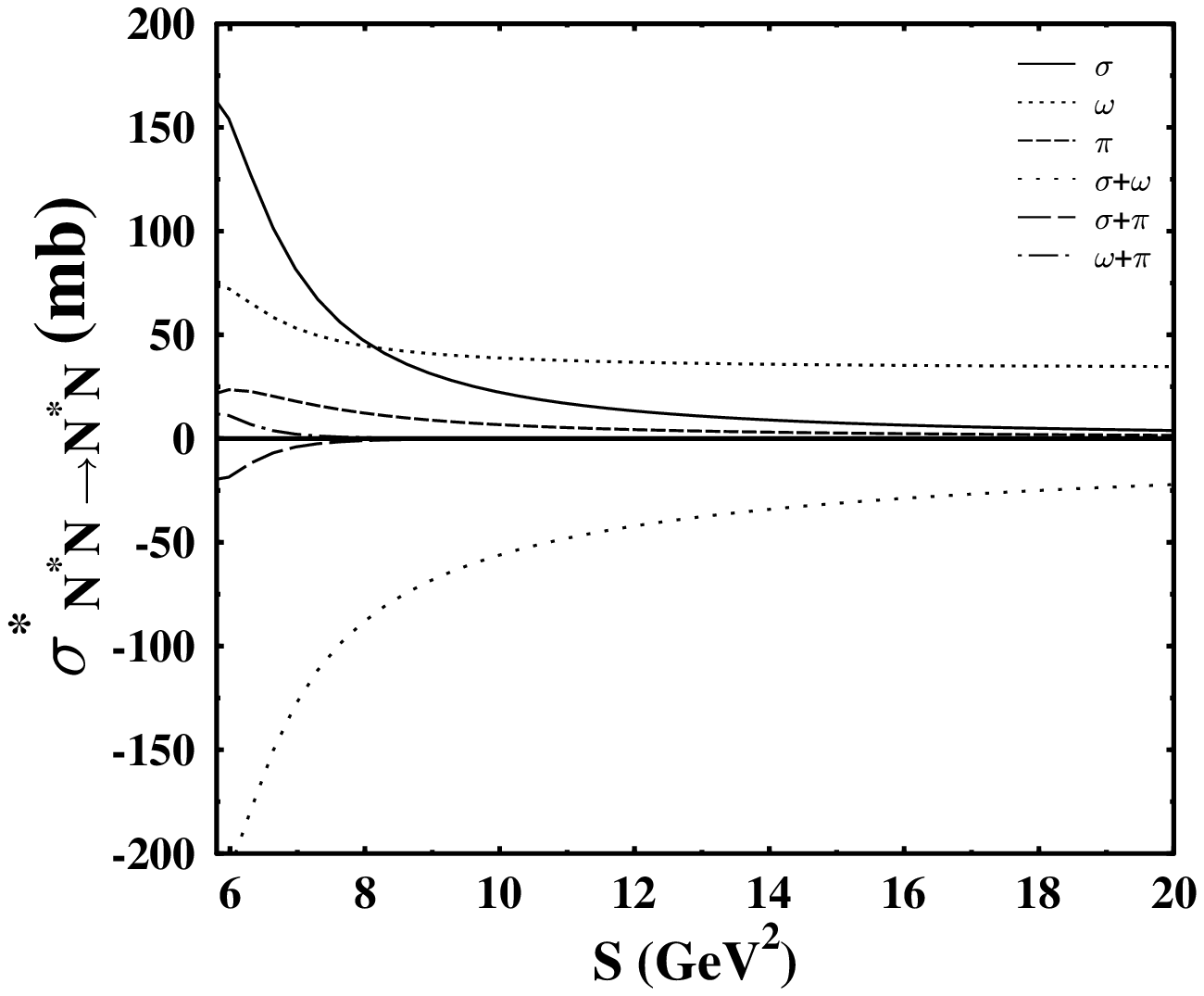,width=15.cm,height=17cm,angle=-90}
\end{figure}
  \newpage
 {\Large Fig. 11}
 \begin{figure}[htbp]
  \vspace{2cm}
 \hskip  0cm \psfig{file=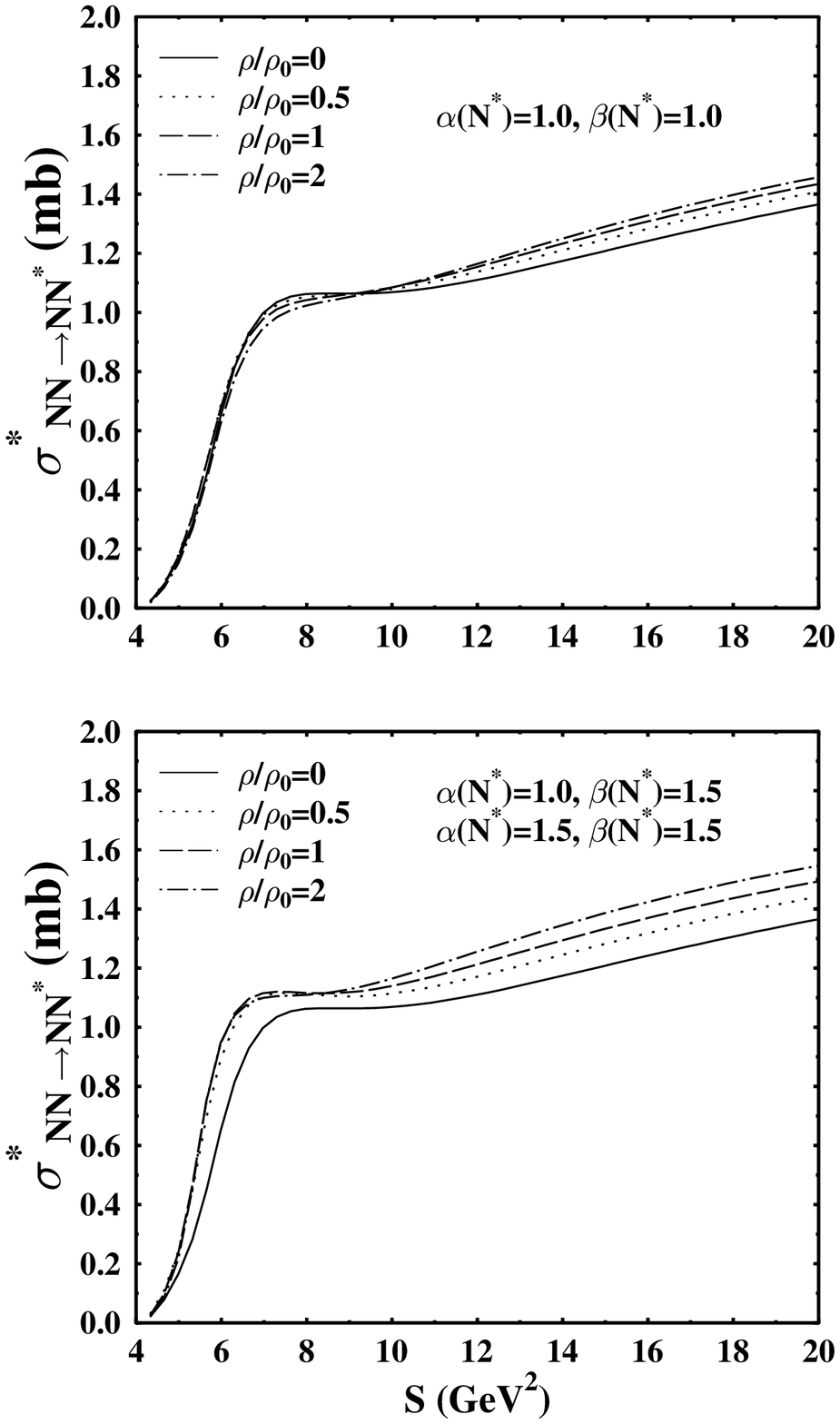,width=15.cm,height=19cm,angle=0}
  \vspace{-5cm}
\end{figure}
  \newpage
 {\Large Fig. 12}
 \begin{figure}[htbp]
  \vspace{0cm}
 \hskip  0cm \psfig{file=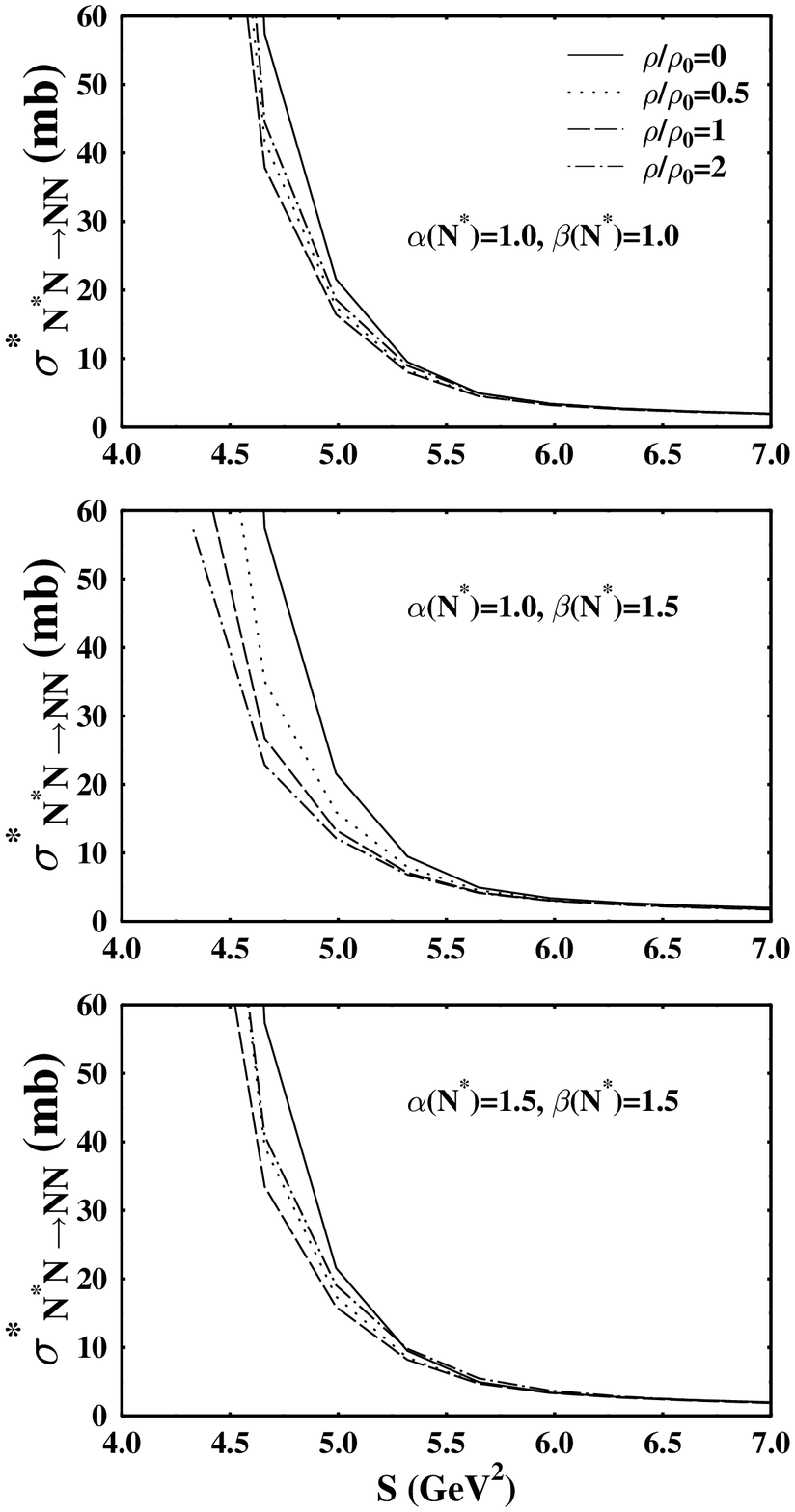,width=15.cm,height=22cm,angle=0}
  \vspace{-5cm}
\end{figure}
  \newpage
 {\Large Fig. 13}
 \begin{figure}[htbp]
  \vspace{0cm}
 \hskip  0cm \psfig{file=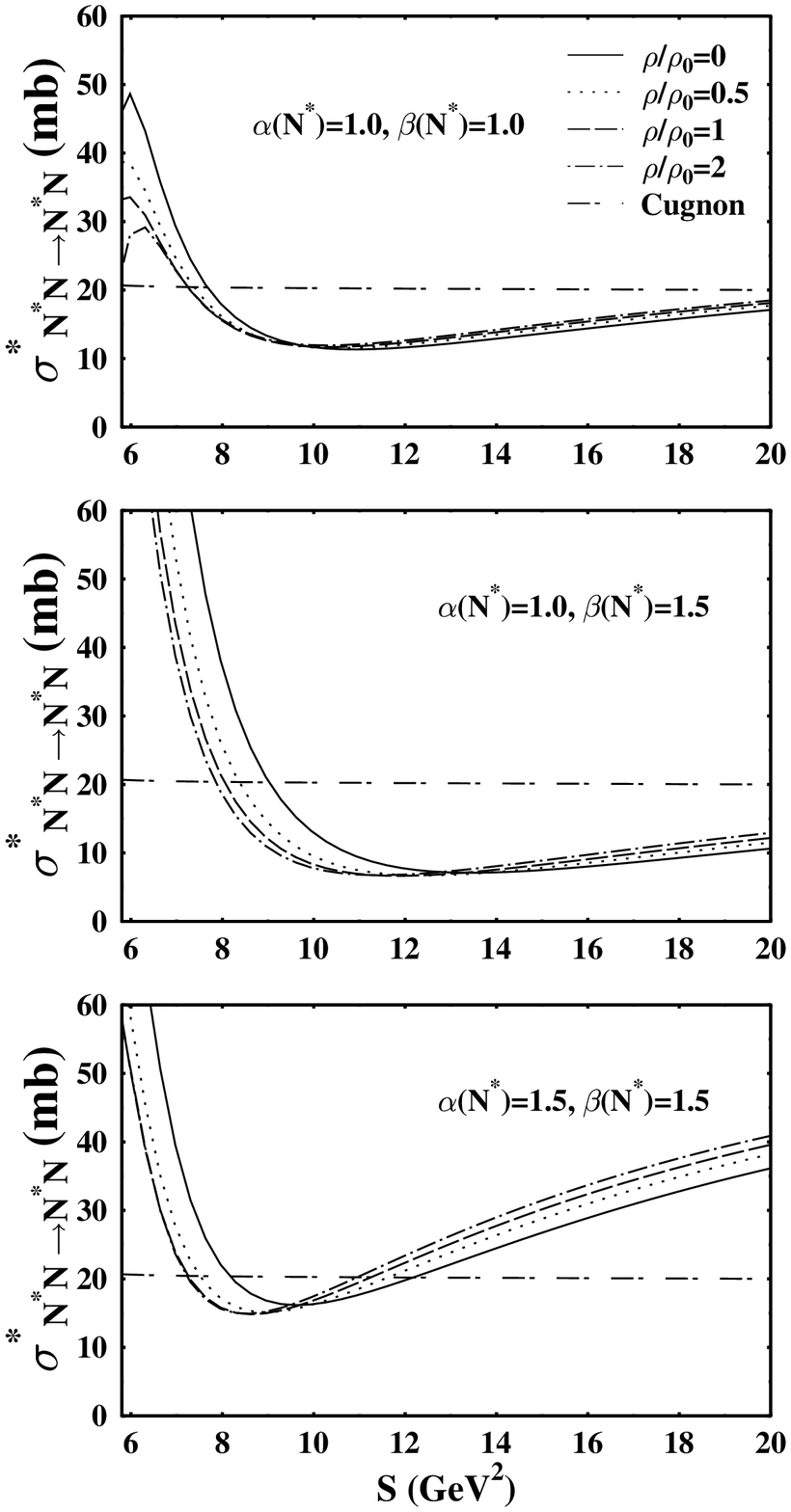,width=15.cm,height=22cm,angle=0}
  \vspace{-5cm}
\end{figure}
       \end{document}